\documentclass[]{aa}  

\usepackage{txfonts}
\usepackage[english]{babel}
\usepackage[utf8]{inputenc}

\usepackage{graphicx}
\usepackage{xcolor}
\usepackage{natbib}
\bibpunct{(}{)}{;}{a}{}{,} 

\usepackage{datetime}

\graphicspath{{figures/}}

\newcommand{\e}[1]{\ensuremath{\times 10^{#1}}} %
\newcommand{\un}[1]{\ensuremath{\ \mathrm{#1}}}

\begin{document}

\title{Thermal and non-thermal emission from reconnecting twisted coronal loops}

\author{R. F. Pinto \inst{1,2}\fnmsep\thanks{\email{rui.pinto@obspm.fr}}, 
  M. Gordovskyy\inst{3},
  P.K. Browning\inst{3},
  \and
  N. Vilmer\inst{1}}

\institute{
  LESIA, Observatoire Paris, CNRS, UPMC, Université Paris-Diderot, 5 place Jules Janssen, 92195 Meudon, France
  \and
  IRAP, Université de Toulouse, UPS-OMP, CNRS, 9 Av. Colonel Roche, BP 44346, F-31028 Toulouse Cedex 4, France
  \and
  Jodrell Bank Centre for Astrophysics, University of Manchester, Manchester M13 9PL, UK
}

\date{(\emph{manuscript version:} \today; \currenttime)}

\authorrunning{Pinto et al.}
\titlerunning{Thermal \& non-thermal emission from twisted coronal loops}

 
\abstract
{
  Twisted magnetic fields should be ubiquitous in the solar corona, particularly in flare-producing active regions where the magnetic fields are strongly non-potential.
  The magnetic energy contained in such twisted fields can be released during solar flares and other explosive phenomena.
  It has been shown recently that reconnection in helical magnetic coronal loops results in plasma heating and particle acceleration distributed within a large volume, including the lower coronal and chromospheric sections of the loops.
  Hence, the magnetic reconnection and particle acceleration scenario involving magnetic helicity can be a viable alternative to the standard flare model, where particles are accelerated only in a small volume located in the upper corona.
}
{
  The key goal of this study is to investigate the links and observational signatures of plasma heating and particle acceleration in kink-unstable twisted coronal loops.
}
{
  We use a combination of MHD simulations and test-particle methods.
  These simulations describe the development of kink instability and magnetic reconnection in twisted coronal loops using resistive compressible MHD, and incorporate atmospheric stratification and large-scale loop curvature. 
  The resulting distributions of hot plasma let us estimate thermal X-ray emission intensities.
  The electric and magnetic fields obtained are used to calculate electron trajectories using the guiding-centre approximation.
  These trajectories combined with the MHD plasma density distributions let us deduce synthetic hard X-ray bremsstrahlung intensities.
}
{
  Our simulations emphasise that the geometry of the emission patterns produced by hot plasma in flaring twisted coronal loops can differ from the actual geometry of the underlying magnetic fields.
  In particular, the twist angles revealed by the emission threads (soft X-ray thermal emission; SXR) are consistently lower than the field-line twist present at the onset of the kink-instability.
  Hard X-ray (HXR) emission due to the interaction of energetic electrons with the stratified background are concentrated at the loop foot-points in these simulations, even though the electrons are accelerated everywhere within the coronal volume of the loop.
  The maximum of HXR emission consistently precedes that of SXR emission, with the HXR light-curve being approximately proportional to the temporal derivative of the SXR light-curve. 
}
{}


\maketitle
%

\section{Introduction}\label{intro}

Solar flares are energetic phenomena commonly understood as fast releases of magnetic energy stored in the solar corona.
The coronal plasma heated during such events produces characteristic emission signatures in the soft X-ray range, and the particles accelerated in the flaring regions produce non-thermal emission tipically in the hard X-ray energy range \citep[see, \emph{e.g.}, the review by][]{benz_flare_2008,fletcher_observational_2011}.
Twisted magnetic flux-ropes can store significant amounts of magnetic free-energy in the solar corona, and are susceptible of developing magneto-hydrodynamical instabilities which lead to the release of this energy.
Coronal loops undergoing a kink instability, in particular, go through an initial exponential growth phase (the linear phase of the instability), until they start reconnecting with the background field \citep{browning_heating_2008}.
They then relax into a state with lower magnetic energy and less twist (hence releasing a fraction of the magnetic free energy stored initially).
It has been shown recently that magnetic reconnection in helical (twisted) magnetic structures may result in impulsive plasma heating capable of producing the main properties of thermal X-ray emission in solar flares \citep{pinto_soft_2015}, and in particle acceleration distributed within large volumes of plasma (including the lower corona and the chromosphere) leading to non-thermal X-ray emission  \citep{gordovskyy_effect_2012,m._gordovskyy_particle_2013,gordovskyy_particle_2014}.
The kink instability, therefore, may be a viable alternative to the standard flare model, which assumes that particles accelerated to high energies in small volumes of plasma near the flaring loop apexes are transported downwards along the loop legs, and that these interact with the denser chromosphere producing hard X-ray emission there and providing hot plasma which would thereafter fill up the loops and emit thermally in soft X-rays.
One long-standing difficulty the kink instability scenario faces is that its triggering requires that the pre-flare loops are strongly twisted \citep[see, \emph{e.g}, discussion in][]{bareford_coronal_2013}, which observations of flaring coronal loops rarely seem to indicate (for both confined and ejective flares).
Notable exceptions exist, however, such as reported by \citet{srivastava_observation_2010}, who measured a total twist angle of about $12\pi$ on a twisted coronal flux-rope before the occurrence of a B5.0 class flare \citep[see also simulations of this event by][]{botha_observational_2012}.
\citet{pinto_soft_2015} pointed out that, however, the morphology of the thermal emission during the initial phases of a flare may lead to an important underestimation of the actual twist of the underlying magnetic field lines.
Furthermore, the kink-instability scenario, along with other scenarios resulting in distributed acceleration/re-acceleration \citep{turkmani_particle_2006,cargill_particle_2006} may allow one to circumvent the so-called number problem \citep{brown_interpretation_1976,brown_local_2009}, one of the main difficulties of the standard flare model.

In the present study, we focus on deducing observational signatures inherent to the triggering of the kink instability in twisted coronal loops, combining thermal and non-thermal effects (plasma heating and particle acceleration) with application to confined flares.
We will consider here loop models consisting of twisted magnetic flux ropes embedded in a gravitationally-stratified background (comprising a hot corona, a transition region and a fraction of the chromosphere), which represent small (or moderate) confined flares.
The origin of such twisted coronal loops can be thought of as the result of photospheric source rotation or \citep[see e.g.][]{brown_observations_2003,dalmasse_photospheric_2014} or of flux-rope emergence through the photosphere  \citep[e.g.][]{luoni_twisted_2011,archontis_emergence_2013,pinto_flux_2013}.
But the detailed analysis of the formation of such twisted coronal loops is out of the scope of this study; we only discuss here the evolution of flux-ropes already twisted and close to the onset of the kink instability.

%
We compare light-curves and time-dependent intensity maps of thermal (free-free continuum) and non-thermal (bremsstrahlung caused by energetic electrons) radiation in the keV and deca-keV ranges.
We investigate how the spatial distribution of the emitted photon fluxes relate to the dynamical and geometrical properties of the simulated loops, how the total continuum emission spectra evolves in time and on how the properties of the emission measures respond to the plasma heating processes occurring in the magnetic loops.


\section{Model description}\label{model}

\begin{figure}[!h]    
  \centering
  {\sf \small \hspace{6ex} Vertical profiles, above footpoint} \\
  \includegraphics[width=.8\linewidth,clip=true,trim=40 410 600 40]{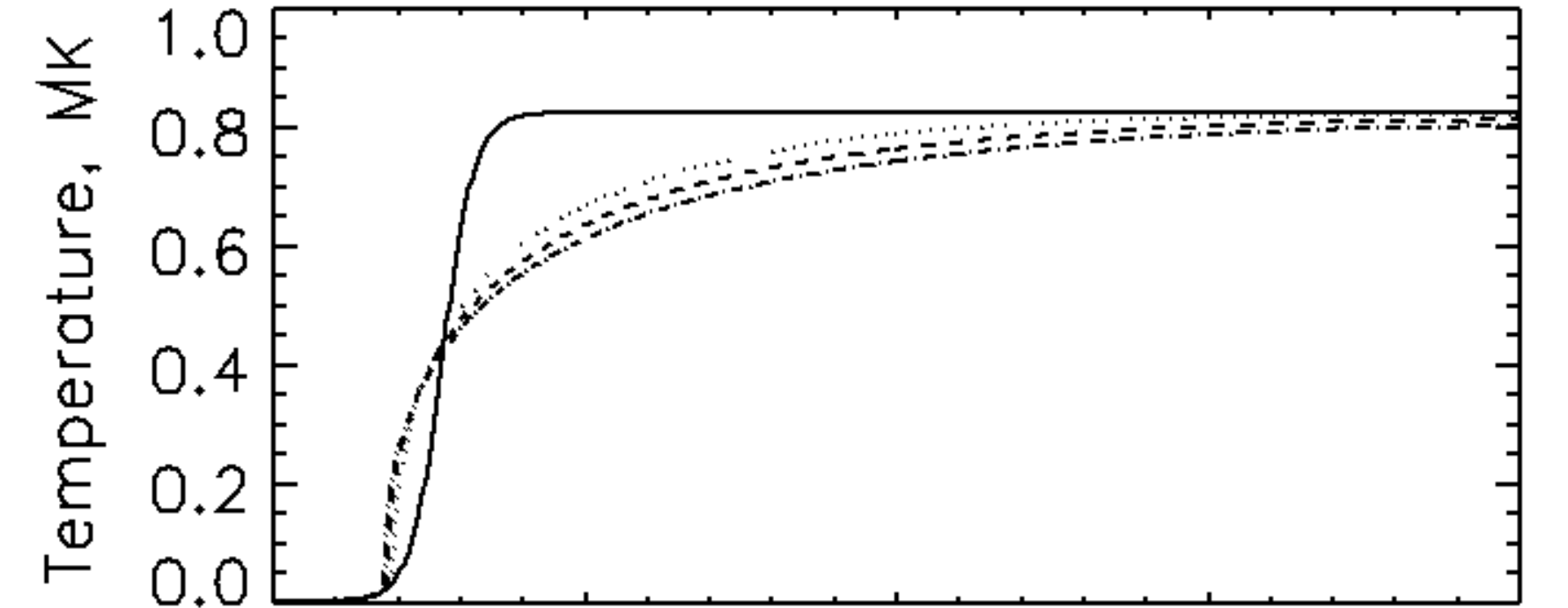} \\
  \includegraphics[width=.8\linewidth,clip=true,trim=40  20 600 360]{initial_strat} \\
  
  \vspace{0.075\linewidth}
  {\sf \small \hspace{6ex} Vertical profiles, crossing loop's apex} \\
  \includegraphics[width=.8\linewidth,clip=true,trim=600 410 20 40]{initial_strat} \\
  \includegraphics[width=.8\linewidth,clip=true,trim=600  20  20 360]{initial_strat} \\

  \caption{
    Plasma temperature $T$ and numeric density $n$ as a function of height during the preparatory thermal relaxation phase at two different positions above the surface.
    Solid lines correspond to the initial state. 
    Dotted, dashed and dot-dashed lines represent the temporal evolution of these quantities during the thermal relaxation phase at equally spaced time intervals after the initial state.
    The two top panels correspond to the $n$ and $T$ profiles along a vertical line right above the loop foot-points, where the magnetic field is mostly vertical.
    The two bottom panels show the same quantities along a vertical line which crosses the centre of the loop, where the magnetic field is mostly horizontal.
  }
  \label{fig:atmos}
\end{figure}

We study the evolution of a twisted coronal loop embedded in a magnetised and gravitationally-stratified background atmosphere.
The coronal loops are twisted magnetic flux-ropes with large-scale curvature (the loops are nearly semi-circular), and with both foot-points anchored in the chromosphere.
The thermo-dynamical evolution of the system is calculated using fully compressible resistive MHD by means of three-dimensional numerical simulations.
The MHD computations provide the basis for the estimation both of the thermal emission by the plasma (in the soft X-ray range), and of the hard X-ray emission produced by charged test-particles (electrons).
The methods employed are described in detail in the following sections (Sects. \ref{model-mhd}, \ref{model-sxr} and \ref{model-hxr}).

\subsection{MHD equations and model parameters}\label{model-mhd}

\begin{figure}
  \centering
	\includegraphics[width=\linewidth]{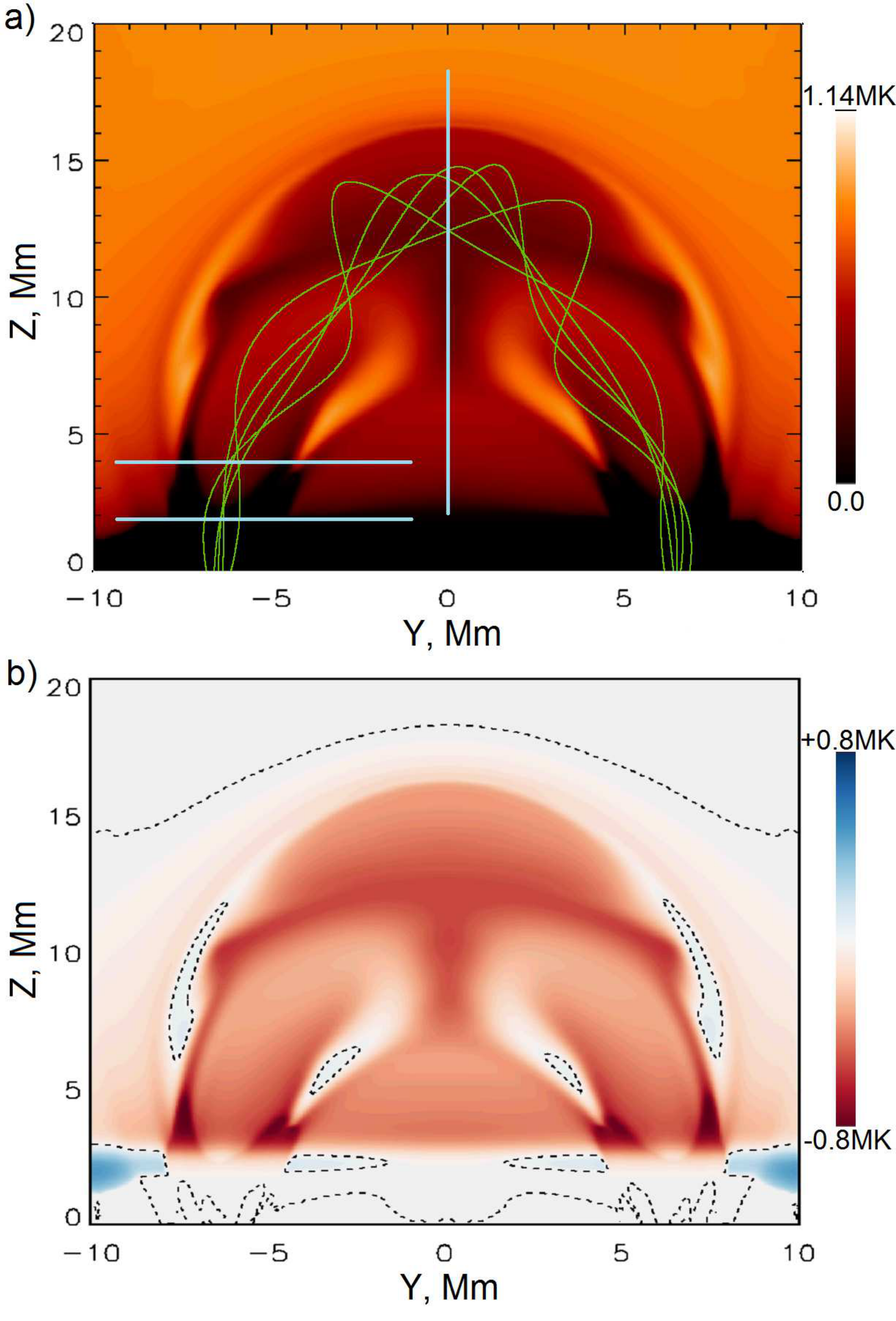}
  \caption{
    Temperature distribution shortly before the onset of the kink instability. 
    Upper panel (a) shows the temperature distribution in the mid-plane $x=0$, while the lower panel (b) show the difference between temperature distribution shortly before the instability and the initial temperature distribution. 
    Selected magnetic field lines are shown in green at panel (a). 
    Light blue lines at panel (a) denote cross-sections at which the temperature and density distributions are shown in Figures~\ref{fig:tem} and \ref{fig:rho}.
    Black dashed lines at panel (b) denote $\Delta T =0$ levels.
  }
  \label{fig:tkink}
\end{figure}

The evolution of the twisted loops is determined by solving the Eqs. (\ref{eq-mhdmain1}-\ref{eq-mhdmain4}) in a 3D $\left(x,y,z\right)$ domain taking into account atmospheric stratification, thermal conduction, and radiative losses.fg

The system is described in terms of magnetic field $\vec{B}$, plasma velocity $\vec{v}$, density $\rho$ and specific energy $\epsilon$: 
\begin{eqnarray}
  \frac{\partial \rho}{\partial t}     &=& - \vec{\nabla} \cdot (\rho \vec{v}), \label{eq-mhdmain1} \\ 
  \frac{\partial \vec{v}}{\partial t} &=& - (\vec{v} \cdot \vec{\nabla}) \vec{v} - \frac 1{\rho} \vec{j} \times \vec{B} - \frac 1{\rho}\vec{\nabla}p - \rho g \vec{z}, \label{eq-mhdmain2} \\ 
  \frac{\partial \epsilon}{\partial t} &=& - (\vec{v} \cdot \vec{\nabla})\epsilon - (\gamma-1)\epsilon\; \vec{\nabla} \cdot \vec{v} \nonumber \\ 
  & & + S_{\mathrm cond} + Q_{\mathrm joule} + Q_{\mathrm visc} - L_{\mathrm rad}, \label{eq-mhdmain3} \\
  \frac{\partial \vec{B}}{\partial t} &=& \vec{\nabla} \times (\vec{v} \times \vec{B})- \vec{\nabla} \times (\eta \vec{j}), \label{eq-mhdmain4} 
\end{eqnarray}
where $\eta$ is plasma resistivity and the current density is $\vec{j} = \vec{\nabla} \times \vec{B} / \mu_0$, and $g$ is the gravitational acceleration, which is constant throughout the whole domain. 
Plasma pressure and temperature can be expressed, respectively, as $p=\left(\gamma-1\right)\rho\epsilon$ and $T =\left(\gamma-1\right)\bar{m}/k_B\epsilon$.
Here $\gamma$ is the ratio of specific heats of the plasma, $\bar{m}$ is average particle mass (which is half the proton mas $m_p$, corresponding to fully-ionised hydrogen plasma).
The electric field, required for the test-particle modelling, can be expressed as 
\begin{equation}
  \label{eq-mhdef}
  \vec{E} = - \vec{v} \times \vec{B} + \eta \vec{j}\ .
\end{equation}
The $S_\mathrm{cond}$, $Q_\mathrm{joule}$, $Q_\mathrm{visc}$ and $L_\mathrm{rad}$ terms in Eq. (\ref{eq-mhdmain3}) account for variations of specific energy due to thermal conduction, Joule heating, viscous heating and radiative losses, respectively.
Thermal conduction is defined as
\begin{equation} \label{eq-cond}
S_{\mathrm cond} = \vec{\nabla} \cdot \left [ \left ( -\kappa_0 T^{5/2} \vec{\hat{B}}\cdot\vec{\nabla} T \right ) \vec{\hat{B}}\right ] + 
\vec{\nabla}\cdot \left[ -\kappa_0 T^{5/2} \frac{B^2_\mathrm{min}}{B^2_\mathrm{min}+B^2} \vec{\nabla} T \right],
\end{equation}
where $B_\mathrm{min}$ is some very small parameter.
Hence, with non-zero magnetic field ($B \gg B_\mathrm{min}$), we get Braginskii conductivity ($\kappa_0 T^{5/2}$) along the magnetic field, but negligible in perpendicular direction, while with negligible magnetic field ($B \ll B_\mathrm{min}$) it is $\kappa_0 T^{5/2}$ and isotropic.
The following characteristic dimensional factors are used: $B_0=4.47\times 10^{-3}\un{T}$, $\rho_0=2\times 10^{-12}\un{kg\cdot m^{-3}}$ and $L_0=10^6\un{m}$, yielding the characteristic values for Alfvén velocity $v_0 =2.82\times 10^6\un{m\cdot s^{-1}}$, time $t_0 = 0.36\un{s}$, temperature 
$T_0=3.23\times 10^8\un{K}$, and electric field $E_0 = 1.26\times 10^4\un{V\cdot m^{-1}}$.
The Joule heating is $Q_{\mathrm joule}=\eta/\rho j^2$, where $\eta$ is the local resistivity $\eta=\eta_\mathrm{back} +\eta_\mathrm{crit}$. 
The background resistivity $\eta_\mathrm{back}$ is constant throughout the simulation domain and represents classical resistivity.
Its value is $10^{-6} \sqrt{\mu_0 \rho_0} L_0^2 B_0^{-1}$ (corresponding to the Lundquist number $10^6$), i.e. this resistivity component does not make any noticeable contribution to the system evolution.
The critical resistivity is defined as 
\begin{equation}
  \label{eq-resist}
  \eta_\mathrm{crit} =
  \left \{
    \begin{array}{lr}
      10^{-3} \sqrt{\mu_0 \rho_0} L_0^2 B_0^{-1}, & j \geq j_\mathrm{crit} \\
      0, & j < j_\mathrm{crit} 
    \end{array}
  \right .  ,
\end{equation}
where $j_\mathrm{crit} = j_1 \rho/\rho_0 \sqrt{T/T_0}$. 
This component represents anomalous resistivity with the threshold similar to that for ion-acoustic instability, when the current drift velocity exceeds local thermal velocity.
The value of $j_1$ constant is chosen so that before the kink instability $\eta_\mathrm{crit}=0$ everywhere in the domain.
The chosen resistivity values have no exact physical justification, as in most other MHD models. 
However, the chosen values appear to be realistic, as they provide realistic reconnection times (tens of seconds).
The radiative losses are calculated using the function from \citet{klimchuk_highly_2008}.
The shock viscosity and corresponding heating rate used in our simulations are described in \citet{arber_staggered_2001}. 
The adopted viscosity values are not representative of real viscosity in the coronal plasma, and the corresponding viscous heating rate is usually lower than the ohmic heating rate due to anomalous resistivity.
However, high velocity gradients during the fast reconnection phase can sometimes result in substantial viscous heating, comparable to or even higher than ohmic heating. 
Ultimately, the viscous effects result in conversion of magnetic energy to internal energy of the plasma and can be considered as an additional resistivity effective at locations with high velocity gradients (see Bareford, et al, submitted, for more detailed description of this effect).
The above equations (\ref{eq-mhdmain1}-\ref{eq-mhdmain4} along with Eqs.~\ref{eq-mhdef},\ref{eq-cond} and \ref{eq-resist}) are solved using the lagrangian remap code LARE3D \citep{arber_staggered_2001}. 
We consider four models: three with small coronal loops with different magnetic field magnitudes (models C, S and V), and one case with a large coronal loop (model Y).
The simulation boxes are cubes with the size of $20\un{Mm}$ in ``small'' models (C, S and V) and $80\un{Mm}$ in the ``large'' model Y; see Table~\ref{tab:models} for details. 
The numerical grid consists of $256\times 256\times 512$ elements with uniform step in each direction (i.e. the  resolution in $z$-direction is twice as high as the resolution in the horizontal directions).

\begin{table}
  \centering
  \begin{tabular}{r c c c c}
    \hline\hline
    \phantom{x}            &   Model C       &  Model S          &      Model V          &         Model Y      \\  
    \hline
    Box size, Mm      &    20           &       20                  &        20                 &         80          \\
    $B_{FP}$, 10$^{-4}T$ &    45          &       90                  &       220                 &         665         \\
    $B_{LT}$, 10$^{-4}T$ &     5       &          10               &           24                 &         67                   \\
    $R_{FP}$, Mm &          0.7       &          0.7               &          0.7                &        2.8              \\
    $R_{LT}$, Mm       &    2.4            &     2.4                    &     2.4                &        9.6                    \\
    ${\it l}$, Mm  &    24.6           &     24.6               &         24.6               &       97.0           \\
    \hline
  \end{tabular}
  \caption{Pre-instability parameters of twisted loops in different models. Magnetic field values $B_{FP}$ and $B_{LT}$ are
absolute magnetic field values at the foot-points and loop-top, respectively; $R_{FP}$ and $R_{LT}$ are the loop cross-section radii
at the foot-points and loop-top, respectively, ${\it l}$ is a loop length.}
  \label{tab:models}
\end{table}

\begin{figure*}[!ht]
  \centering
  
  \begin{minipage}{.6\linewidth}
    \centering
    \includegraphics[width=\linewidth]{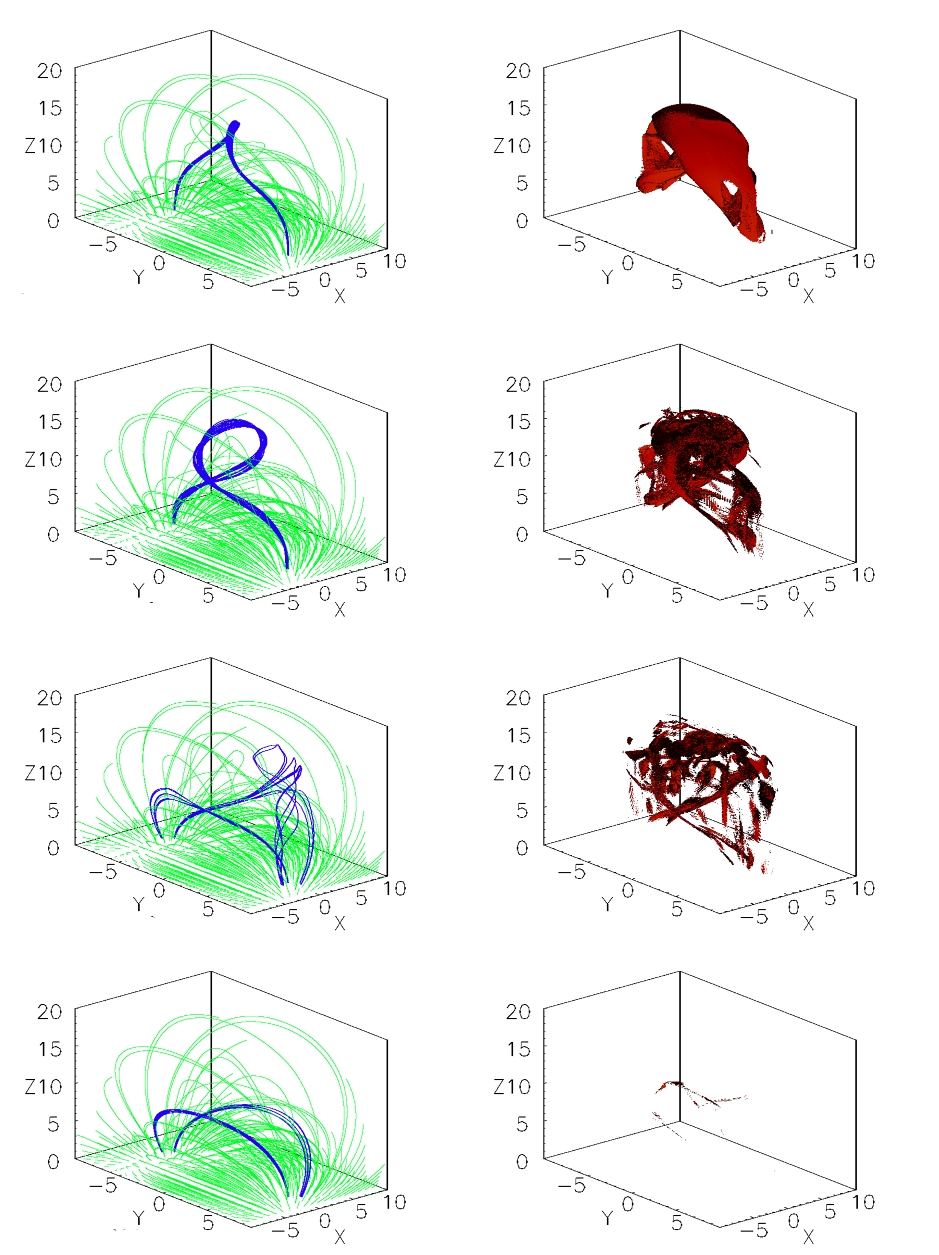}
    \caption{
      Magnetic field topology and current distribution at 4 representative instants of the simulations (from top to bottom: $t=0,\ 30,\ 120,\ 196\un{s}$ after the onset of the kink instability;  model S), illustrating the onset and growth of the kink instability, the reconnection episode that follows, and the relaxation of the coronal loop towards a non-twisted state.
      The lines on the left column are magnetic field-lines, with the blue lines having seed points very close to flux-rope's foot-point centres.
      The axis are labelled in units of $\un{Mm}$.
      The right column shows iso-surfaces of current density $j = j_{crit}$ (see Eq. \ref{eq-resist} and the discussion which follows).
    }
    \label{fig:blines_current}
  \end{minipage}
  \hspace{0.04\linewidth}
  \begin{minipage}{.35\linewidth}
    \centering
    \includegraphics[width=.9\linewidth]{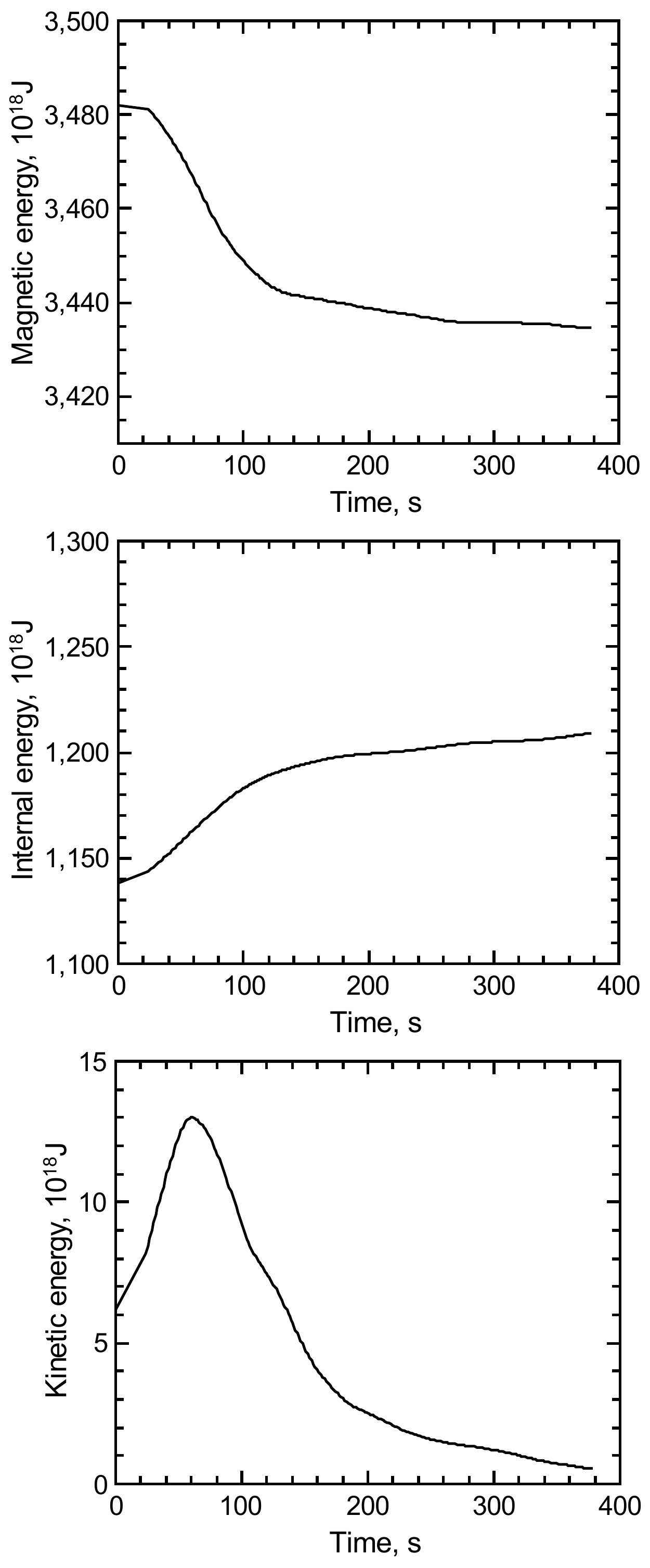}
    \caption{
      Evolution of the magnetic, internal and kinetic energy in model S after the onset of the kink instability.
    }
    \label{fig:energetics}
  \end{minipage}
\end{figure*}

\begin{figure}[!h]
  \centering
  \includegraphics[width=0.85\linewidth]{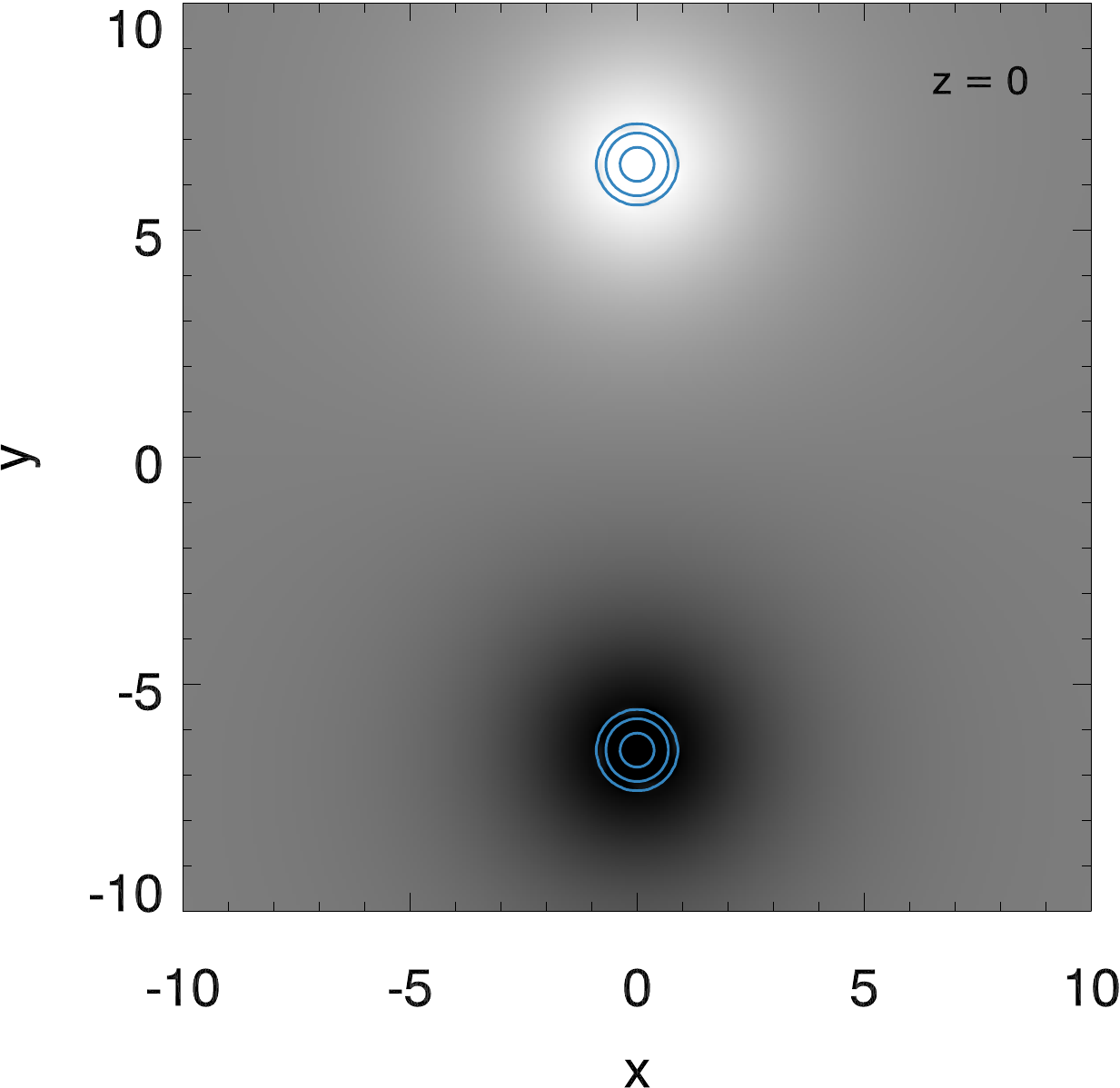}
  \caption{ 
      Foot-point twisting profile and vertical magnetic field amplitude at the lower boundary (for model S).
      The grey-scale shows the amplitude of the vertical magnetic field $B_z$ (white and black representing positive and negative polarity) and the blue lines are streamlines of the surface motions (see Eq. \ref{eq-rota}).
      The axis are labelled in units of$\un{Mm}$.
      }
    \label{fig:twisting}
\end{figure}

All the models undergo an initial preparation phase (consisting of a thermal relaxation stage, and a footpoint twisting stage).
The initial magnetic field geometry is that of a potential field arcade in a gravitationally-stratified atmosphere. 
As in \citet{gordovskyy_particle_2014}, the initial atmosphere is constructed of two isothermal regions: a lower region with thickness of about $2\un{Mm}$ representing the chromosphere with temperature $\lesssim 10^4\un{K}$ and density exponentially decreasing from $5\times 10^{13}$ to
$\sim 10^{9}$~cm$^{-3}$, and an overlaying corona with temperature $\sim 0.8\un{MK}$ and nearly constant density $\sim 10^{9}\un{cm^{-3}}$. 
This idealised initial configuration is not in thermal equilibrium, however, due to the presence of magnetic field-aligned thermal conduction.
We first let the system relax until it reaches quasi thermal equilibrium. 
Figure \ref{fig:atmos} shows how the vertical profiles of density and temperature evolve during the relaxation process. 
It can be seen that the temperature transition around $z=2-4\un{Mm}$ becomes gradually smoother. 
The temperature distribution in the vertical direction tends approximately to $T\left(z\right) \sim \left(z+const.\right)^{2/7}$ where the magnetic field-lines are vertical.
The equilibrium temperature distribution will, however, be considerably different where the field is nearly horizontal (as in the middle of the domain, across the centre of the loop arcades). 
Heat transfer during the relaxation phase results in some evaporation: plasma from nearly vertical field in the transition region ($z=2-3 \un{Mm}$) flows upwards, noticeably increasing the loop plasma density up to $z=10\un{Mm}$. 
After about $2000 t_0$, temperature and density distributions become nearly stationary (the resulting atmosphere is thermally relaxed).

Foot-point twisting applied to the initially untwisted potential field in order to build the unstable twisted loop configurations.
The twisting phase does not represent an actual rotation of photospheric foot-points (an effect which is often observed, but out of scope of this study). 
We can use here rotation velocities considerably higher than those observed in the photosphere in order to speed up the simulations, as long as these remain substantially lower than the local Alfvén speed.
Under these conditions, the twisting phase should not have noticeable dynamical effects on the kink instability and reconnection \citep[see, for more detail, ][Bareford, et al, submitted]{gordovskyy_particle_2014}.
The foot-point centres are located at $\left[0.0,\pm 6.5\un{Mm},0.0\right]$ in models C, S and V, and at $[0.0,\pm 26\un{Mm},0.0]$ in model Y. 
Foot-points are rotated with the following profile
\begin{equation}\label{eq-rota}
\omega(r) = \omega_0 \frac{1-\tanh\left[-(r-R_{fp})/L_{wall}\right]}2,
\end{equation}
where $r$ is the distance from a foot-point centre, $\omega_0 \approx 0.02\un{s^{-1}}$ is the frequency of rotation, $R_{fp}$ is the foot-point radius, which is $0.7\un{Mm}$ in models C, S and V, and $2.8\un{Mm}$ in model Y, and $L_{wall}=0.13\un{Mm}$ is a scaling factor.
  Taking model S as a reference case, the foot-point rotation period is about $300\un{s}$, with a corresponding velocity at the edge of the foot-points of $1.4\e{4}\un{m/s}$.
  The coronal Alfvén speed is close to $7\e{5} \un{m/s}$, and the Alfvén travel time along the whole loop is $\sim 35\un{s}$. 
  Hence, foot-point twisting motions do not generate transients which propagate into the corona, and the coronal field has ample time to adjust and remain in near-equilibrium as it is being twisted.
  The other models (V and Y) have stronger magnetic fields and higher corresponding Alfvén speeds, for which quasi-steadiness is even better achieved.
  Chromospheric Alfvén speed is, of course, much lower, and approaches the footpoint rotation speed. 
  However, the chromosphere has a plasma $\beta \gg 1$, with the sound speed remaining considerably higher than the twisting speed throughout the whole chromospheric layer (up to one order of magnitude higher for model S). 
  These reasons altogether ensure that the system evolves through a sequence of equilibria (even though small amplitude dynamical features, such as waves can be seen during the driving period).
  But most importantly, the forcing (twisting) time-scale is always higher than the kink instability growth-time.

  Figure \ref{fig:twisting} shows the streamlines of the rotating motions applied at the surface together with the vertical magnetic field distribution $B_z$.
  These circular vortical motions are applied at the zones of maximal $B_z$ (the foot-points), and thus are spatially coincident with the contours of the vertical field, which are approximately circular there \citep[\emph{cf.} discussions by][]{torok_evolution_2003,aulanier_equilibrium_2005}.

Such foot-point rotation effectively results in continued helicity injection with the total twist increasing by $\sim 0.015\un{\pi/s}$, i.e after about $500\un{s}$ the total twist is of about $8\pi$.
At this stage, the twisted loop arcade is slightly thicker and taller than initially (due to the increase in magnetic pressure within the magnetic loop) but its large-scale structure stills maintains its quasi-circular symmetry.
The system is close to the onset of the kink instability (but still in mechanical equilibrium), and we define this state as the initial condition ($t=0$) hereafter.
The configuration of the magnetic flux-rope, the plasma density and temperature at this stage are represented in Fig. \ref{fig:tkink}, and the corresponding pre-instability parameters are listed in Table \ref{tab:models}.


\subsection{Synthetic soft X-ray emission from a model loop}
\label{model-sxr}

We calculate the thermal X-ray continuum emission from the model's number density $n$ and temperature $T$ at each point of the domain, as in \citet{pinto_soft_2015}.
The plasma emissivity at a given photon energy $h\nu$ is defined as
\begin{equation}
  \label{eq:emissivity}
  \epsilon\left(h\nu, T\right) = \epsilon_0 n^2 T^{-1/2} g_{ff}\left(h\nu, T\right) \exp{\left( -\frac{h\nu}{k_b T}\right)}\ ,
\end{equation}
where $g_{ff}\left(h\nu, T\right)$ is the Gaunt factor for free-free bremsstrahlung emission.
The coefficient $\epsilon_0$ is $6.8\e{-38}$ if the emissivity is to be expressed in units of $\mathrm{erg\cdot cm^{-3}\cdot s^{-1}\cdot Hz^{-1}}$ \citep{tucker_radiation_1975}.
The Gaunt factor is approximated by the piece-wise approximation
\begin{equation}
  \label{eq:g_ff}
  g_{ff}\left(h\nu, T\right) =
\left\{
     \begin{array}{lr}
       1, & h\nu \lesssim k_bT \\
       \left(\frac{k_bT}{h\nu}\right)^{0.4},  & h\nu > k_b T
     \end{array}
   \right. 
\end{equation}
The photon flux density emitted at the photon energy $h\nu$ is defined as
\begin{equation}
  \label{eq:photonflux}
  I\left(h\nu, T\right) = I_0 \frac{\mathrm{EM}}{h\nu \sqrt{k_bT}} g_{ff}\left(h\nu, T\right) \exp{\left( -\frac{h\nu}{k_b T}\right)}\ ,
\end{equation}
where $\mathrm{EM} = n^2V$ is the emission measure of a finite volume $V$ of plasma of density $n$ and temperature $T$.
The coefficient $I_0$ is $1.07\e{-42}$ for a photon flux measured at a distance of $1\un{AU}$ in units of $\mathrm{photons\cdot cm^{-2}\cdot s^{-1}\cdot keV^{-1}}$.
The total photon flux over a given spectral band is computed by integrating Eq. (\ref{eq:photonflux}) over the corresponding range of values of $h\nu$.
We compute the photon flux at different photon energies for each individual grid cell (\emph{i.e}, volume element), each one having well-defined values for the emission measure, density and temperature.
As the corona is optically thin to X-ray radiation, the total flux emitted is obtained by adding the individual contributions over the whole loop (or over a region of interest).

We estimate the distributions of the differential emission measure $\mathrm{DEM}\left(T\right)$ in our simulations by computing the emission measure of the plasma regions whose temperature lies within successive temperature intervals at a given time and normalising it by the temperature interval width $\Delta T$.
This distribution then quantifies the amount of plasma at different temperatures, independently of the temperature interval considered.
That is, the differential emission measure are defined here as a function of temperature as 
\begin{equation}
  \label{eq:dem_t}
  \mathrm{DEM}\left(T\right) = \sum_k n^2_k \cdot \frac{\delta V_k}{\Delta T_k} \ ,
\end{equation}
where the index $k$ runs through all the plasma elements (grid-cells in the simulations) which lie within the temperature interval $\left[T,  T+\delta T\right]$.
The values of $n_k$ and $\delta V_k$ are the number density and the volume of each element.
In other words, we first compute a temperature histogram with a given temperature bin size $\Delta T$.
Then, we verify which grid-cells have a temperature $T$ within each of the bins and we sum over all the corresponding individual $\mathrm{EM}$.
Variations in density for plasma at a given temperature are implicitly accounted for.
Total emission measures $\mathrm{EM}$ can be obtained by integrating Eq. (\ref{eq:dem_t}) in respect to temperature, in the relevant interval of temperatures.

\begin{figure}[!t]
  \centering
	\includegraphics[width=\linewidth]{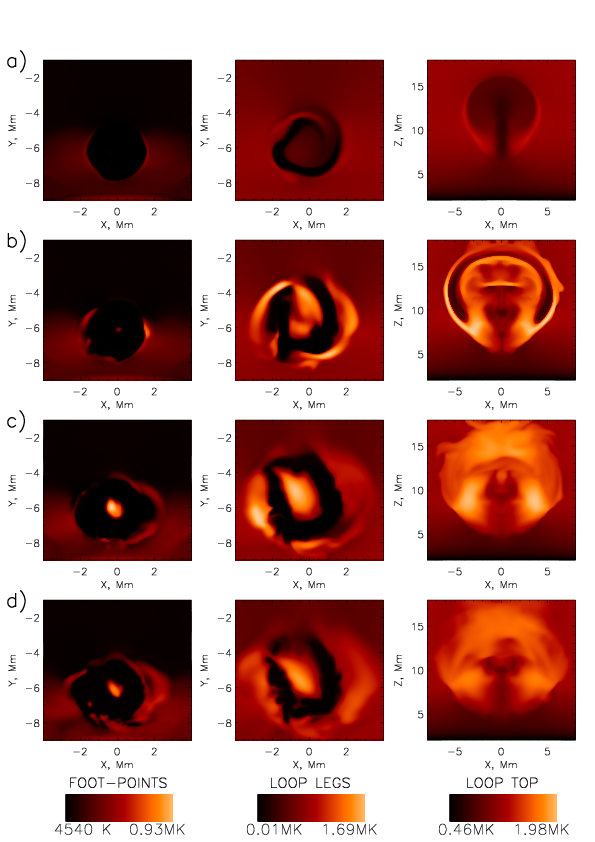}
  \caption{
    Temperature distributions during and after the kink instability.
    Three loop cross-sections are shown (corresponding to the blue lines in  Fig.~\ref{fig:tkink}): left foot-point ($z = 1.9$~Mm) (left column), left loop leg ($z = 4.25$~Mm) (middle column) and vertical plain crossing the loop top (right column).
    Times shown are (after onset of the kink instability): panel (a) -- 0~s , (b) -- 81~s, (c) -- 196~s, (d) -- 377~s.}
 \label{fig:tem}
\end{figure}
\begin{figure}[!t]
  \centering
	\includegraphics[width=\linewidth]{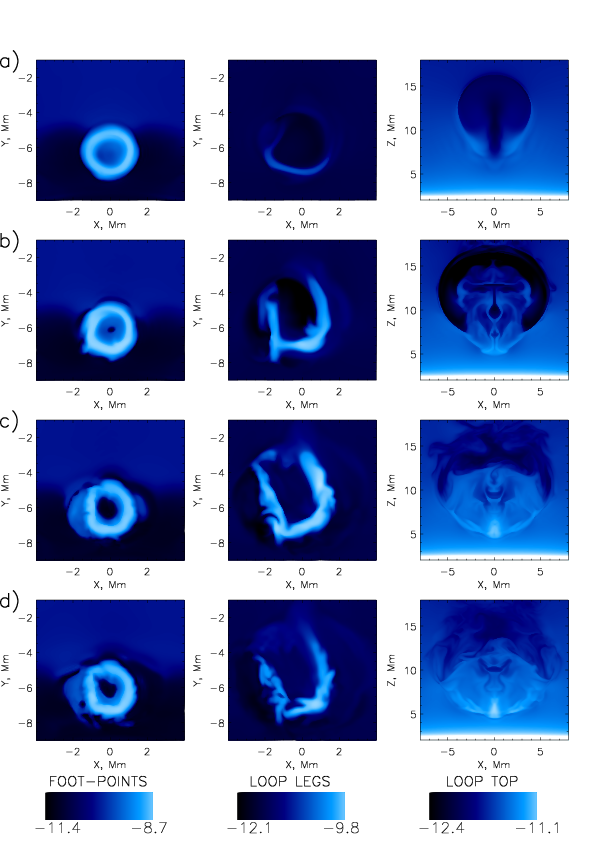}
  \caption{Logarithm of density distributions. Locations and times are the same as in Figure~\ref{fig:tem}.}
 \label{fig:rho}
\end{figure}

\subsection{Test-particle trajectories and synthetic hard X-ray bremsstrahlung}
\label{model-hxr}

Electron distribution functions are calculated using the GCA test-particle code based on first-order guiding-centre approximation \citep{gordovskyy_particle_2010-1,gordovskyy_particle_2011}:
\begin{eqnarray}
\frac {d \bf r}{dt} &=& {\bf u} + \frac{(\gamma v_{||})}\gamma {\bf b} \\
{\bf u} &=& {\bf u}_E + \frac mq \frac {(\gamma v_{||})^2}{\gamma \kappa^2 B} [{\bf b} \times ({\bf b}\cdot{\bf \nabla}){\bf b}]  + \\
           & &  \frac mq \frac \mu {\gamma \kappa^2 B} [{\bf b} \times ({\bf \nabla} (\kappa B))]\\
\frac{d (\gamma v_{||})}{dt} &=& \frac qm {\bf E}\cdot {\bf b} - \frac \mu \gamma ({\bf b} \cdot {\bf \nabla}(\kappa B)) + \\
           & & (\gamma v_{||}) {\bf u}_E\cdot (({\bf b}\cdot{\bf \nabla}){\bf b}) + \left[\frac{dv_{||}}{dt}\right]_{\mathrm coll}\\
\gamma &=& \sqrt{\frac{c^2 +(\gamma v_{||})^2 + 2 \mu B}{c^2 - u^2}}\\
\frac{d\mu }{dt} &=& \left[\frac{d\mu}{dt}\right]_{\mathrm coll}.
\end{eqnarray}
Here $\vec{r}$, $\vec{u}$ and $v_{||}$ are the particle gyro-centre position, transversal drift velocity ($\perp \vec{B}$) and longitudinal velocity ($||\ \vec{B}$), respectively; $\vec{b}=\frac{\vec{B}}B$ is the magnetic field direction vector, 
$\vec{u}_E = \frac {\vec{E}\times\vec{b}}{B}$ is the $E \times B$ drift component; $\mu$ is the magnetic moment per mass 
unit $\mu = \frac{\gamma^2 u_{\rm g}^2}{2B}$, where $u_{\rm g}$ is the particle gyration velocity. The relativistic coefficients are defined as 
$\gamma = (1-v^2/c^2)^{-1/2}$ and $\kappa = \sqrt{1-u_E^2/c^2}$, where the absolute particle velocity is $v=\sqrt{v_{||}^2+u^2+u_{\rm g}^2}$.
Coulomb collisions (introduced through $\left[\frac{dv_{||}}{dt}\right]_{\mathrm coll}$ and $\left[\frac{d\mu}{dt}\right]_{\mathrm coll}$) are taken into account through averaged
deceleration and pitch-angle deflection of particles with mass $m$ and charge $\pm e$ in fully-ionised hydrogen plasma \citep[see \emph{e.g}][]{emslie_collisional_1978}. Thus, 
\[
\frac{dv}{dt} = -a\frac{2 n}{m^2 v^2},
\]
with 
\[
a= \frac {2\pi e^4}{(4\pi \varepsilon_0)^2} \Lambda \left( \frac {m_e+m_p}{m_e m_p} \right),
\]
where $\Lambda$ is the Coulomb logarithm, and quasi-random changes of particle pitch-angles with probabilities, and, hence, 
\begin{equation}\label{eq-dvdtcoll}
v(t)+\Delta v = \sqrt[3]{v^3(t)-\frac{2an}{m^2}\Delta t}.
\end{equation}
Corresponding quasi-random angular deflections are calculated as
\begin{equation}\label{eq-dmudtcoll}
\Delta\theta = 1.3 \frac{an}{m^2 v^3} \sqrt{\mu} \mathcal{N} \sqrt{\Delta t} ,
\end{equation}
where $\mathcal{N}$ is a random number from normal distribution with the width of $1.0$. Tests show that, statistically, this formula provides a good approximation for pitch-angle diffusion as per 
$\frac{\partial}{\partial t} = \frac{an}{mv^3} \frac {\partial}{\partial \mu}\left( (1-\mu^2) \frac{\partial}{\partial \mu}\right)$, where $\mu= \cos\theta=v_{||} / v$ is the pitch-angle cosine \citep{landau_statistical_1937}.
Therefore, collisional terms can be written using Eqs.~\ref{eq-dvdtcoll}-\ref{eq-dmudtcoll}:
\begin{eqnarray}
\left[\frac{d\mu}{dt}\right]_{\mathrm coll} &=& \sqrt{1-\mu^2}\frac{\Delta \theta}{\Delta t}\\
\left[\frac{dv_{||}}{dt}\right]_{\mathrm coll} &=& \mu\frac{\Delta v}{\Delta t} + v \left[\frac{d\mu}{dt}\right]_{\mathrm coll}.
\end{eqnarray}

Initially, the box is randomly filled with $2^{20}$ test-electrons which are uniformly distributed in space and in respect to pitch-angle cosine $\mu$, while in respect to absolute velocity $v$ particles have a Maxwellian distribution corresponding to $T=0.83\un{MK}$ (the temperature of the coronal background; see Fig. \ref{fig:atmos}). 
Thermal bath conditions are applied on all six boundaries of the domain: particles are free to leave the domain, and for every particle leaving the domain one new particle is generated randomly from a Maxwellian distribution (at the thermal bath's temperature) and is injected at the same point of the boundary.

Hard X-ray bremstrahlung intensities are calculated using the Kramers formula \citep[see \emph{e.g}][]{kontar_nonuniform_2002}:
\[
I(\epsilon_{ph})=Q_0 \int \limits_{\epsilon_{ph}}^{\infty}f_{beam}(\epsilon)n\frac 1{\epsilon_{ph} \epsilon}d\epsilon,
\]
where $I(\epsilon_{ph})$ is the rate of emission of photons with energy $\epsilon_{ph}$, $\epsilon$ is an electron energy, $f_{beam}$ is the differential beam electron density ($dn/dE$), $n$ is the space and time-varying electronic density of the ambient plasma, and $Q_0$ is the constant from the Kramers bremsstrahlung cross-section formula.

The test particle approach is widely used to study particle acceleration in flares, but is not a fully self-consistent approach since it neglects the electromagnetic fields generated by the test particles themselves. 
However, in the configurations studied here, the fraction of the released magnetic energy carried by the accelerated particles is relatively small - typically around $5\%$ \citep{gordovskyy_particle_2014}.
Furthermore, due to the fragmented nature of the current sheet, energetic electrons are farily uniformly distributed within the loop volume, so we expect that their effects on the fields is fairly small. Hence, the test particle approach is quite well justified.



\section{Magnetic fields, currents and hot plasma}
\label{sec:magfield}

\begin{figure*}
  \centering
  \textsf{\hspace{0.02\linewidth} Model S \hspace{0.25\linewidth} Model V \hspace{0.25\linewidth} Model Y} \\

  \includegraphics[width=.31\linewidth,clip=true,trim=0 50 0 0]{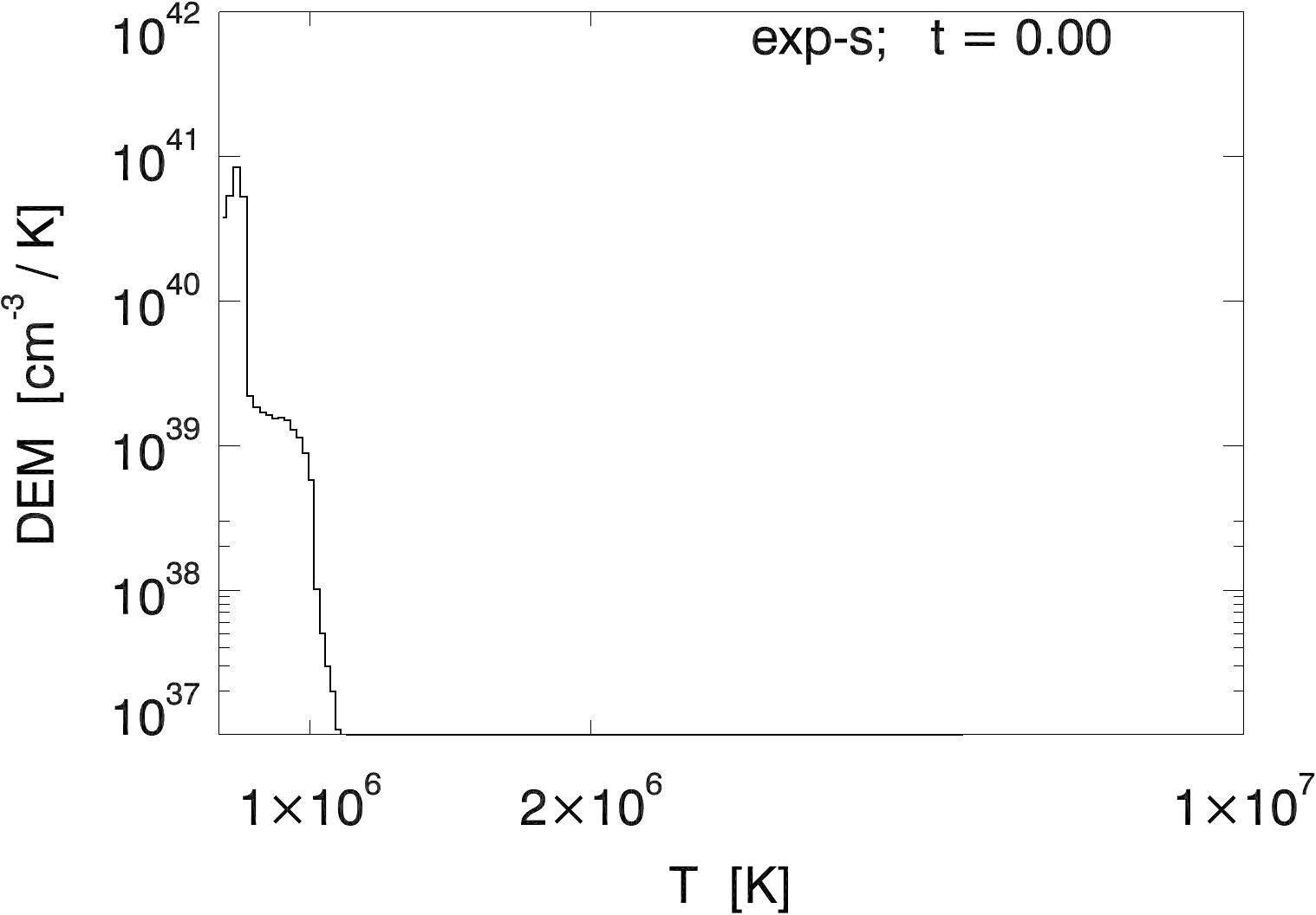} \hspace{0.01\linewidth}
  \includegraphics[width=.31\linewidth,clip=true,trim=0 50 0 0]{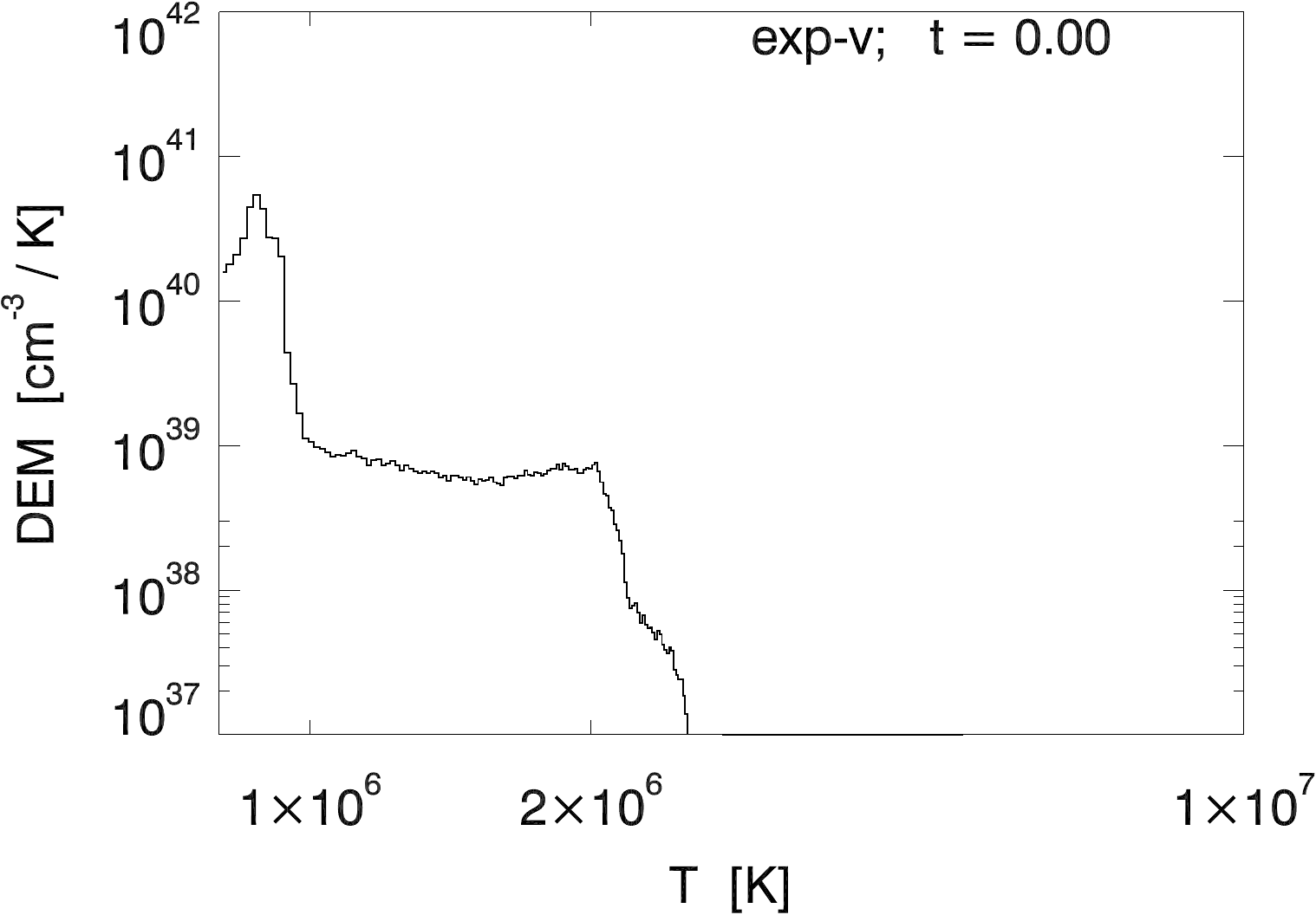} \hspace{0.01\linewidth}
  \includegraphics[width=.31\linewidth,clip=true,trim=0 50 0 0]{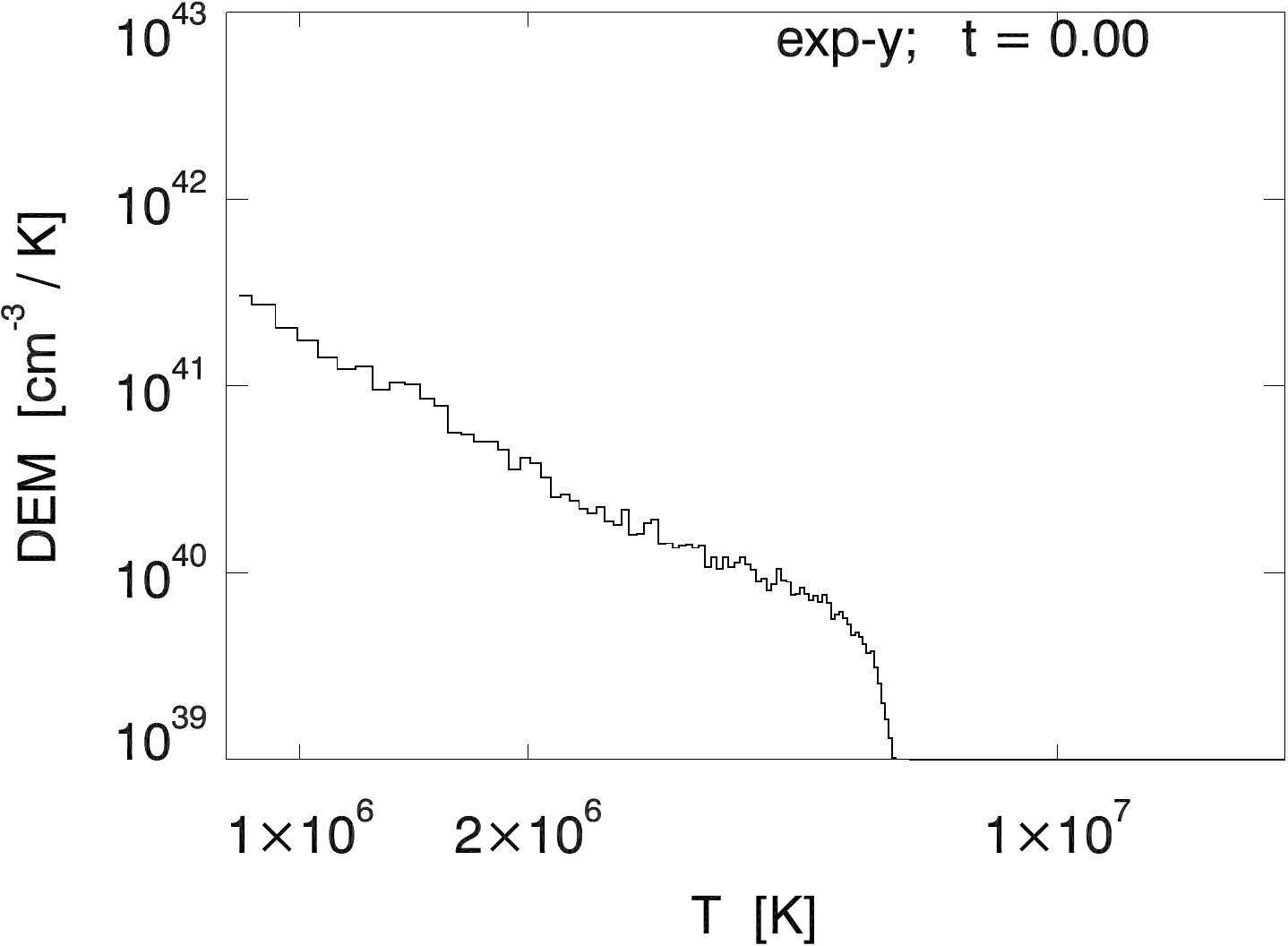} \\

  \includegraphics[width=.31\linewidth,clip=true,trim=0 50 0 0]{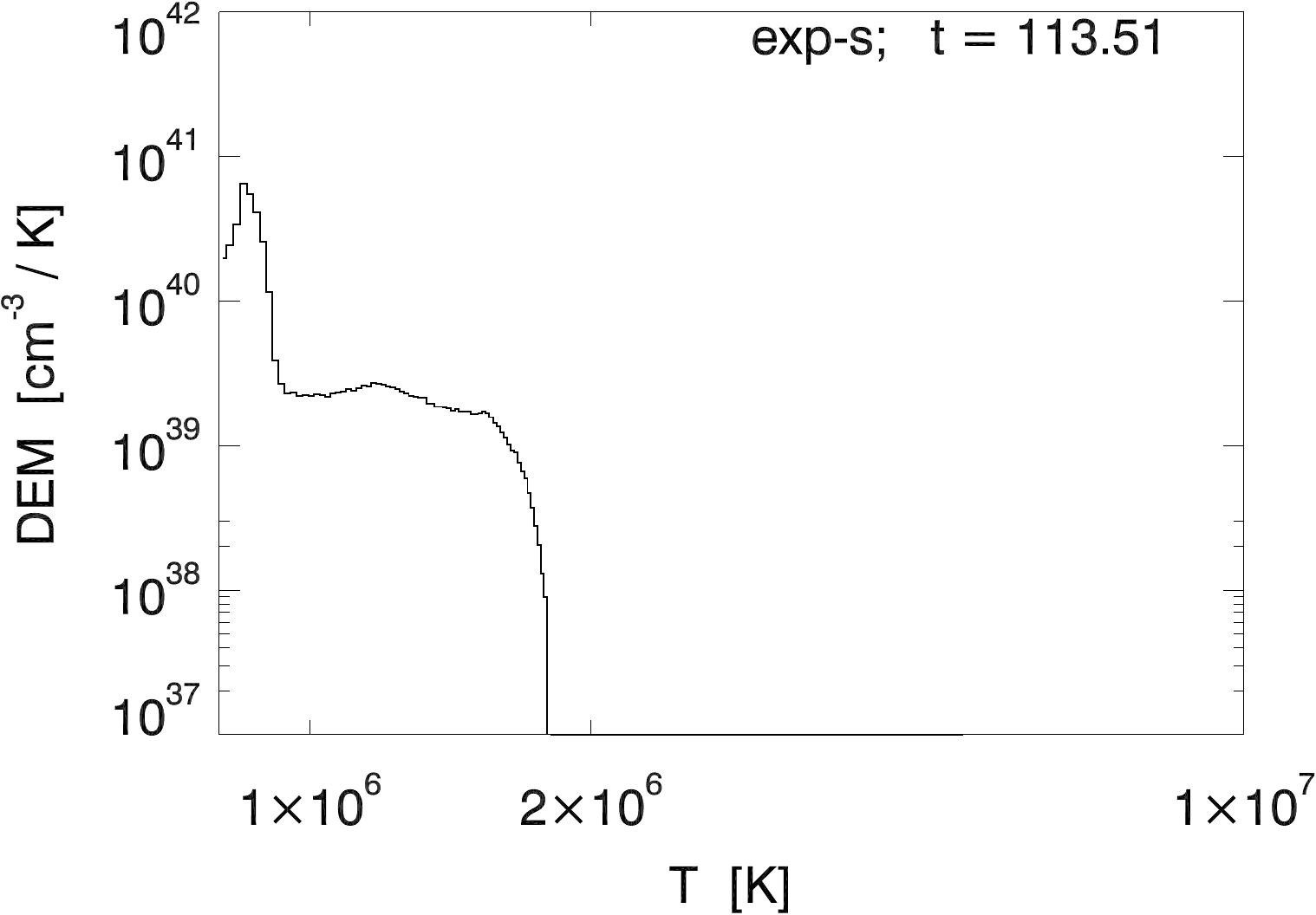} \hspace{0.01\linewidth}
  \includegraphics[width=.31\linewidth,clip=true,trim=0 50 0 0]{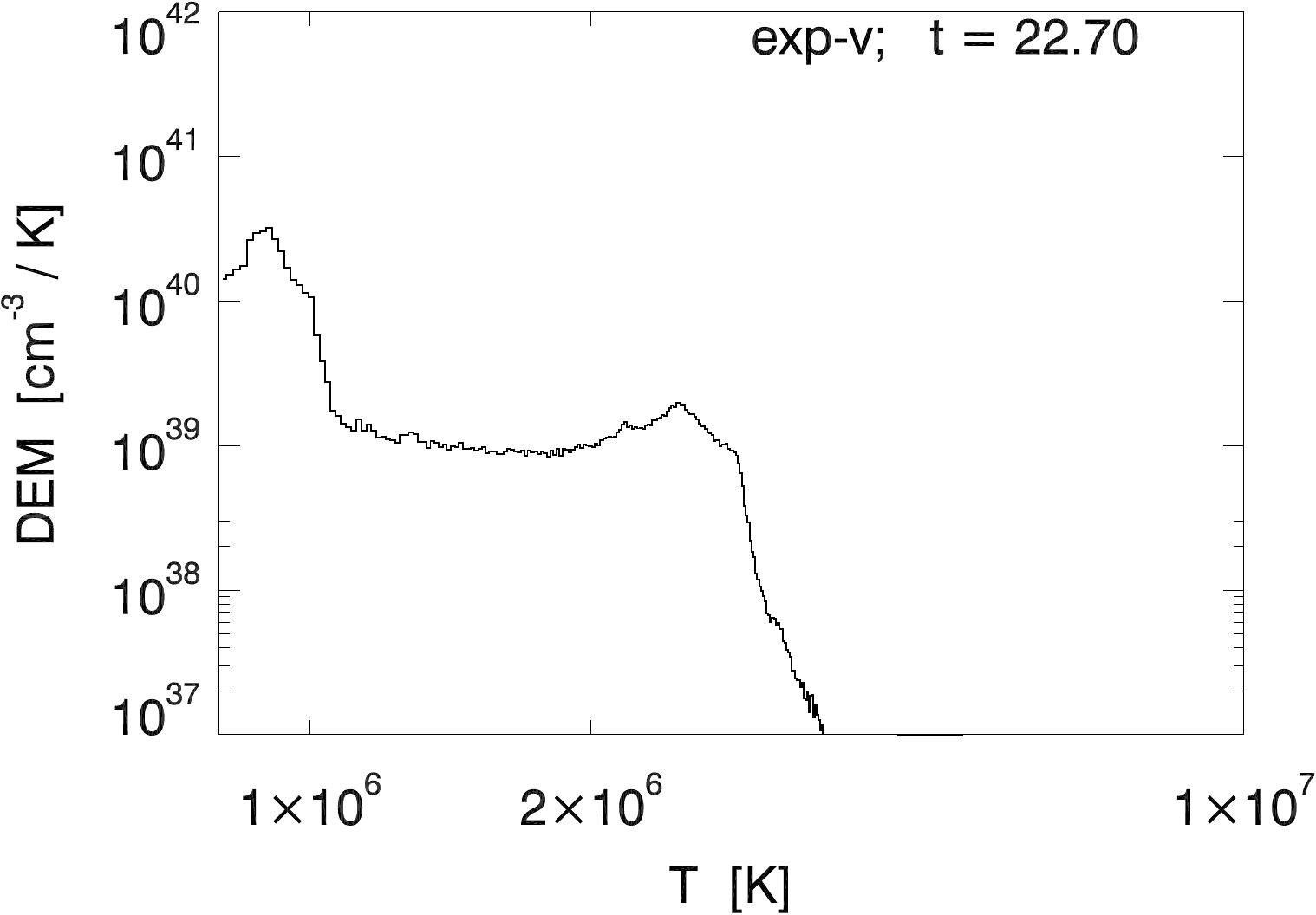} \hspace{0.01\linewidth}
  \includegraphics[width=.31\linewidth,clip=true,trim=0 50 0 0]{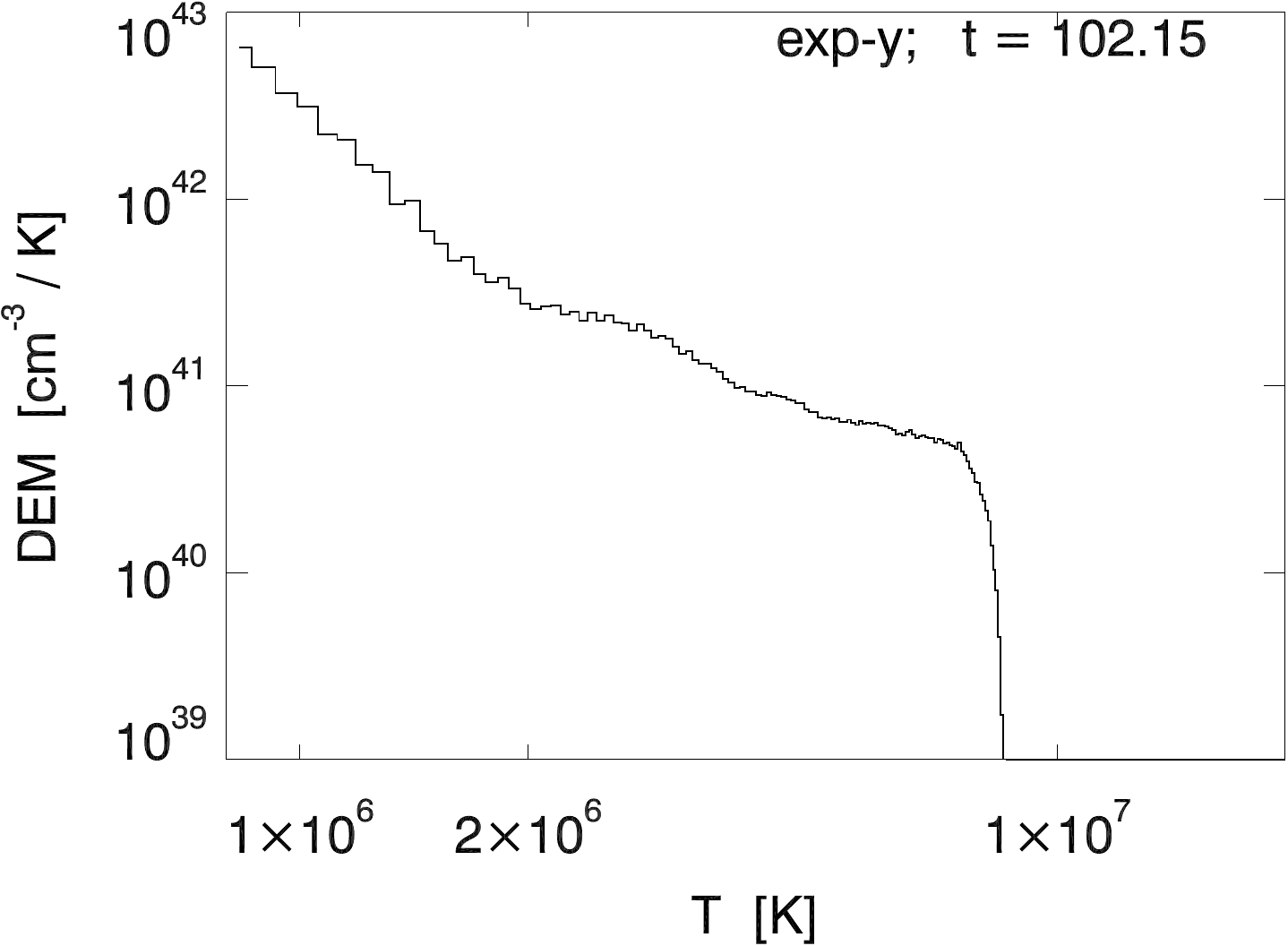} \\

  \includegraphics[width=.31\linewidth,clip=true,trim=0 00 0 0]{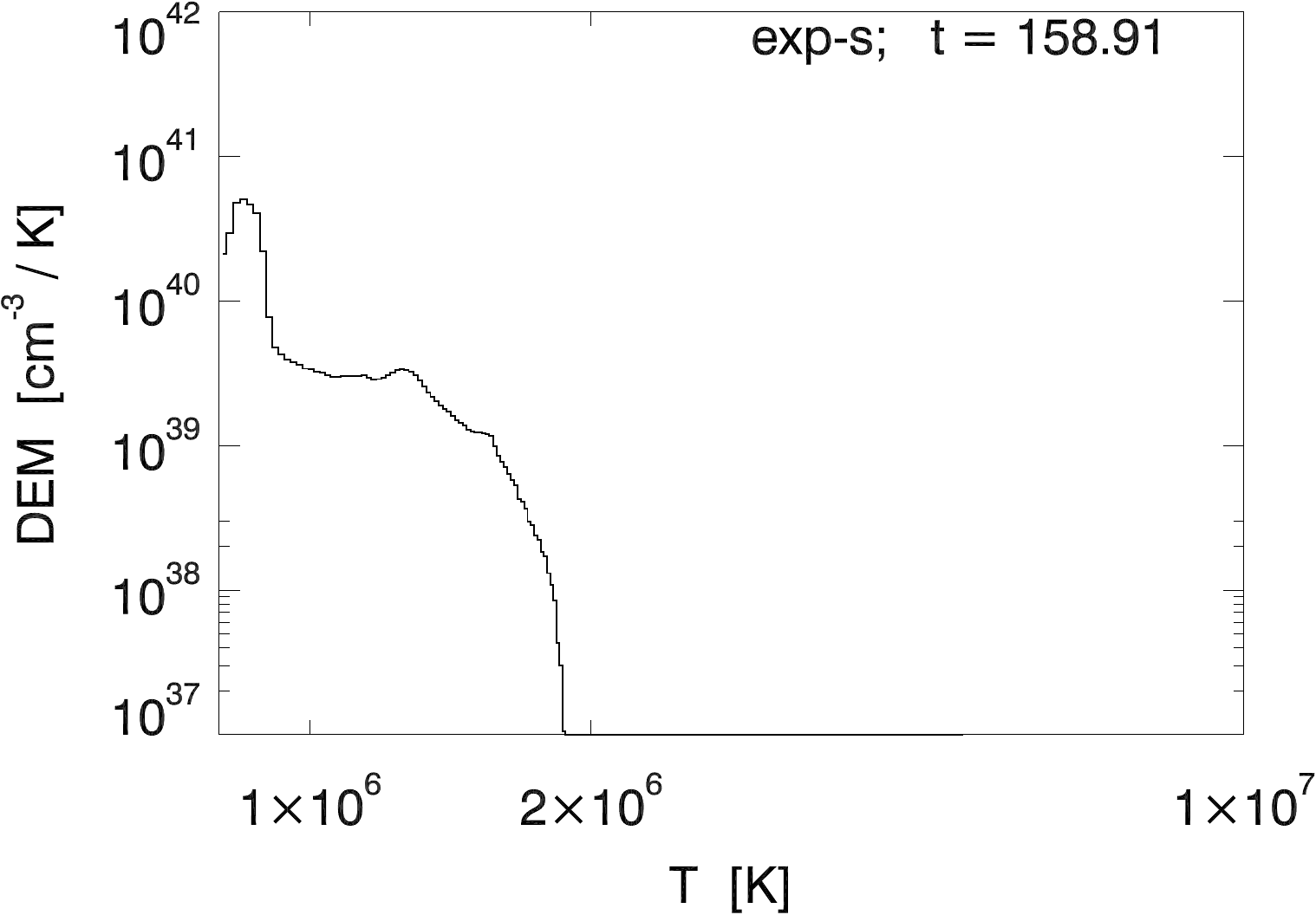} \hspace{0.01\linewidth}
  \includegraphics[width=.31\linewidth,clip=true,trim=0 00 0 0]{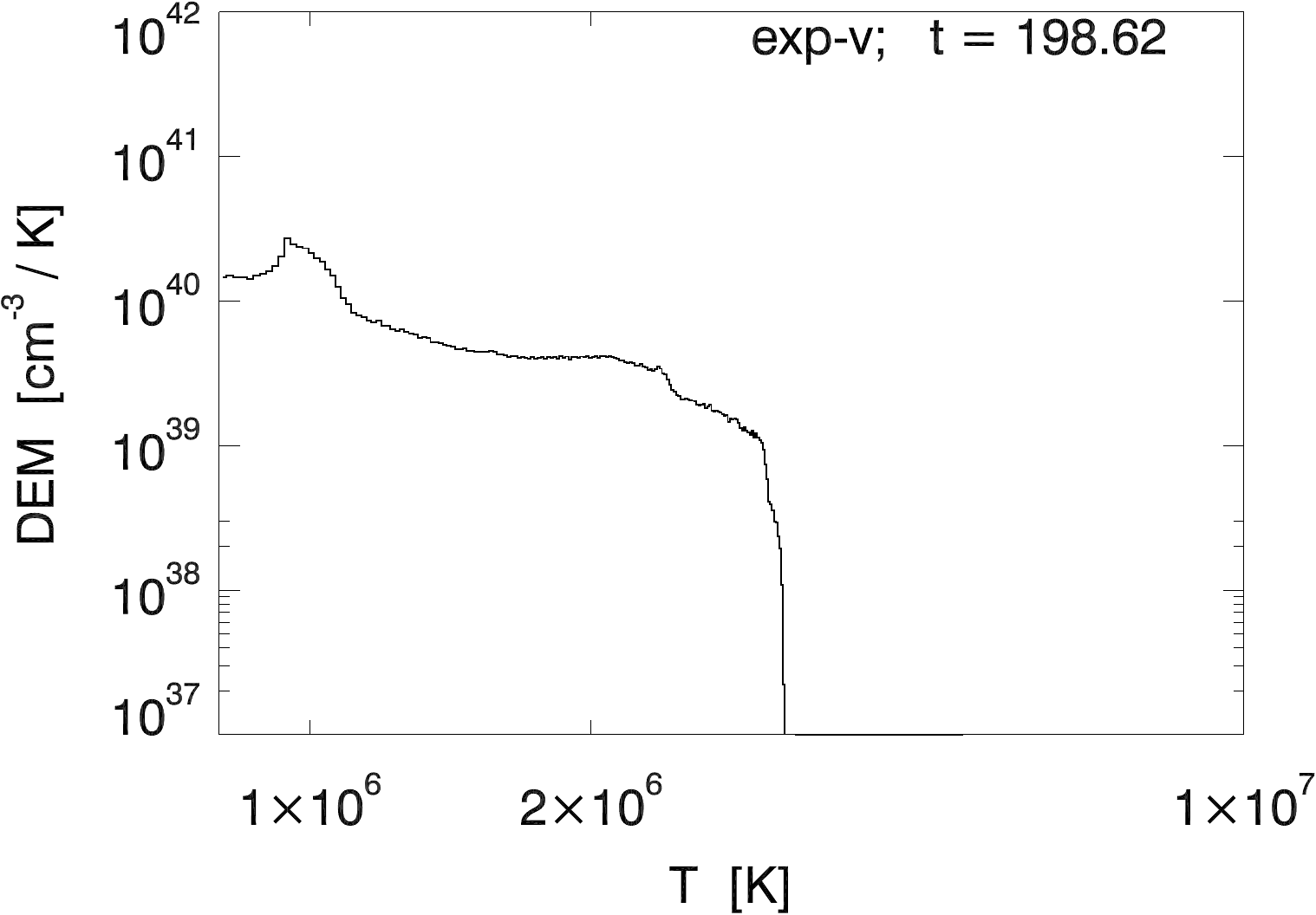} \hspace{0.01\linewidth}
  \includegraphics[width=.31\linewidth,clip=true,trim=0 00 0 0]{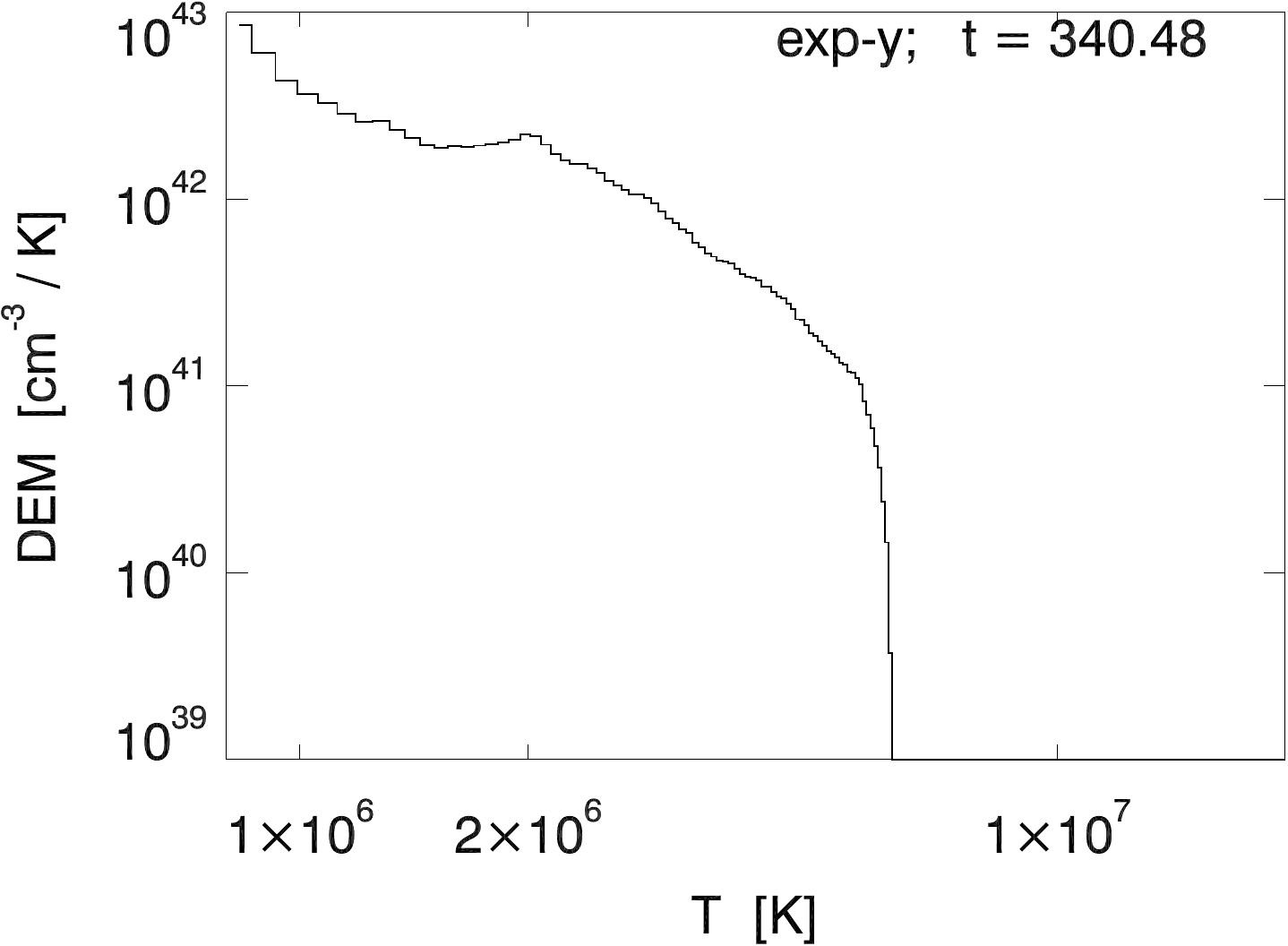} 

  \caption{
    Temporal evolution of the DEM for cases S, V and Y (left, middle and right columns, respectively).
    Time runs from the top to the bottom rows (as indicated in the inset legend on each individual panel, where time is indicated in units of $t_0$).
    For each one of the cases represented, the first row corresponds to the linear phase of the kink-instability, the second row represents the period of maximum thermal emission, and the third row the beginning of the relaxation phase.
    The upper tail which forms in the DEM distribution corresponds to the plasma heated after the development of the kink-instability in the magnetic flux-rope.
    This corresponds to a \emph{hot plasma component}, which is particularly distinguishable in case V.
    The maximum temperature reached depends mostly on flux-rope's magnetic field amplitude, approximately as $\sim B^2$ \citep[\emph{cf.}][]{pinto_soft_2015}.
  }
  \label{fig:dem}
\end{figure*}

The kink instability in the present MHD simulations develops when the total twist reaches around $7-9\pi$. 
It leads to the formation of a helical current sheet around the twisted flux-rope (see Fig. \ref{fig:blines_current}), in a way similar to that described by  \citet{brown_observations_2003,browning_heating_2008,hood_coronal_2009,gordovskyy_magnetic_2012,gordovskyy_particle_2014,pinto_soft_2015}.
This, in turn, switches on the anomalous resistivity (where and when the current density becomes stronger than $j_{crit}$; see Eq. \ref{eq-resist}), and leads to the start of the reconnection in the corona. 
There also are very strong currents near the loop foot-points, but the critical current threshold $j_{crit}$ in the lower atmosphere is much higher, and hence the resistivity there remains equal to the much lower background resistivity.
Hence, the reconnection and current dissipation occur predominantly in the coronal part of a loop. 
The current density distribution evolves gradually into a more fragmented pattern thereafter, until it fades away.
Magnetic reconnection occurs both within the twisted loop, and between twisted flux-rope field-lines and the ambient field.

The reconnection episode is followed by a magnetic relaxation phase lasting about $100-200\un{s}$ in the smaller models, and about $300-400\un{s}$ in the large-scale models. 
During this time, the magnetic energy is reduced by $\sim50-70~\%$ of the free magnetic energy in the system (Fig.~\ref{fig:energetics}).
The amount of free magnetic energy is defined here as the difference between the magnetic energy just before kink instability and the initial magnetic energy of the potential (untwisted) magnetic configuration. 
Most of this energy (around $80-90~\%$) is converted directly to thermal energy due to Ohmic dissipation, while a small fraction is released in form of plasma kinetic energy. 
The latter is then also converted to thermal energy due to viscous effects, mostly around shocks. 
A more detailed description of the velocity field in reconnecting twisted loops is given by Gordovskyy, et al (\emph{submitted}).

Ohmic dissipation is much enhanced in the strong helical current, leading to plasma heating.
%
Figures \ref{fig:tem} and \ref{fig:rho} show the spatial distribution of the plasma temperature and density at different stages of the evolution of the system.
Initially, the plasma temperature and density are everywhere equal to those of the initial gravitationally-stratified background (\emph{cf.} Fig. \ref{fig:atmos}).
During the thermal relaxation stage, thermal conduction increases the temperature and decreases the density in the transition region, while reducing the temperature and increasing the density in the lower corona.
Foot-point twisting leads to an increase in magnetic pressure inside the twisted loop, making it expand.
This expansion, in turn, results in a slight drop in plasma density and temperature inside the loops (by adiabatic expansion).
Just before the kink instability occurs, and near the loop-top, the temperature is about $40-50\%$ lower than the temperature in the initial gravitationally-stratified atmosphere (see Figures \ref{fig:tkink} and \ref{fig:tem}(a)), while the density is $50-60\%$ lower (Fig.~\ref{fig:rho}(a)).
The picture is more complicated closer to the foot-points: plasma around the expanding twisted flux-tube is being pushed vertically rather than horizontally due to the strong vertical density gradient. 
This results in accumulation of mass in a cylindrical shell around the loop's legs.
The density in these shells is about $2.5$ times higher than outside, while the temperature is about $1.5-2$ times higher due to adiabatic compression.
After the reconnection starts, the high temperature distributions near the loop-top resemble the current density distributions \citep[e.g.][]{hood_coronal_2009}, starting with a helical shell, which then gradually becomes more fragmented and fades away. 
In contrast, closer to foot-points the hot plasma can be seen on the central axis of the loop. 
The density near the loop top evolves from a regular structure towards a fragmented one.
Also, a substantial increase in plasma density after onset of reconnection can be observed in the loop's legs, which is due to plasma evaporation from the lower atmospheric layers.

Figure \ref{fig:dem} shows the $\mathrm{DEM}\left(T\right)$ of the coronal plasma at different instants of the simulation for three different models (S, V and Y).
The DEM distribution extends into higher temperatures during the linear phase of the kink instability in all the models, forming a distinctive upper tail.
The total emission measure of this hotter plasma component keeps increasing until the peak phase is reached, and a clear hot peak is sometimes visible (especially in model V).
The bulk of this hotter plasma reaches temperatures about twice the temperature of the background plasma in these simulations.
The DEM distribution flattens out afterwards, during the magnetic relaxation phase, but the maximum temperature remains higher than initially for a long period of time.
Overall, the temporal evolution of the DEM analysed here is very similar to those discussed by \citet{pinto_soft_2015} in respect to a simpler model of a kink-unstable flaring loop.
They related the formation and the dissipation of hot flare plasma components in the DEM to observations by, \emph{e.g}, \citet{reale_evidence_2009}, \citet{battaglia_rhessi_2012} and \citet{sylwester_solar_2014}.
The only remarkable differences are that in our case the flare plasma reaches lower maximum temperatures and that the hot plasma component that forms during the course of the flare is less pronounced (because the ratio background to flare plasma is also higher here).

\section{SXR and HXR emission}
\label{sec:emission}

\begin{figure*}
  \centering

  \begin{minipage}{.45\linewidth}
    \centering
    {\sf Emissivity ($2\un{keV}$)} \\
    \includegraphics[width=0.45\linewidth,clip=true,trim=0 0 0 80]{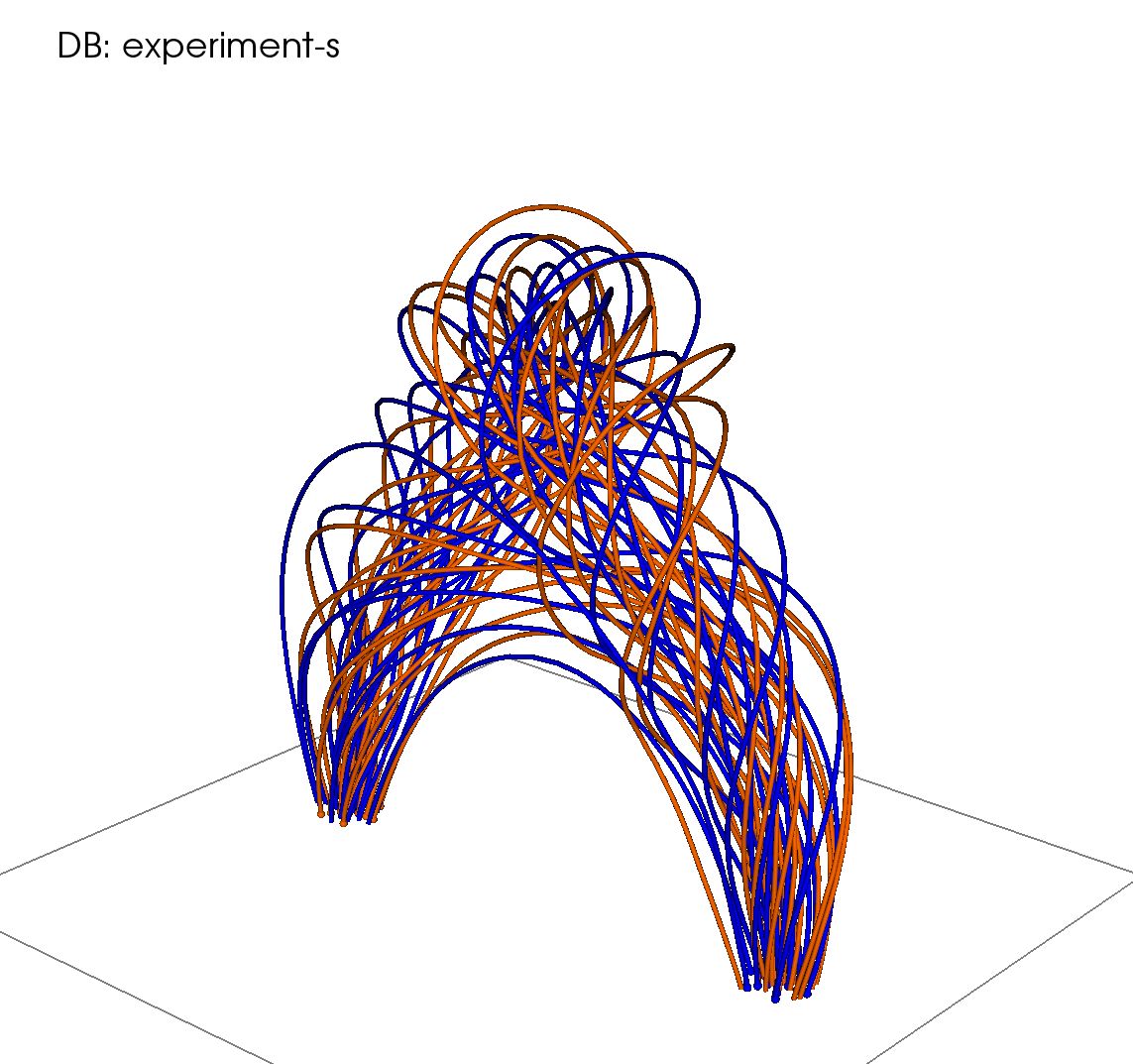} 
    \includegraphics[width=0.45\linewidth,clip=true,trim=0 0 0 80]{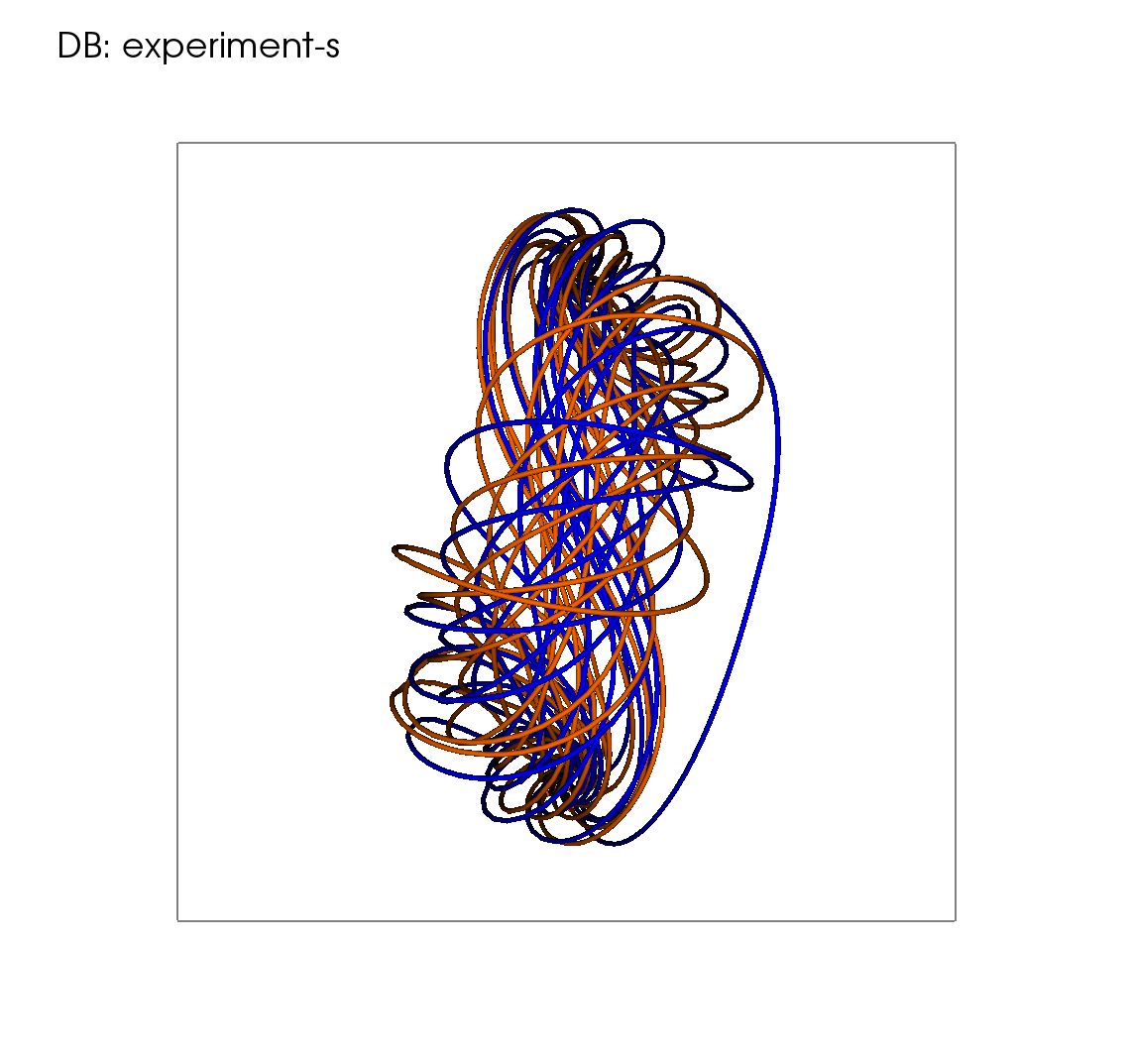}\\
 
    \includegraphics[width=0.45\linewidth,clip=true,trim=0 0 0 80]{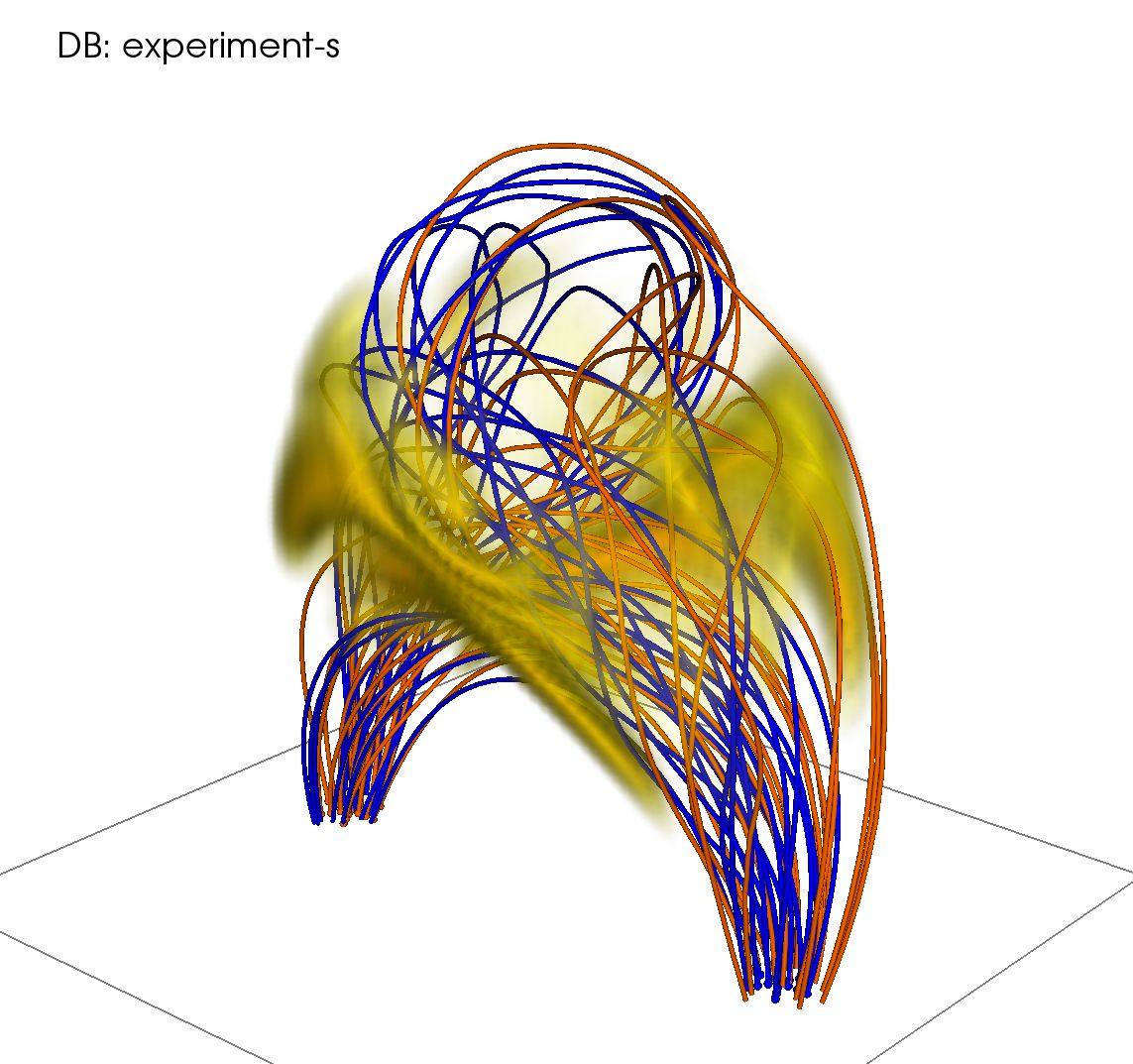} 
    \includegraphics[width=0.45\linewidth,clip=true,trim=0 0 0 80]{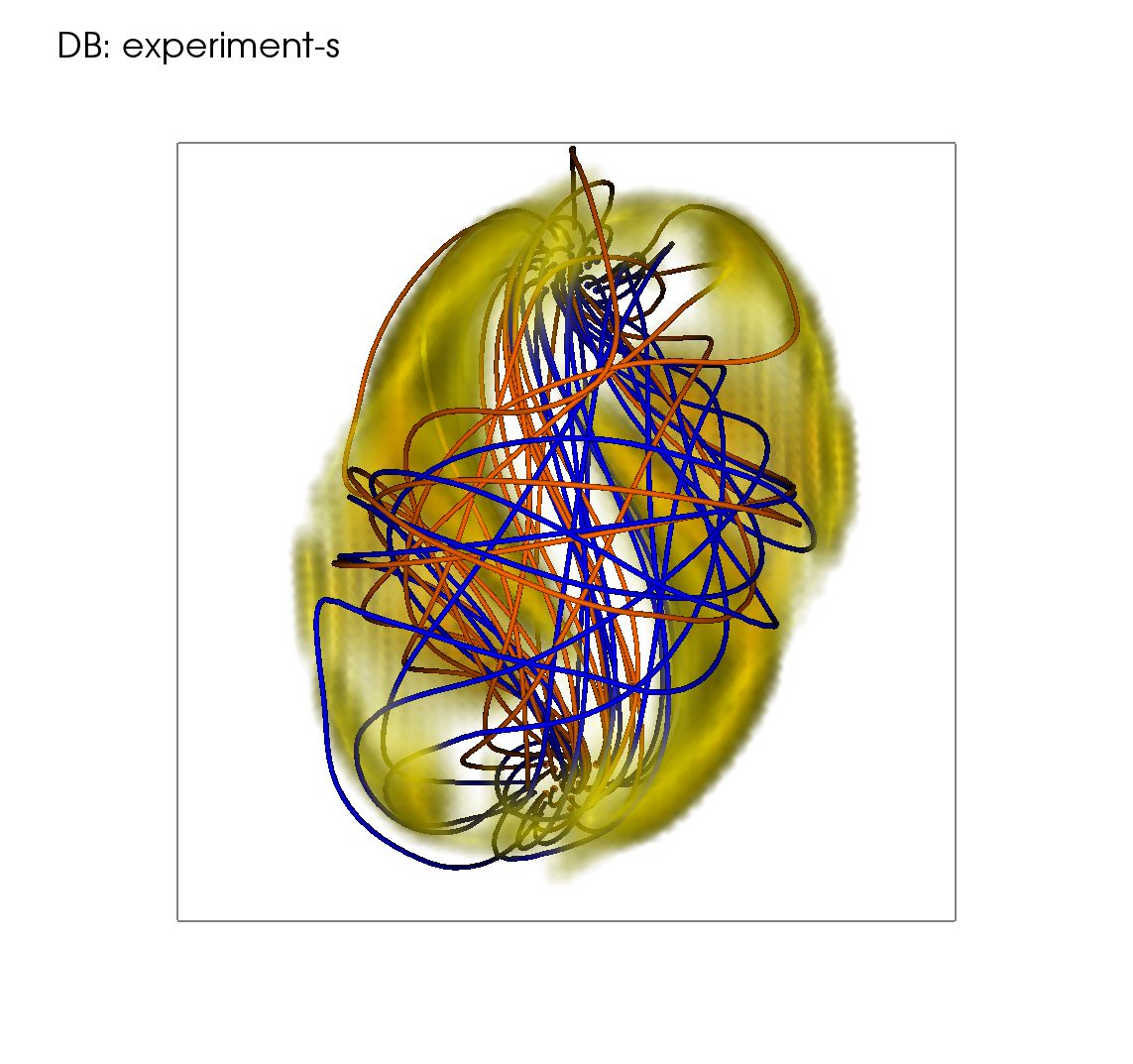}\\

    \includegraphics[width=0.45\linewidth,clip=true,trim=0 0 0 80]{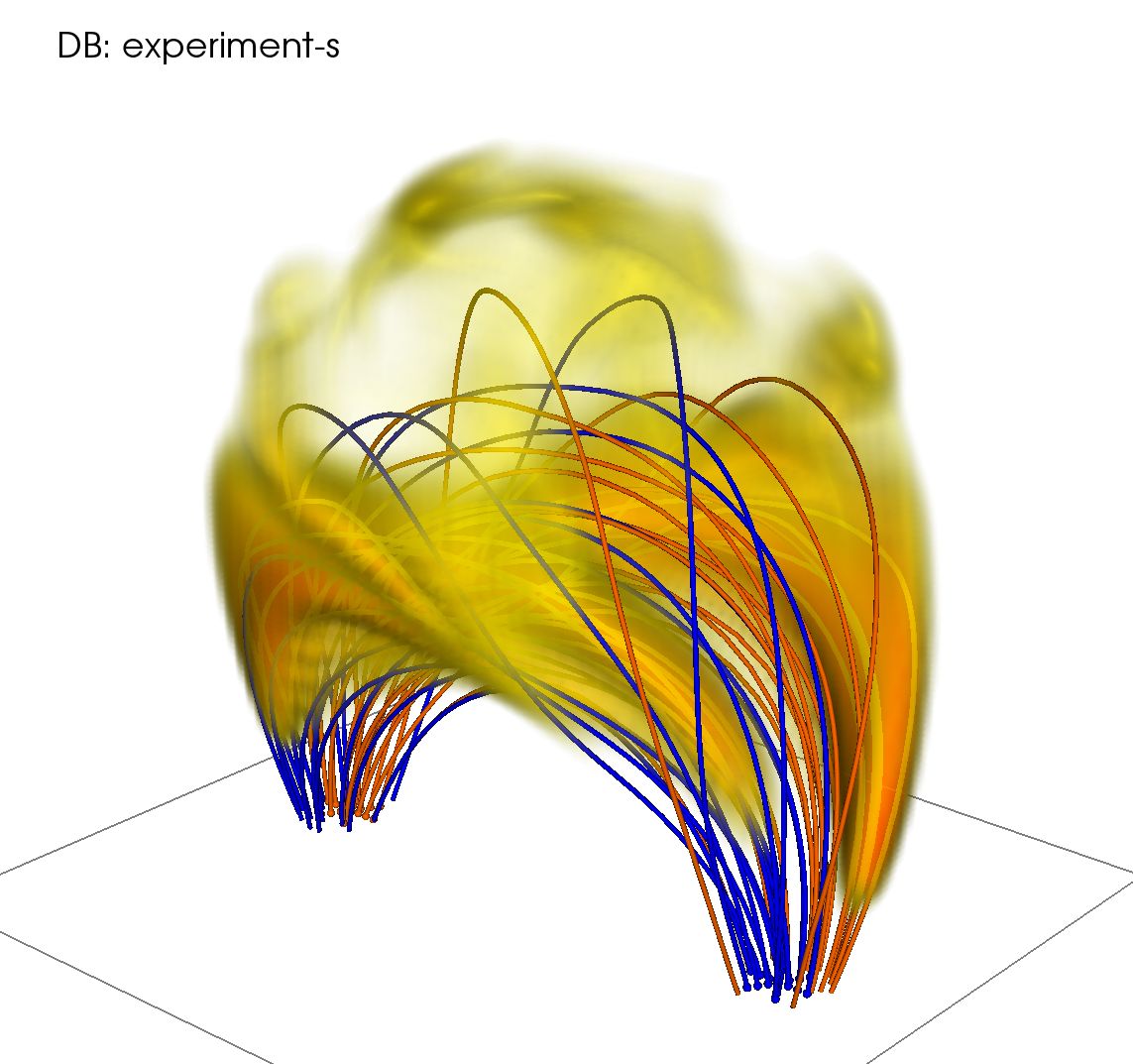} 
    \includegraphics[width=0.45\linewidth,clip=true,trim=0 0 0 80]{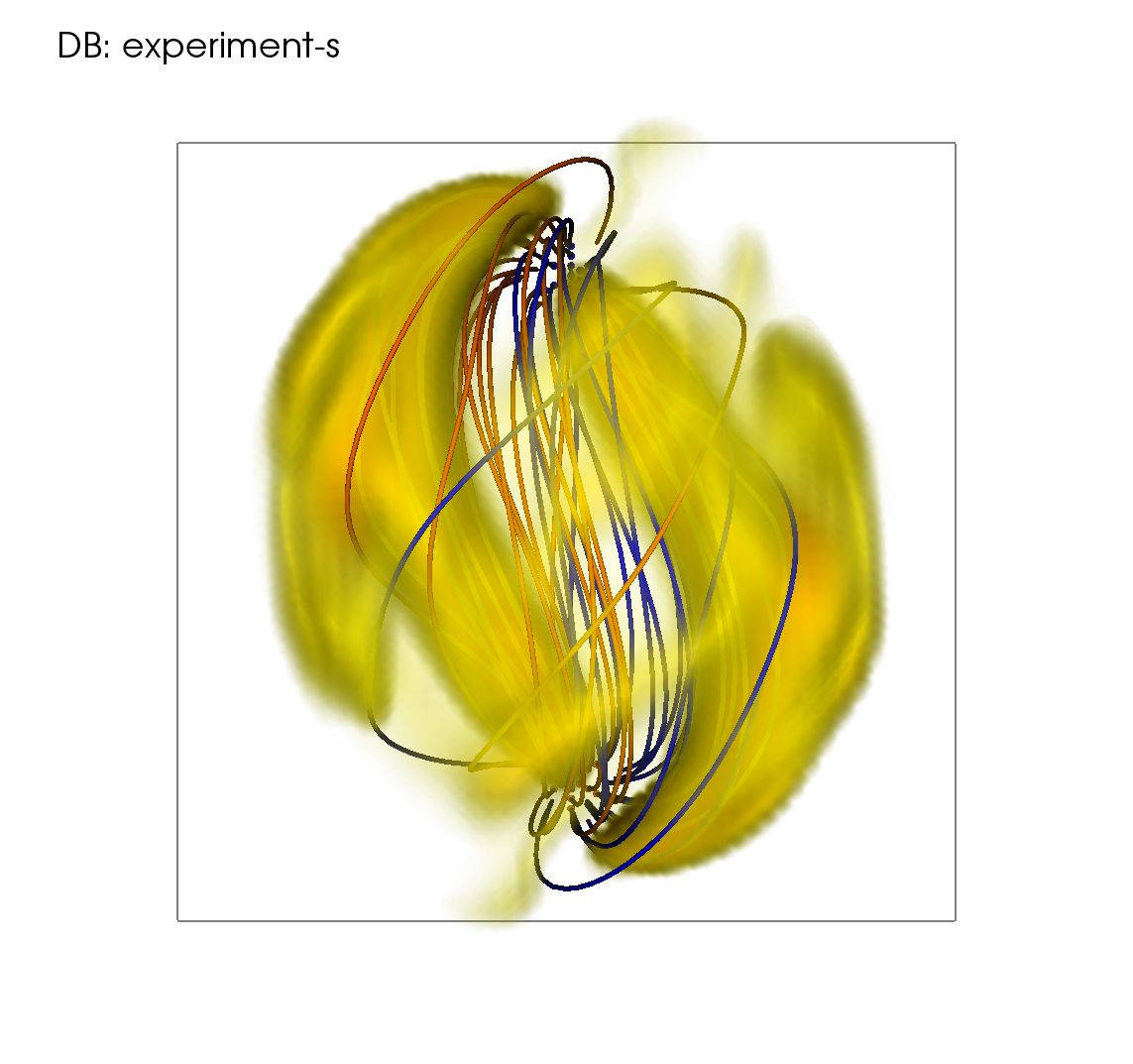}\\

    \includegraphics[width=0.45\linewidth,clip=true,trim=0 0 0 80]{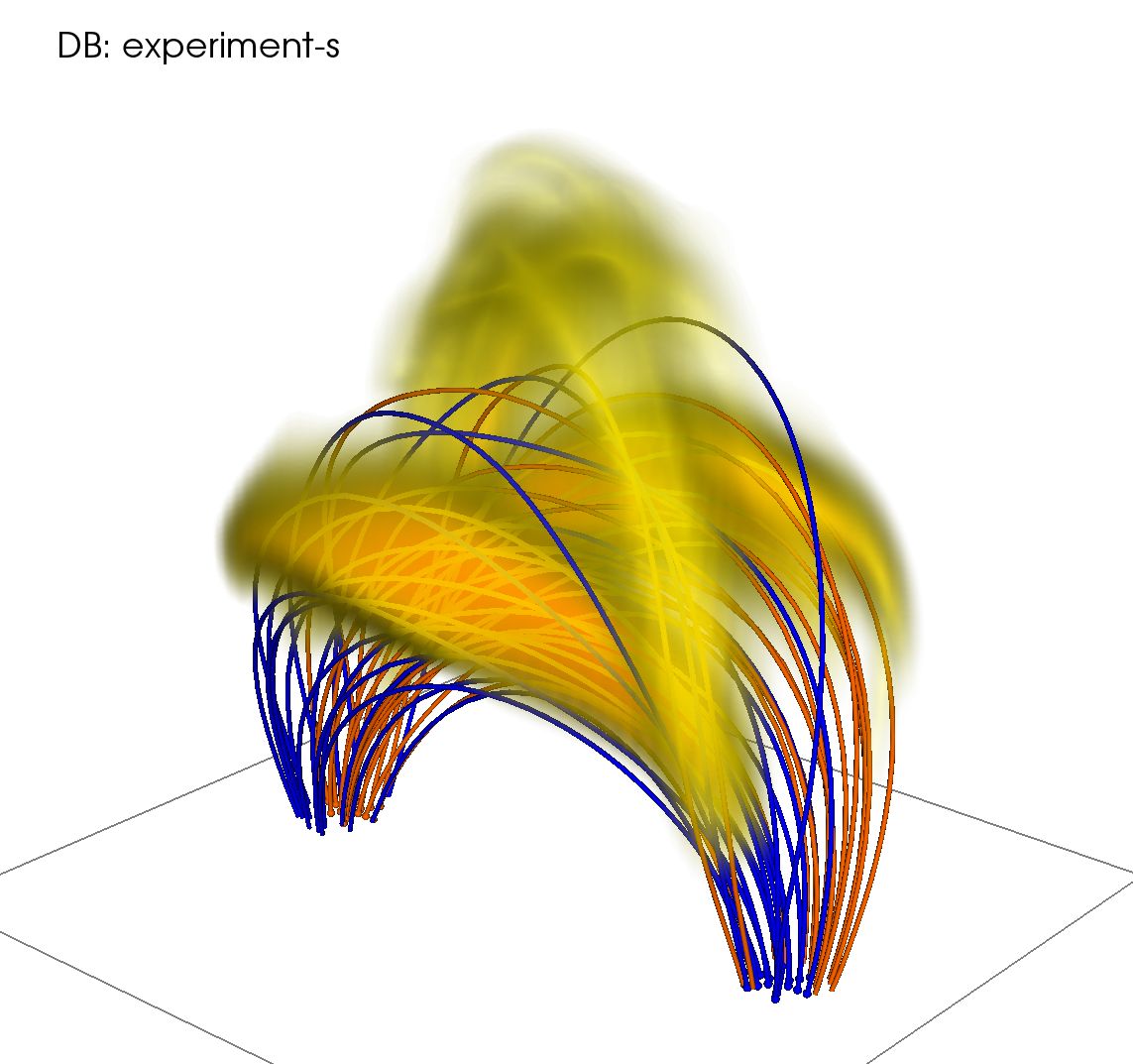} 
    \includegraphics[width=0.45\linewidth,clip=true,trim=0 0 0 80]{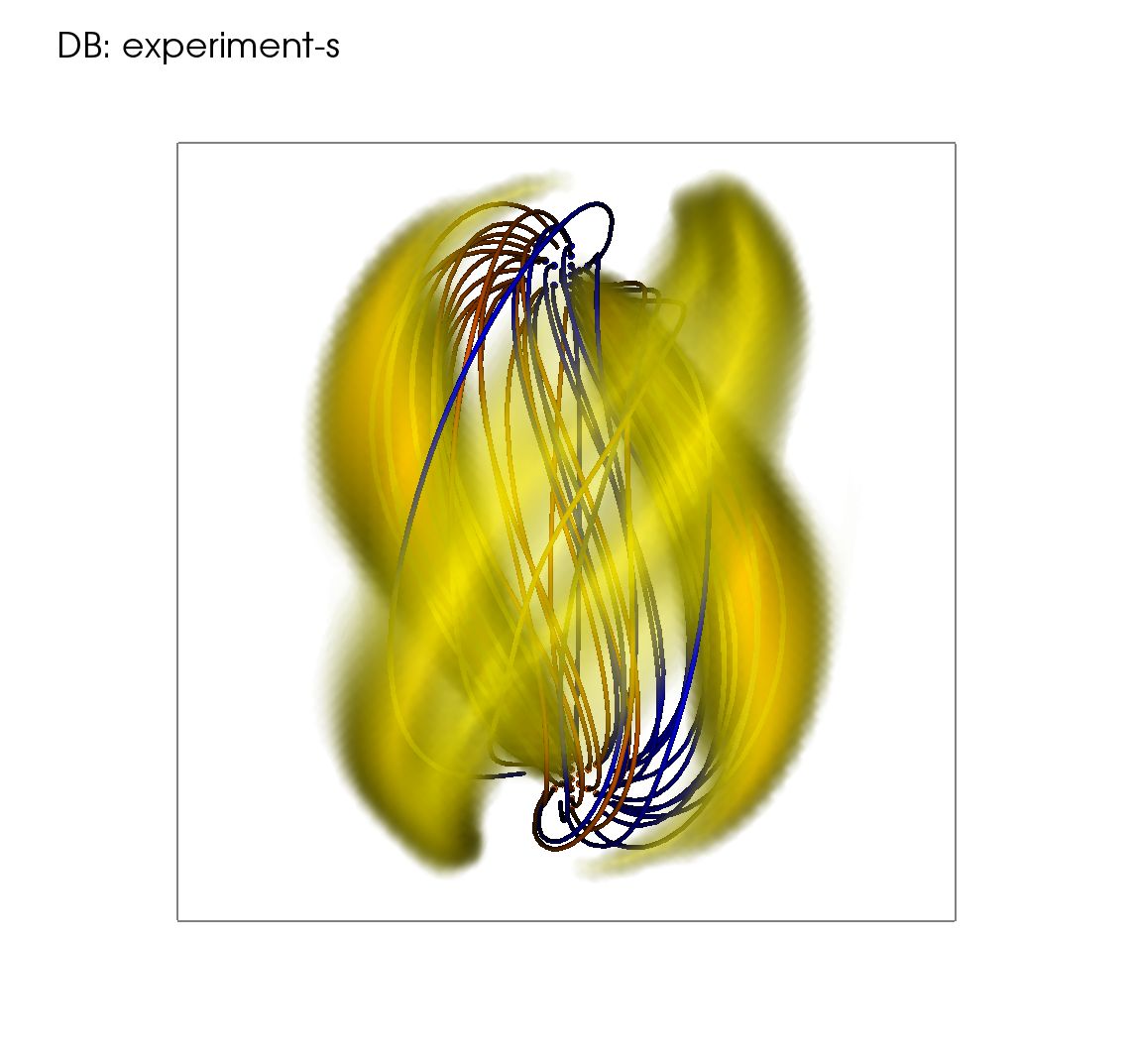} \\

    \bigskip
    
    \flushright
    \includegraphics[width=.55\hsize,clip=true,trim=0 6 0 3]%
    {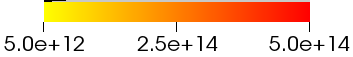} 
  \end{minipage}
  \hspace{0.025\linewidth}
  \begin{minipage}{.45\linewidth}
    \centering
    {\sf Temperature} \\
    \includegraphics[width=0.45\linewidth,clip=true,trim=0 0 0 80]{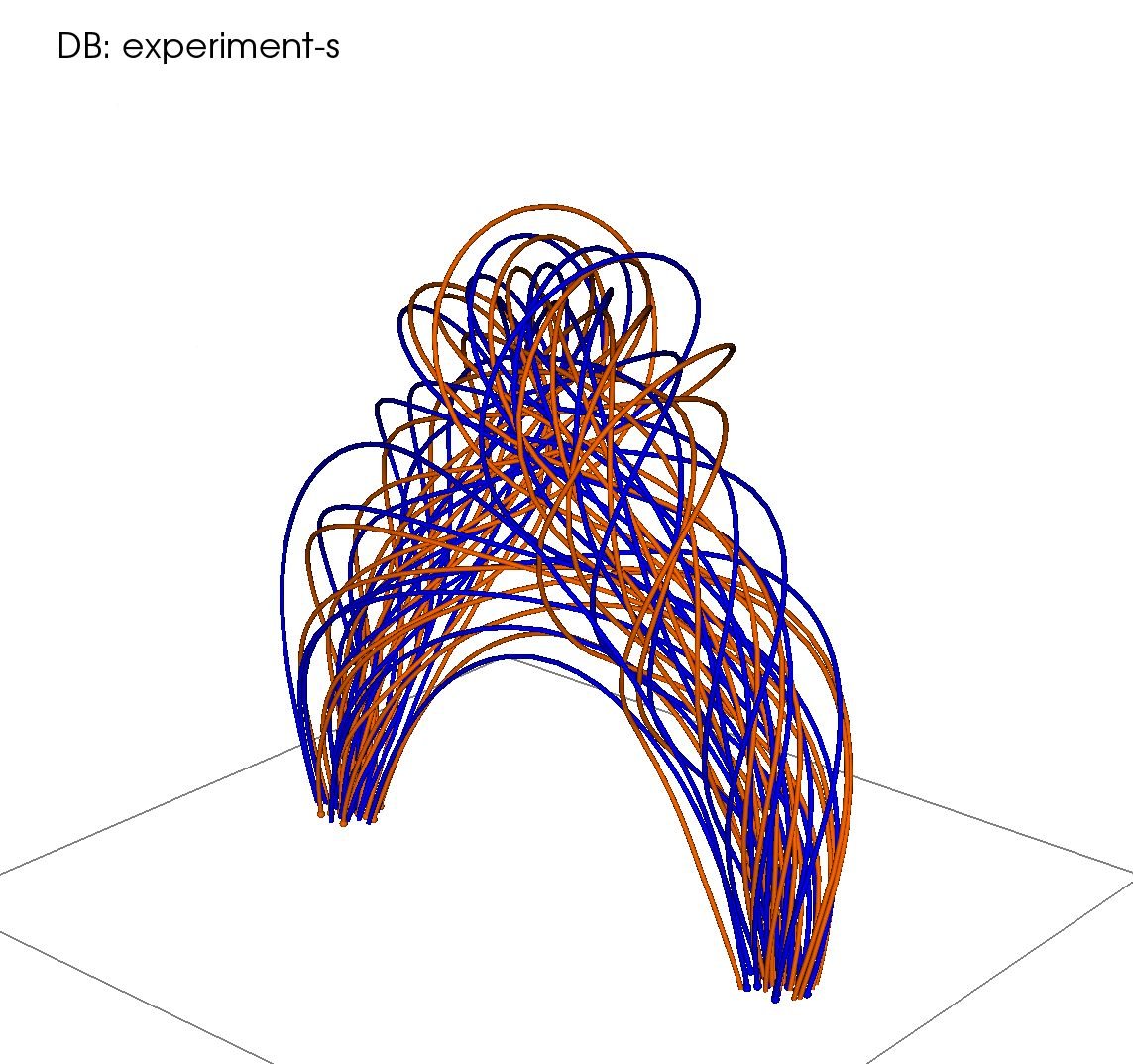} 
    \includegraphics[width=0.45\linewidth,clip=true,trim=0 0 0 80]{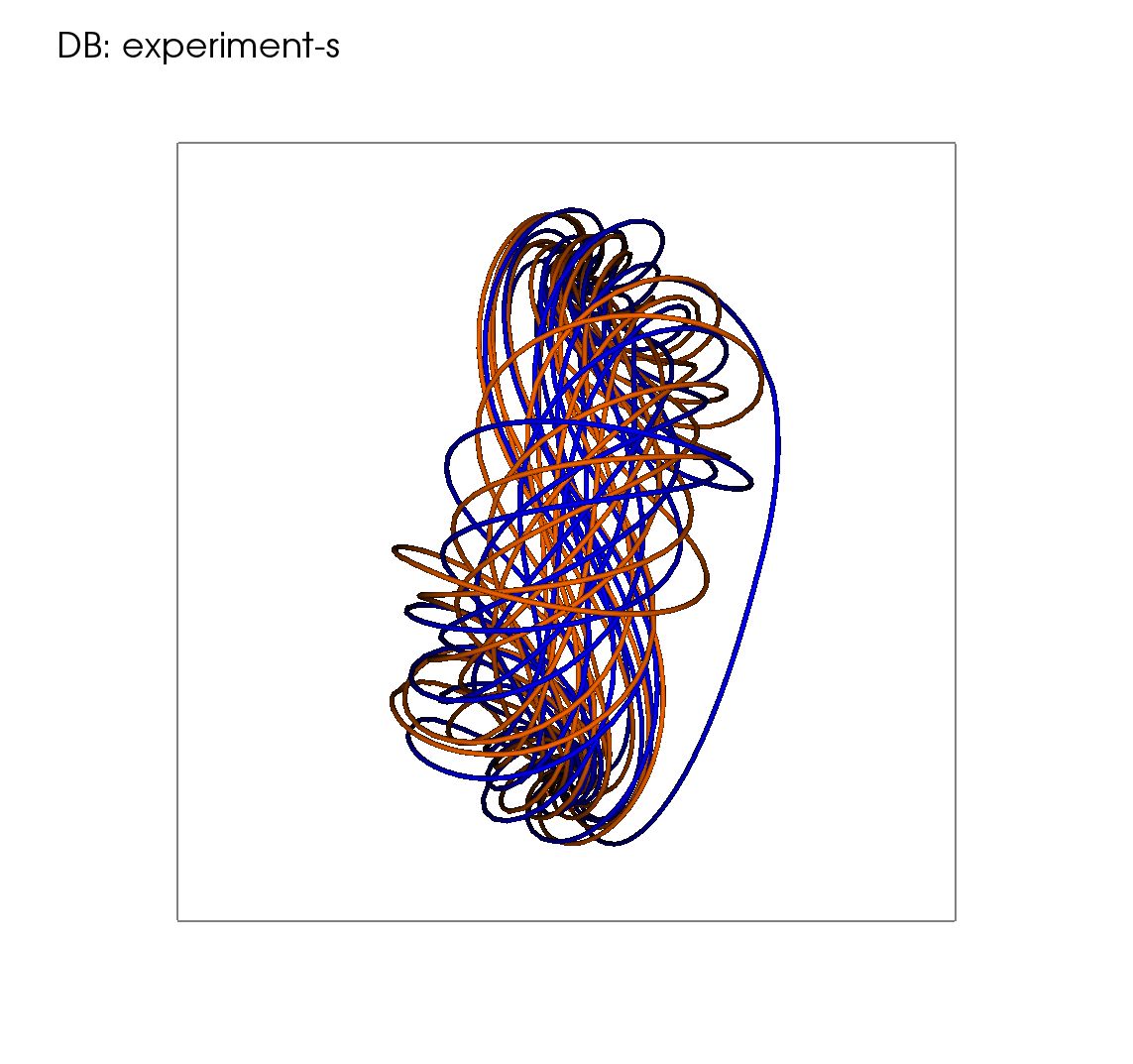}\\
 
    \includegraphics[width=0.45\linewidth,clip=true,trim=0 0 0 80]{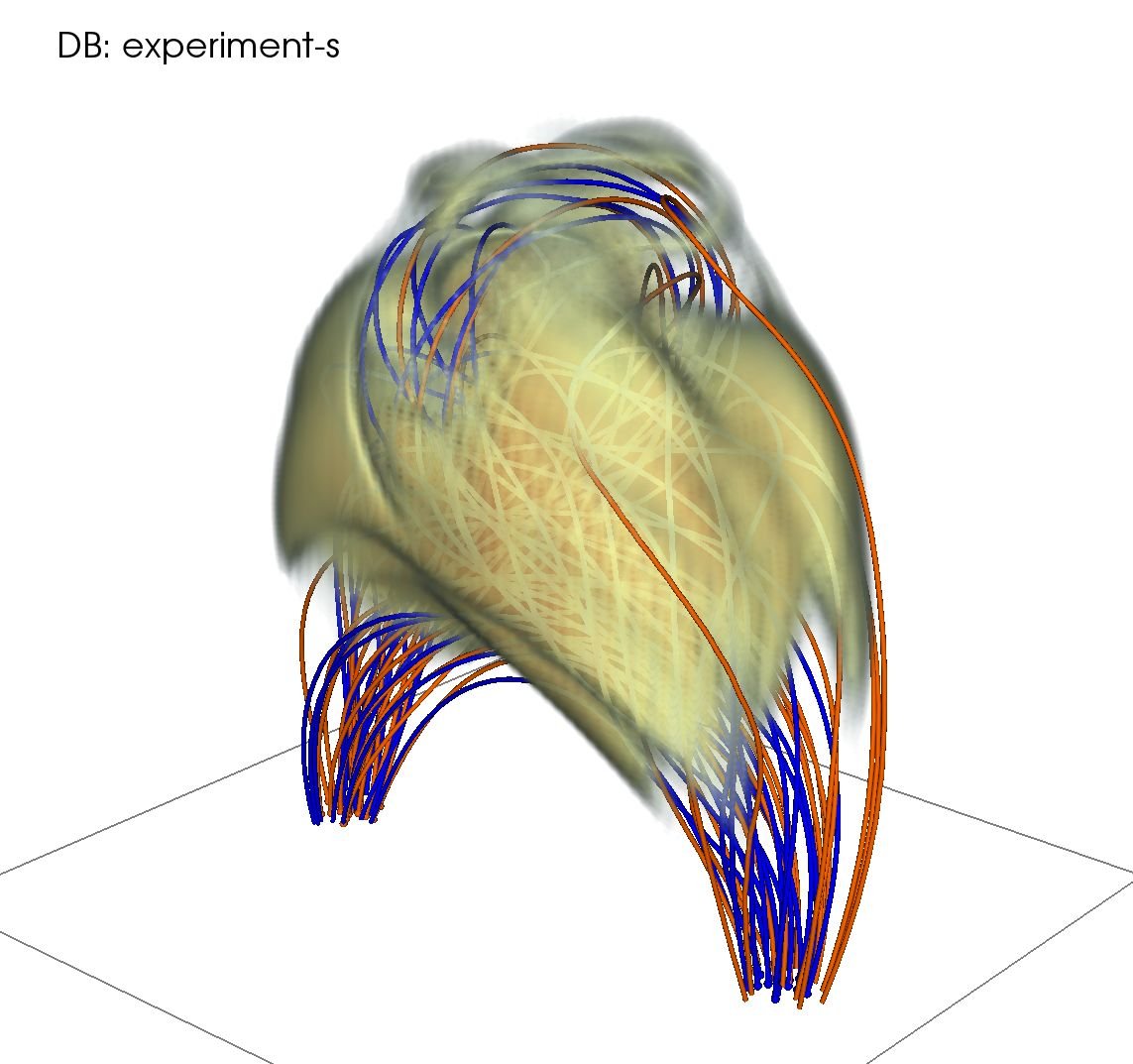} 
    \includegraphics[width=0.45\linewidth,clip=true,trim=0 0 0 80]{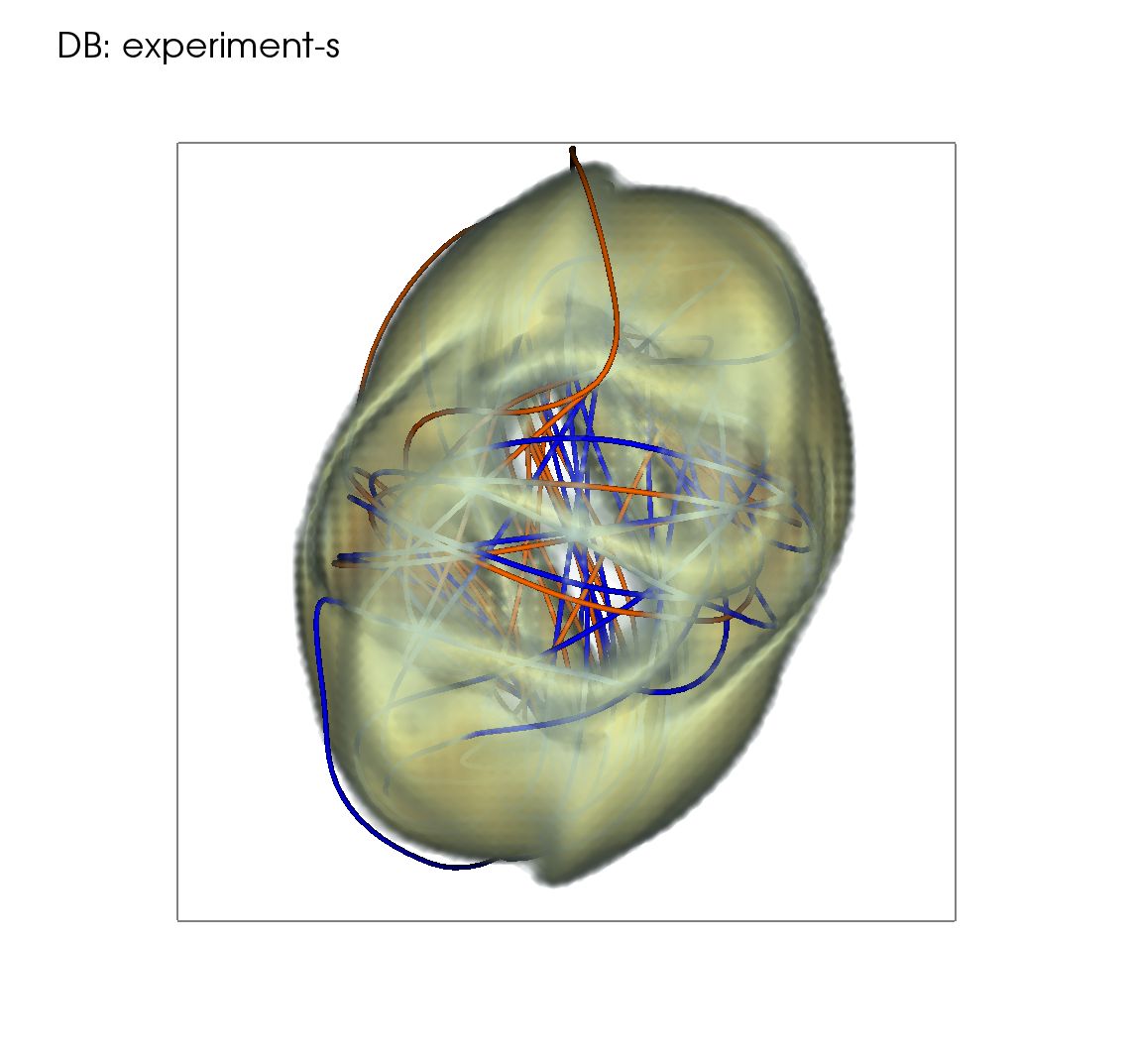}\\

    \includegraphics[width=0.45\linewidth,clip=true,trim=0 0 0 80]{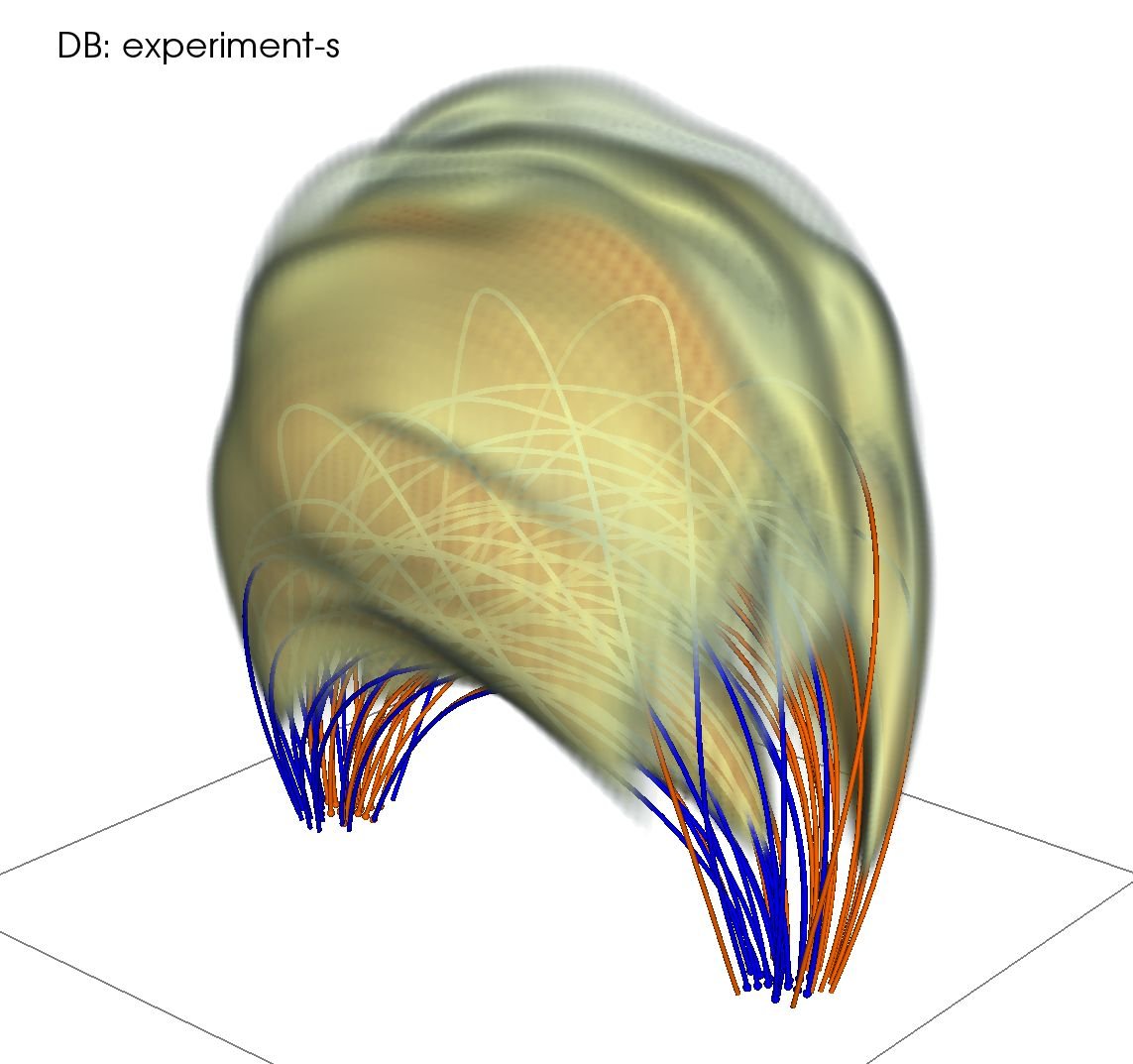} 
    \includegraphics[width=0.45\linewidth,clip=true,trim=0 0 0 80]{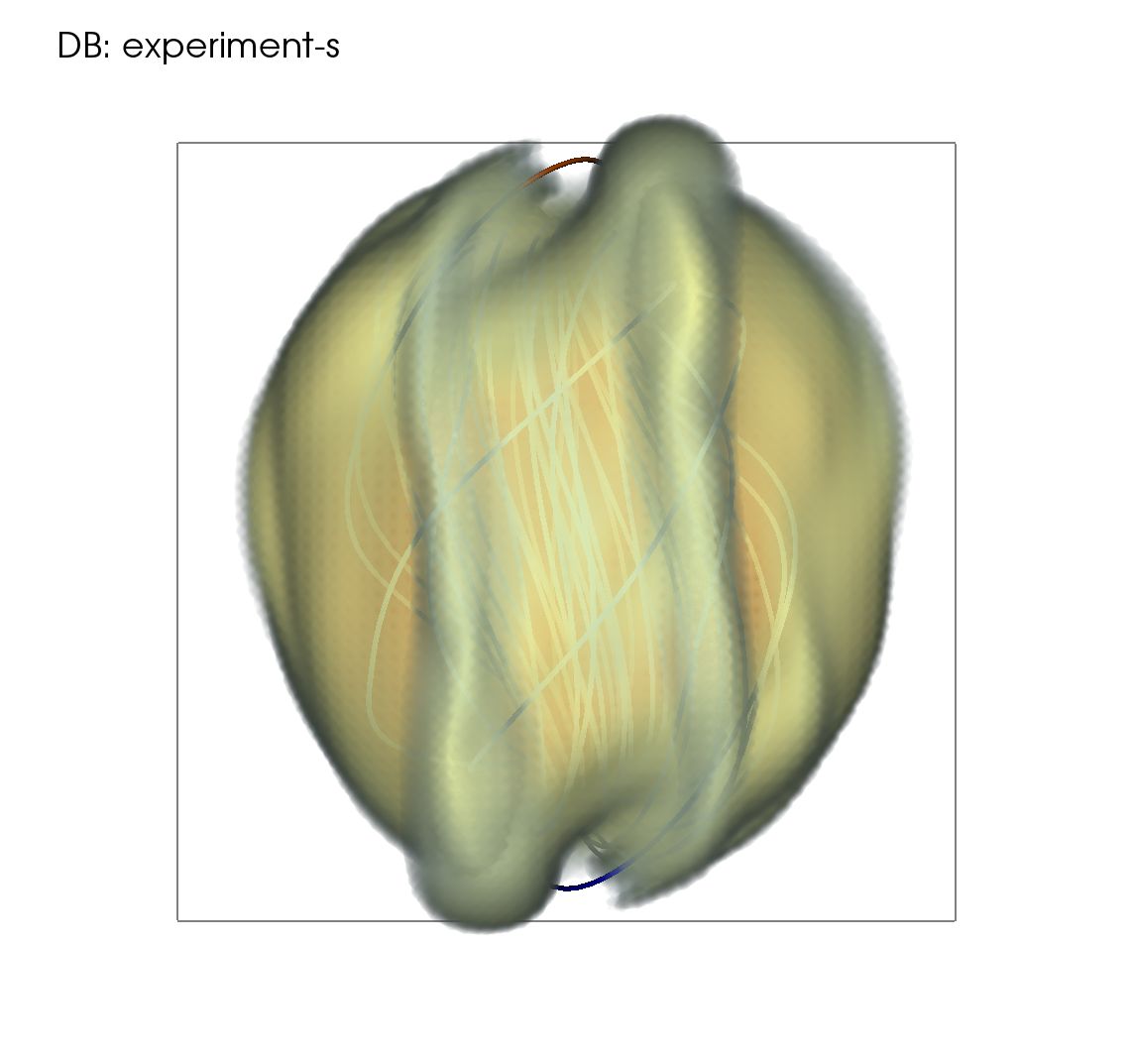}\\

    \includegraphics[width=0.45\linewidth,clip=true,trim=0 0 0 80]{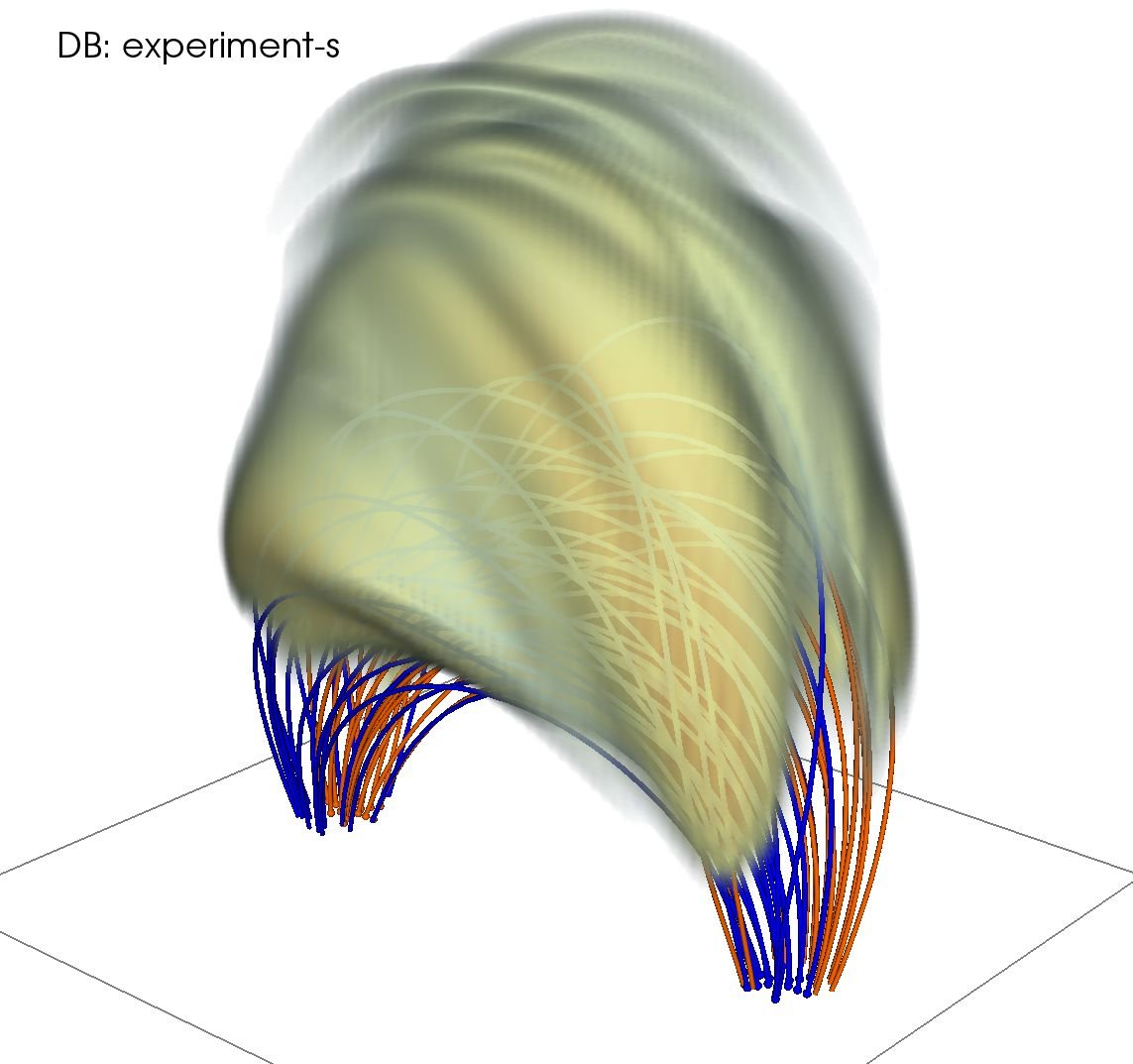} 
    \includegraphics[width=0.45\linewidth,clip=true,trim=0 0 0 80]{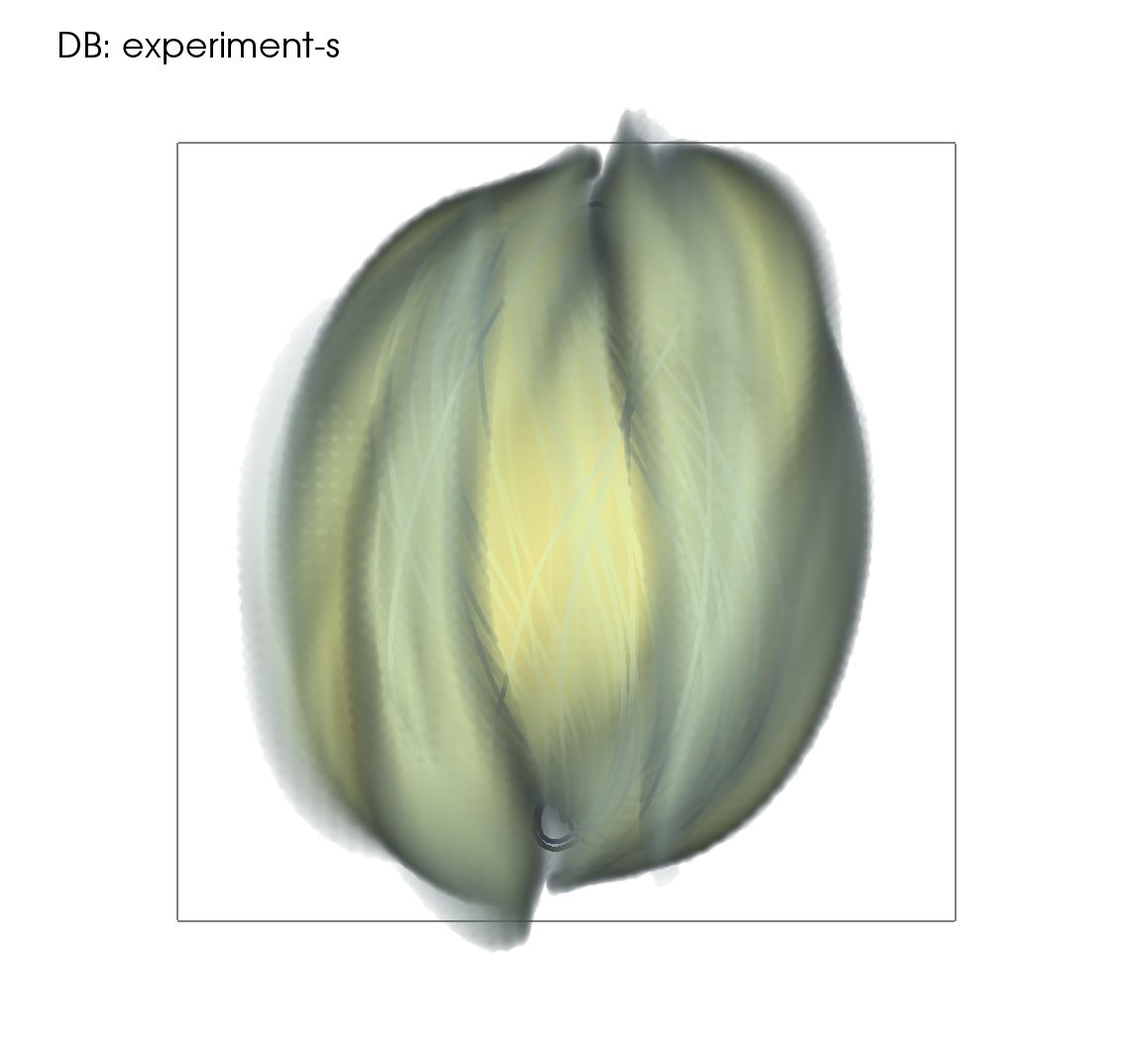} \\

    \bigskip
    
    \flushright
    \includegraphics[width=.55\hsize,clip=true,trim=0 0 0 0]%
    {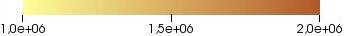} 
  \end{minipage}

  \caption{
      Volume renderings of the thermal continuum emissivity at $2\un{keV}$ (left) and of the plasma temperature (right) at different instants for model S.
      The left and right columns of each plot show, respectively, a perspective view and a top view of the same loop.
      The colour-scales represent emissivities ($\epsilon$; see eq. \ref{eq:emissivity}) between $5\e{12}$ and $5\e{14} \un{erg\cdot s^{-1}\cdot cm^{-3}\cdot Hz^{-1}}$ and plasma temperatures between $1$ and $2\un{MK}$.
      The orange and blue lines represent magnetic field-lines rooted, respectively, at the left and at the right foot-points.
      The instants represented are, from top to bottom, $t=34,\ 80,\ 136$ and $227\un{s}$.
      The grey line delimits the bottom face of the numerical domain (a square of dimensions $20\times 20 \un{Mm}$).  
  }
  \label{fig:thermal_vol}
\end{figure*}

The combination of the variations in plasma density and temperature give rise to the soft X-ray thermal emission patterns displayed in Figs. \ref{fig:thermal_vol}.
The figure shows volume renderings of the thermal continuum emissivity at $2\un{keV}$ at several instants ($t=34,\ 80,\ 136$ and $227\un{s}$, respectively, from top to bottom) together with renderings of the plasma temperature.
The left and right columns of each subplot show the same loop from different points of view (side/perspective and top views).
The orange and blue lines represent magnetic field-lines rooted at the left and at the right foot-points, respectively.
On the left subplot, the yellow/orange volumes represent the plasma emissivity $\epsilon$ at $2\un{keV}$ (see Eq. \ref{eq:emissivity}), ranging from $\epsilon = 5\e{12}$ (light yellow) to $\epsilon = 5\e{14} \un{erg\cdot s^{-1}\cdot cm^{-3}\cdot Hz^{-1}}$ (dark orange).
On the right subplot, the light yellow/dark orange volumes represent temperatures ranging from $1\un{MK}$ to $2\un{MK}$.
It is notorious that the continuum thermal plasma emissivity does not exactly follow the evolution of the plasma temperature.
  The difference is due to the variations in plasma density which result from the turbulent dynamics of the kinking loop -- note the dependence on density in Eq. (\ref{eq:emissivity}) --, as was also the case in \citet{pinto_soft_2015}.

\begin{figure*}
  \centering
  \includegraphics[width=.85\linewidth,clip=true,trim=0 410 0 380]{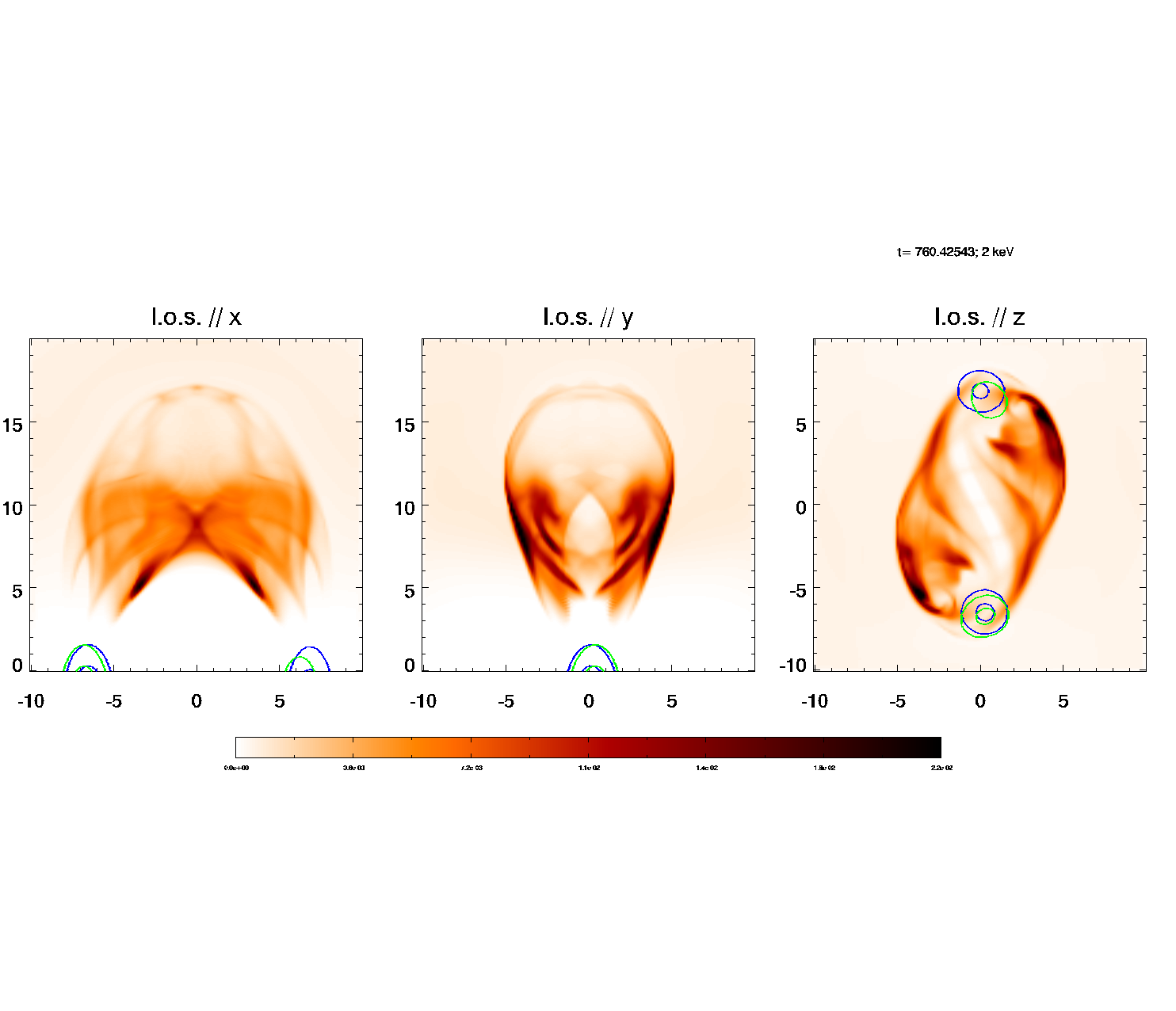} \\
  \includegraphics[width=.85\linewidth,clip=true,trim=0 410 0 420]{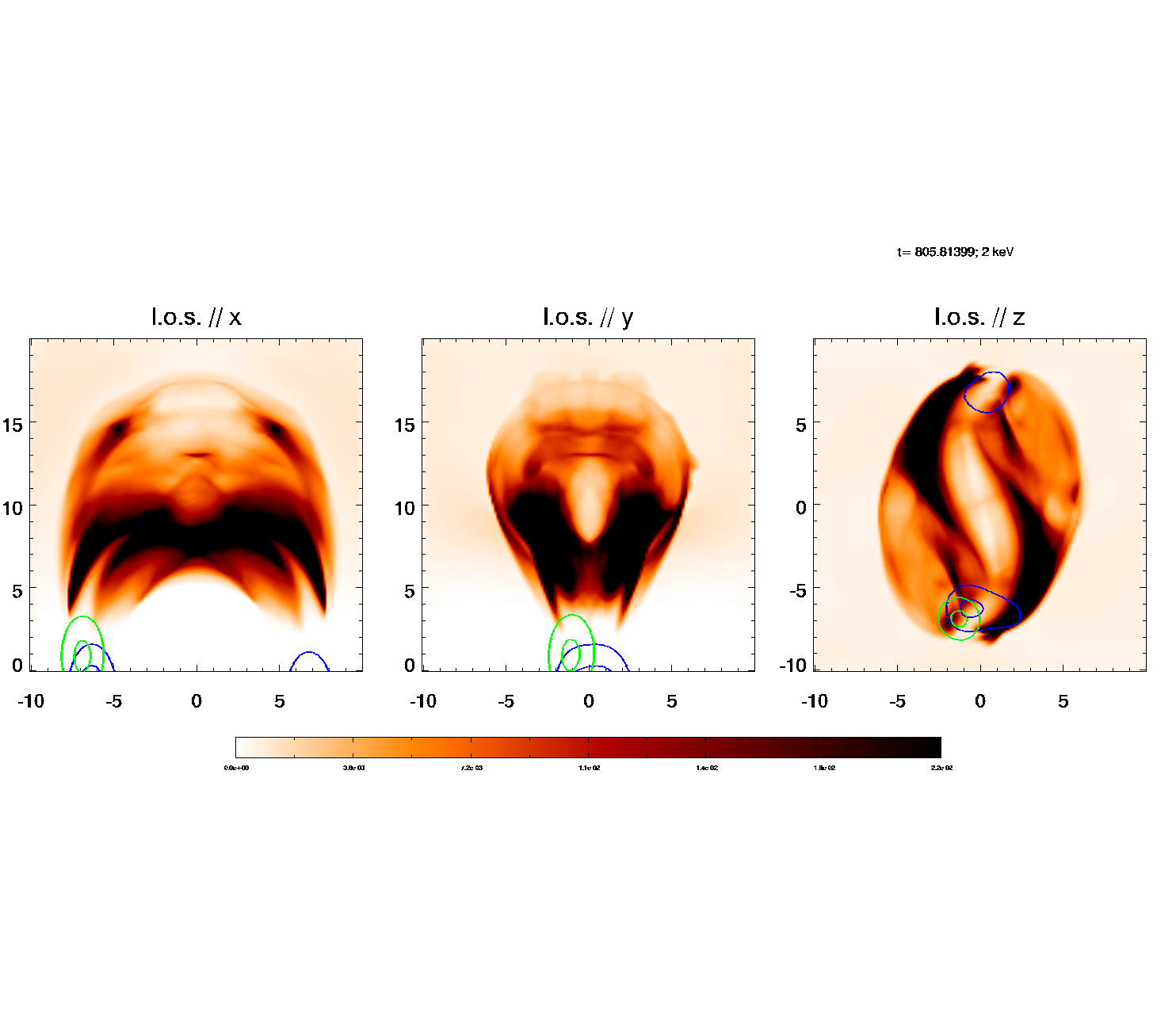} \\
  \includegraphics[width=.85\linewidth,clip=true,trim=0 410 0 420]{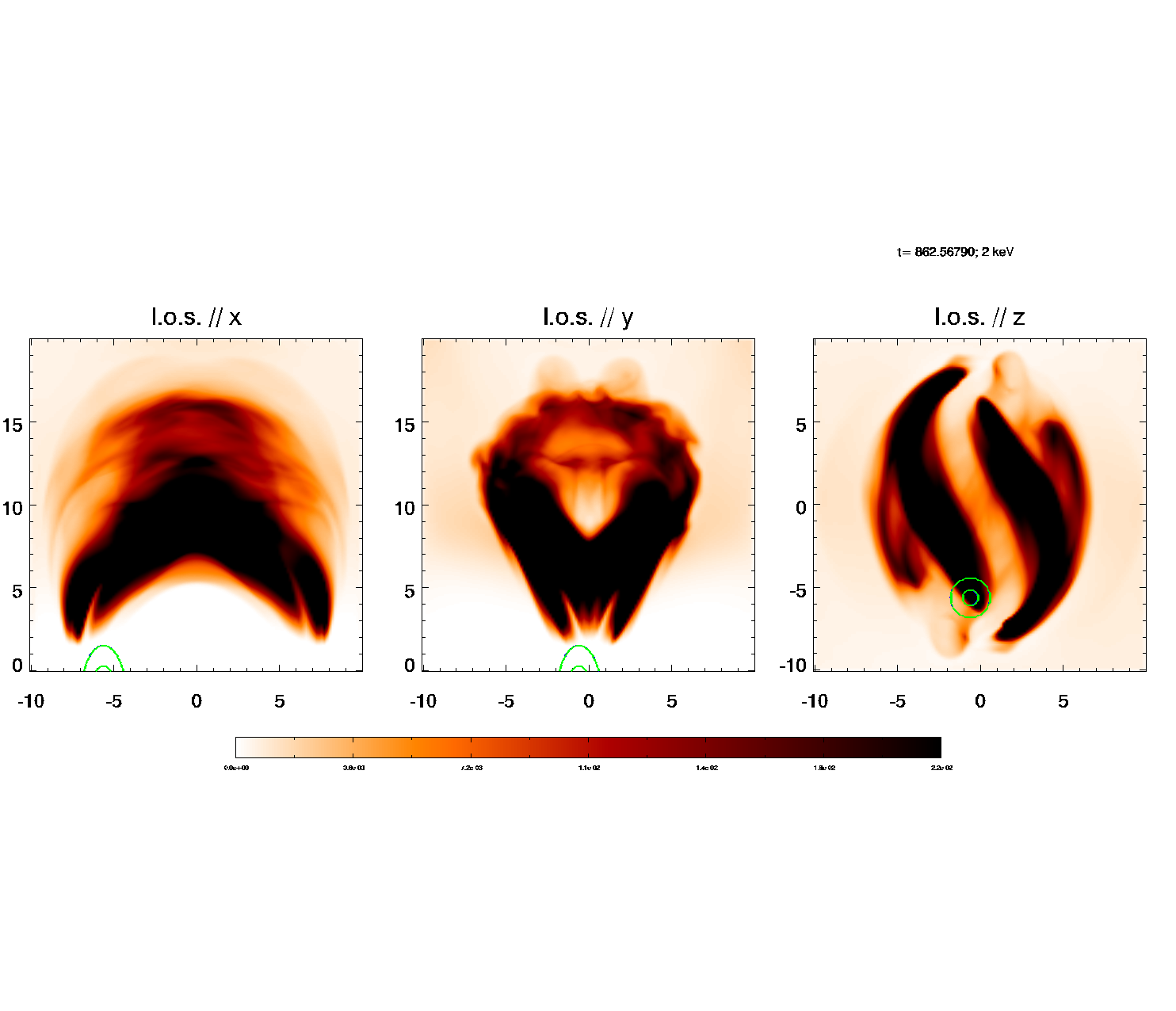} \\
  \includegraphics[width=.85\linewidth,clip=true,trim=0 410 0 420]{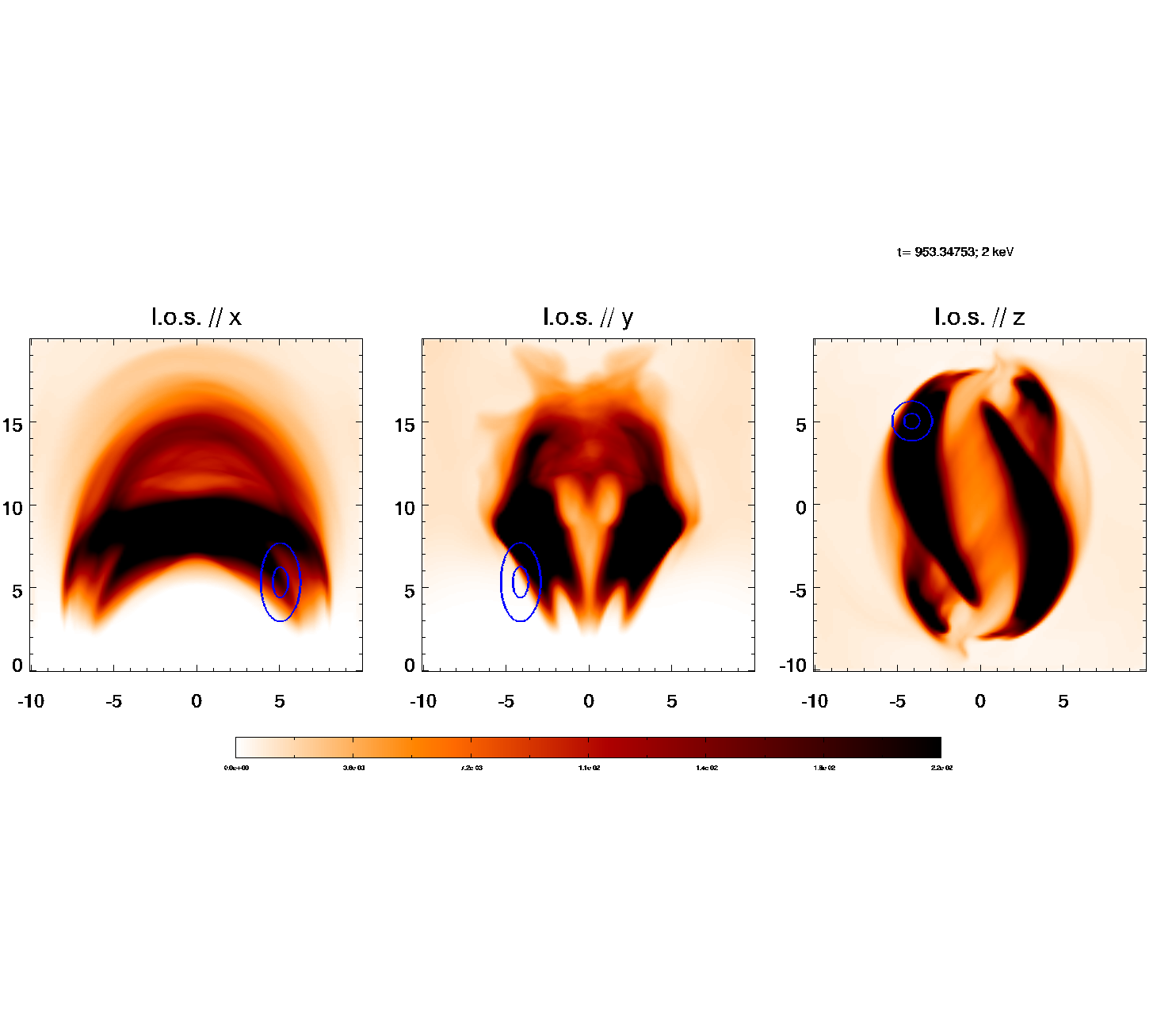} 
  \caption{
    Thermal continuum photon flux emitted at $2\un{keV}$ (colour-scale) and hard X-ray emission at $5$ and $12\un{keV}$ (blue and green contours, respectively, at $90\%$ and $75\%$ of the peak emission at those photon energies) integrated along the line-of-sight in three different directions (from left to right: side, edge and top views), for our reference case (model S).
    The thermal and non-thermal photon fluxes are normalised to arbitrary units for easier comparison.
    The axis are in units of $\un{Mm}$.
    The instants represented are, from top to bottom, $t=34,\ 80,\ 136$ and $227\un{s}$ (as in Fig. \ref{fig:thermal_vol}).
  }
  \label{fig:thermal_los}
\end{figure*}

The first traces of thermal emission appear at mid-height and around the loop legs.
These occur in the form of two separate sheet-like emission structures placed around the boundaries of the loop which are pushed against the background medium as the flux-ropes writhes.
Thermal emission then fills-up the whole flux-rope's volume for a short period of time (at about the maximum of emission).
The total thermal emission fades out afterwards as the loop's plasma cools down globally during the relaxation phase.
At this point, the thermal emission concentrates into a few separate field-aligned threads, most notably at mid-heights and crossing the top part of the loop (see, especially, the top views in Fig. \ref{fig:thermal_vol}).
As the figure shows, thermal emission is only clearly discernible well past the onset of the kink instability, when the flux-rope has already lost an important fraction of its initial twist. 
Therefore, the amount of twist which can be measured by tracing out the emission threads is much lower than the value of the initial magnetic twist.
Both quantities differ by a factor $4$ in our simulations (the emission threads show a twist of about $2\pi$ while the initial magnetic twist is $\sim 8\pi$).
This situation is analogous to that described by \citet{pinto_soft_2015} for a simplified model of a kink unstable coronal loop (without large-scale curvature).
Note that the thermal emission shows a structure with some writhe, which is in the sense of the original field-line twist.
Fig. \ref{fig:thermal_los} represents the thermal photon flux at the same photon energy ($2\un{keV}$) and at the same times as Fig. \ref{fig:thermal_vol}, but integrated along the line-of-sight for three different viewpoints (side, front and top of the loop), with the darker tones representing stronger photon flux.
The contour lines show the locations of the peaks of non-thermal hard X-ray bremsstrahlung (HXR) emission at $5$ and $12\un{keV}$ (blue and green contours, respectively, at $90\%$ and $75\%$ of the peak emission).
The hard X-ray emission spatial profiles are smoothed with a Gaussian filter with $\sim 1\un{Mm}$ width, roughly corresponding to the resolution of modern instruments (we are not trying to reproduce a particular instrument's performance, however).
It is interesting to note that, depending on the line-of-sight (l.o.s), the thermal emission may appear as coming from an unique coronal loop or from two independent sources.
This happens because the emission structures which are aligned with the l.o.s. are much enhanced in respect to the emission structures oriented orthogonally, as the former occupies a shorter depth along the l.o.s than the latter.
The top views (rightmost columns) on Fig. \ref{fig:thermal_vol} and on Fig. \ref{fig:thermal_los} show this effect very clearly.
The emission threads which cross the loop above its apex are well visible in the volume renderings in Fig. \ref{fig:thermal_vol}, but nearly disappear when integrated along a vertical line of sight (Fig. \ref{fig:thermal_los}).
This effect further hinders the possibility of measuring the actual levels of magnetic field-line twist directly from the morphology of the thermal emission threads.

\begin{figure*}
  \centering
  \includegraphics[width=\linewidth]{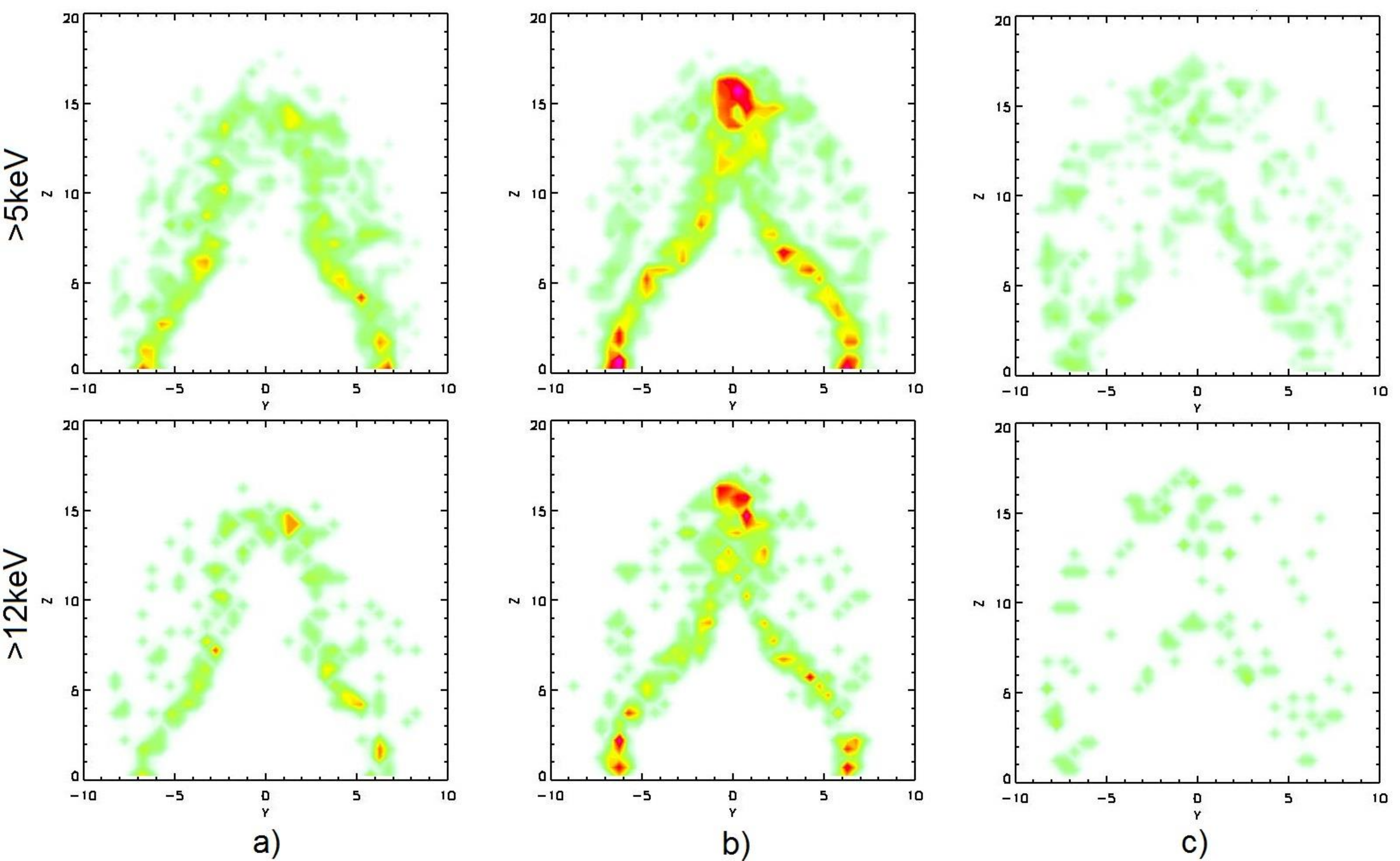}
  \caption{
    Spatial distribution of energetic electrons in the model S seen from the loop side (\emph{cf.} the first column of Fig. \ref{fig:thermal_los}) at different instants.
    Left column (a), middle (b) and right (c) columns correspond to $35\un{s}$, $81\un{s}$, $104\un{s}$ after onset of the kink instability.
    The axis indicate distances in $Mm$.
    Red and green colours indicate high and low electron densities, respectively.
    Upper and lower panels correspond to particles with energies $>5\un{keV}$ and $>12\un{keV}$, respectively. 
}
  \label{fig:edistr}
\end{figure*}


\begin{figure*}
  \centering

  \emph{SXR, HXR light-curves (normalised amplitudes)}\\ \smallskip

  \textsf{\hspace{0.02\linewidth} Model C \hspace{0.25\linewidth} Model S \hspace{0.25\linewidth} Model V} \\

  \includegraphics[width=.31\linewidth,clip=true,trim=22 0 0 22]{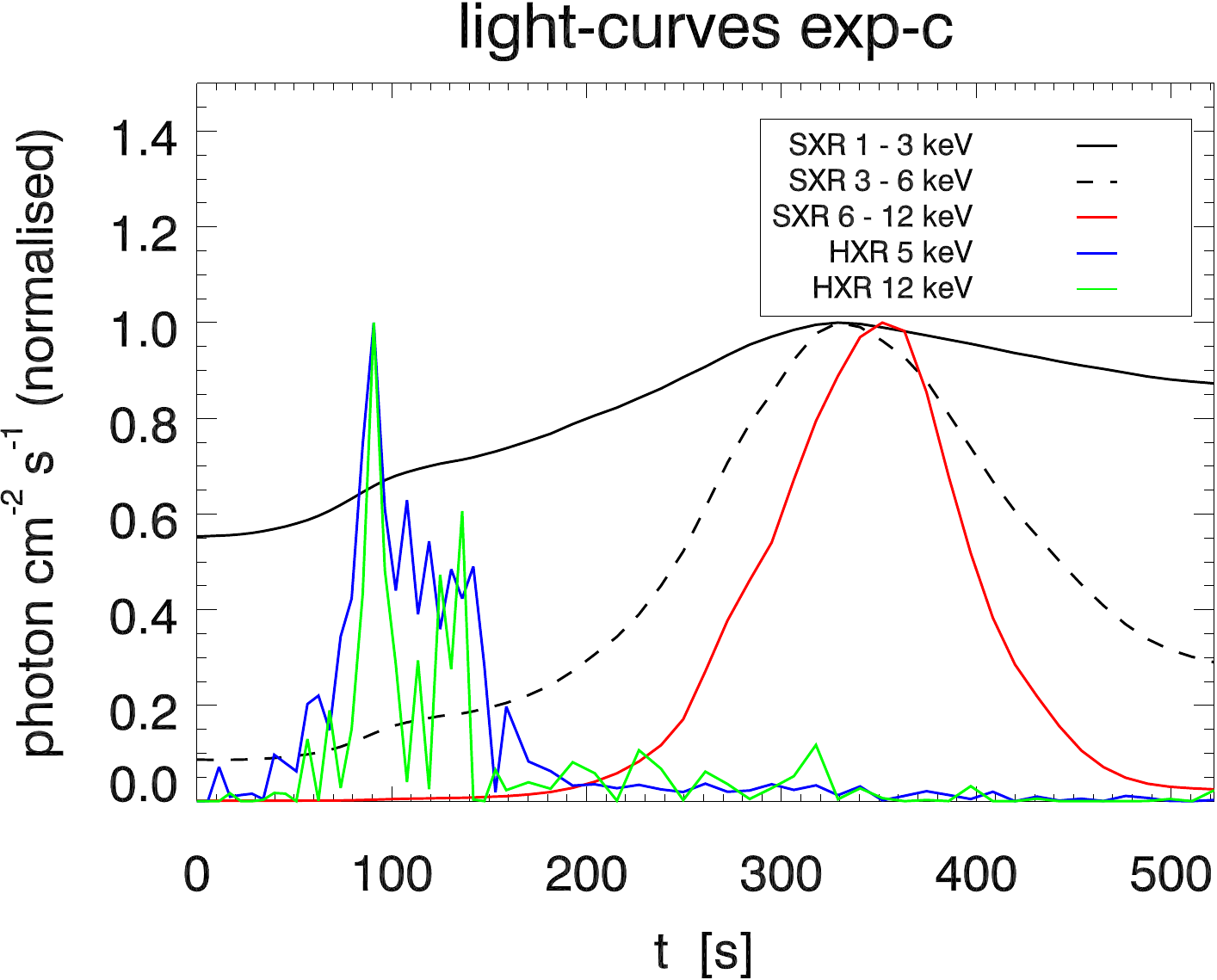}  \hspace{0.02\linewidth}
  \includegraphics[width=.31\linewidth,clip=true,trim=22 0 0 22]{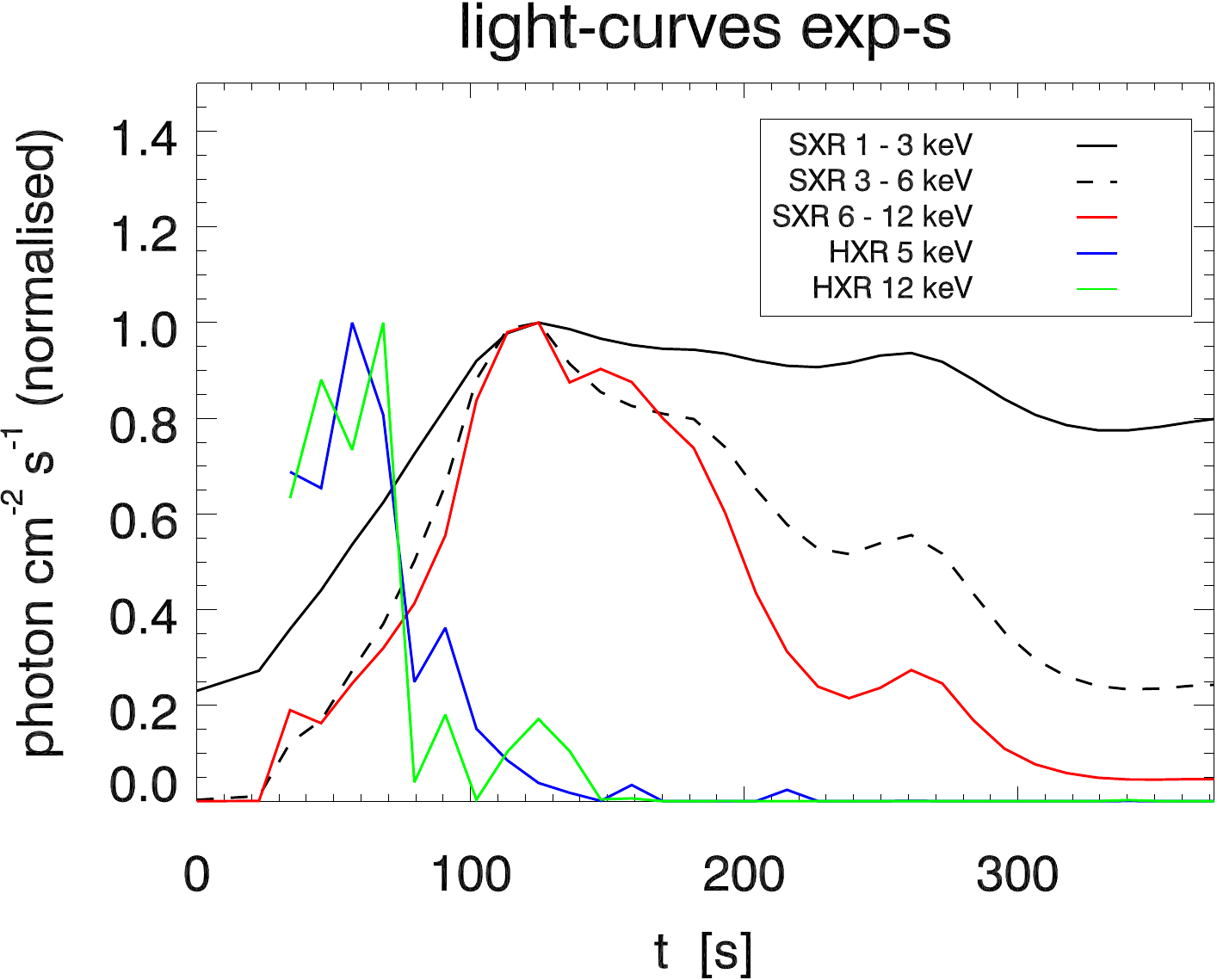}  \hspace{0.02\linewidth}
  \includegraphics[width=.31\linewidth,clip=true,trim=22 0 0 22]{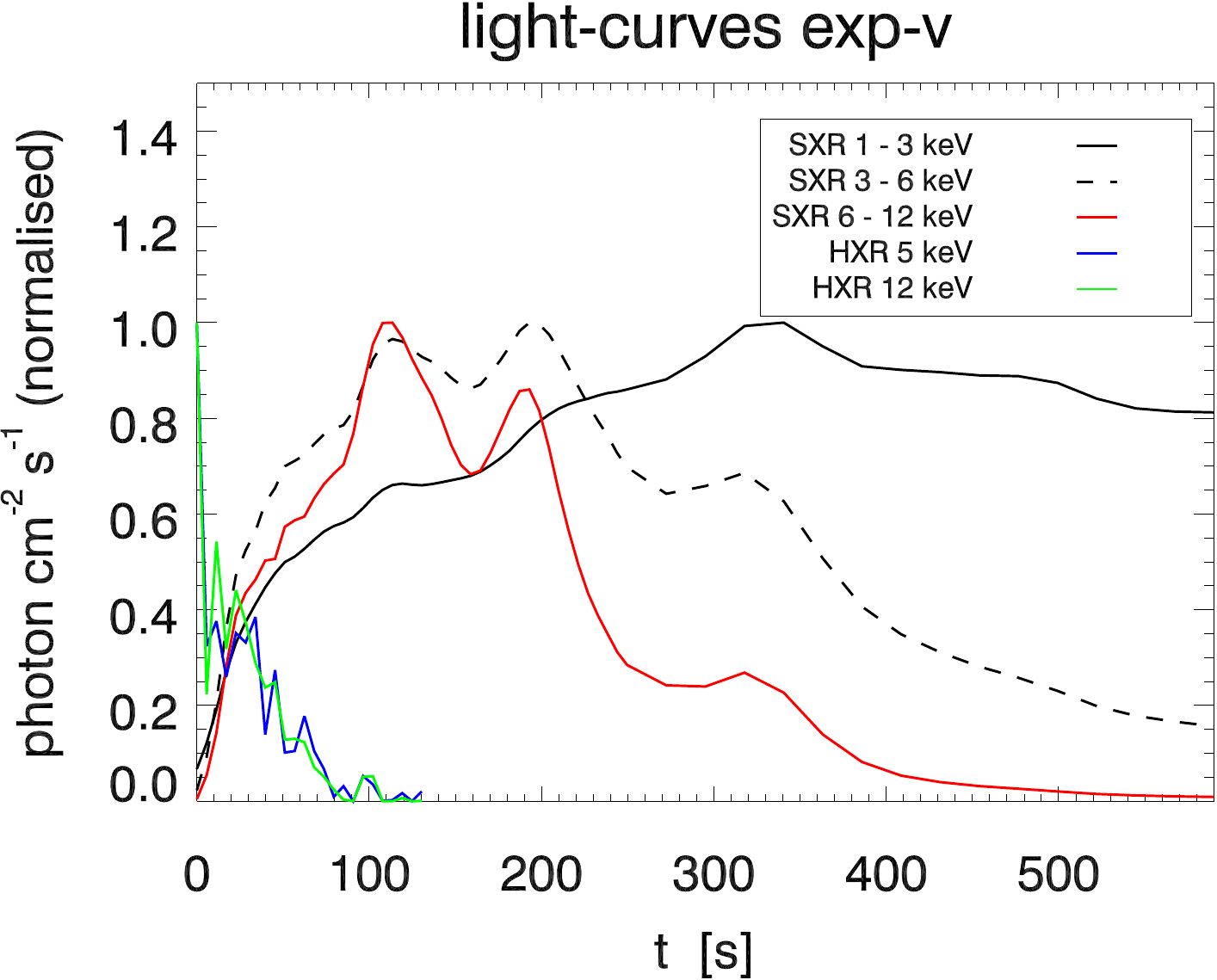} \\ \vspace{3ex}

  \emph{$d/dt$(SXR), HXR (normalised amplitudes)} \\ \smallskip

  \textsf{\hspace{0.02\linewidth} Model C \hspace{0.25\linewidth} Model S \hspace{0.25\linewidth} Model V} \\
  
  \includegraphics[width=.31\linewidth,clip=true,trim=22 0 0 22]{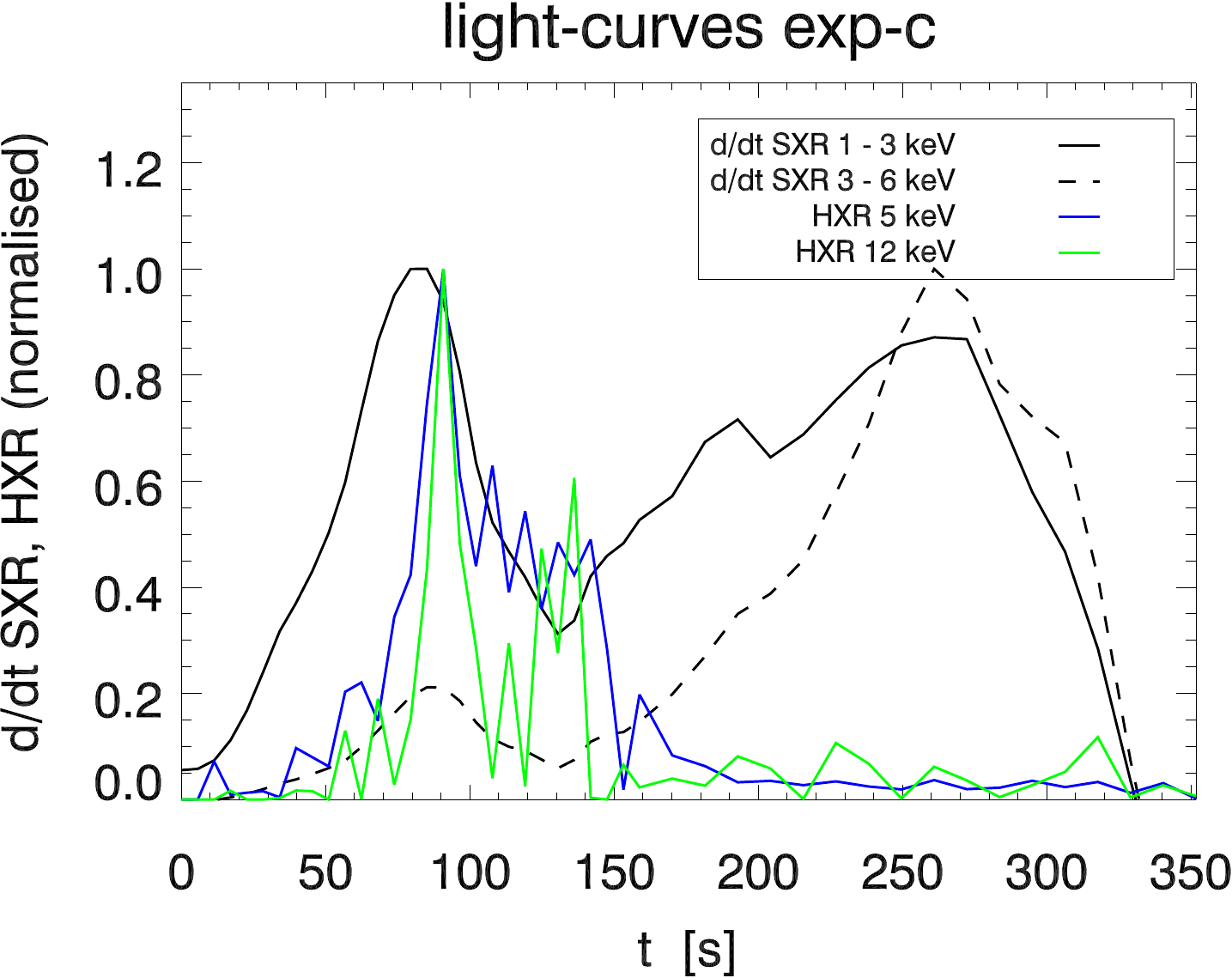}  \hspace{0.02\linewidth}
  \includegraphics[width=.31\linewidth,clip=true,trim=22 0 0 22]{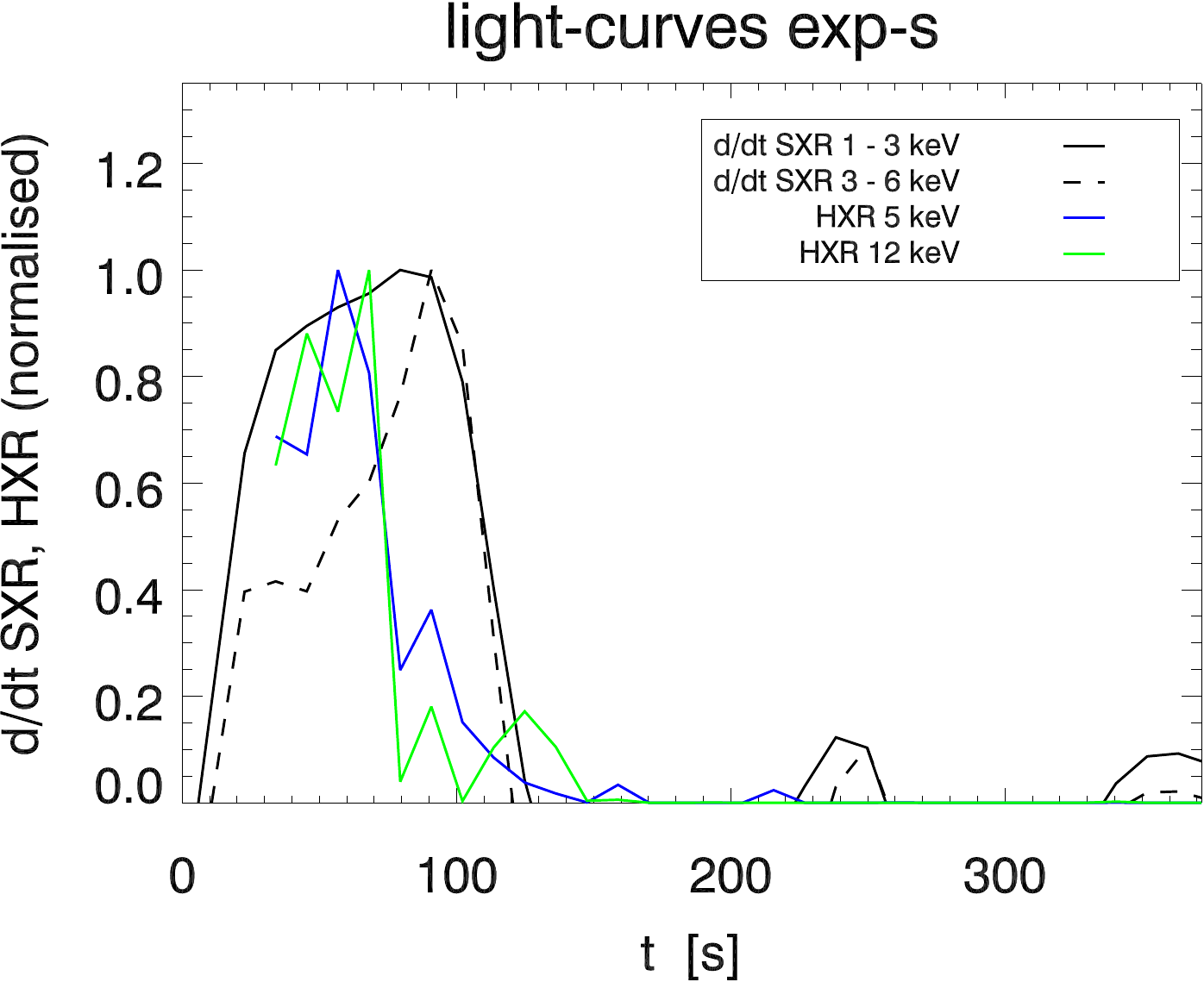}  \hspace{0.02\linewidth}
  \includegraphics[width=.31\linewidth,clip=true,trim=22 0 0 22]{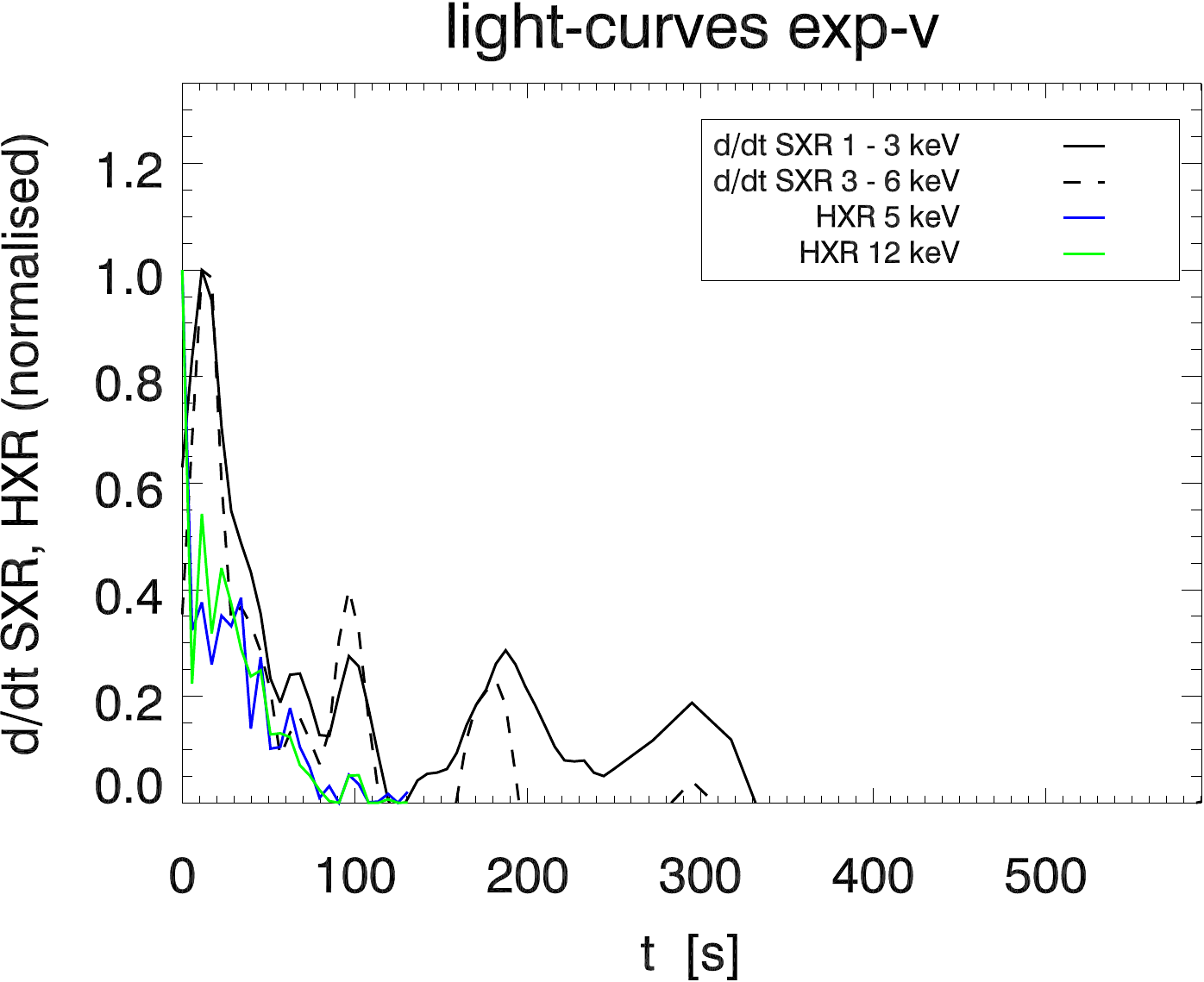}

  \caption{
    The top row shows light-curves of the soft X-ray emission (black and red lines) and hard X-ray emission (blue and green lines) in models C, S and V at different photon energy bands (see the inset legend).
    The bottom row shows the temporal derivative of the soft X-ray lightcurves together with the hard X-ray lightcurves during the first $150\un{s}$ for the same models.
    The curves are all normalised to their peak value for easier representation.
    The peaks in the higher energy SXR light-curves (red lines) tend to precede the lower energy ones (black lines); this especially visible in models C and S.
    The decay is also faster at higher SXR photon energy bands.
    The HXR light-curves (blue and green lines) tend to peak at the moment when the SXR growth rate is at its highest (this is especially visible in model S), such that the temporal derivatives of the SXR curves correlate well with the corresponding HXR curves.
    A few post-flare oscillations are visible in all the SXR light-curves represented
  }
  \label{fig:lightcurves}

\end{figure*}

Generally, the non-thermal electron behaviour we observe here is similar to that described in \citet{m._gordovskyy_particle_2013}. 
However, electrons appear to lose their energies faster. 
Most likely this happens because of the higher density closer to the footpoints.
Thus, in the present model thermal conduction in the initial atmosphere leads to an increase of the density by factor of 2 around  $z\approx 3-5\un{Mm}$ over footpoints and by factor of $\sim 3$ around $z\approx 2-5\un{Mm}$ in horizontal field (see Fig.~\ref{fig:atmos}).
In addition, dense plasma is also accumulated around loop' legs during the twisting (Fig. \ref{fig:rho}).
Similarly, hard X-ray bremsstrahlung emission evolves slightly faster than in previous simulations. 
Although there are plenty of energetic electrons in all the coronal loop, the bulk of the emission is coming from very flat footpoint sources with the height of $1-2 \un{Mm}$ (comparable to the resolution of these intensity maps).
But the situation could, in principle, be different in coronal loop configurations with higher coronal densities (HXR emission peaks could eventually appear at higher coronal heights, as is sometimes observed in solar flares).
There HXR footpoint emission occasionally shows slight asymmetries, but these disappear quite quickly.
On the overall, the HXR emission is essentially similar and simultaneous on both footpoints.
The area of HXR footpoints initially increases with time due to the twisted loop reconnection with ambient field \citep[see][]{gordovskyy_magnetic_2012}.
This increase is not monotonic and, at some point, the footpoint areas begin to shrink.
This happens due to two different factors: numerical undersampling due to the limited test-particle number, and modulation of the HXR intensities. 
Figure \ref{fig:edistr} show the spatial distribution of electrons with energies above $5$ and $12\un{keV}$ at different moments of the temporal evolution of model S (the same as in Fig. \ref{fig:thermal_los}).
Electron distribution functions are calculated by reducing test-particles on a $40\times 40 \times 40 \times 40$ $\left[x,y,z,\epsilon\right]$ grid with uniform spatial step and exponential energy step to reduce undersampling at higher energies.
It can be seen that it is rather similar to the distribution of the parallel electric field in the domain (see Fig. \ref{fig:blines_current}), and hence that energetic electrons are distributed within most of the flaring loop volume, in spite of the HXR emission in Fig. \ref{fig:thermal_los} being much more concentrated in the lower layers.
Note that the HXR emission spatial patterns are not a direct reflection of the electron acceleration mechanisms, but rather of the interaction between the accelerated electrons with the highly stratified coronal plasma.
A substantial amount of energetic particles accumulates, nonetheless, near the loop footpoints. 
This is because the magnetic field convergence reduces the parallel component of the particle velocities.
The magnetic trapping is even more effective due to collisional pitch-angle scattering moving particles out of the loss-cone.
Since all the magnetic field lines of the twisted loop are tied to the lower boundary (i.e., the photosphere), virtually all energetic particles leaving the domain do so through the lower boundary.

Figure \ref{fig:lightcurves} shows the light-curves of the total thermal continuum emission integrated in three photon energy bands between $1$ and $12\un{keV}$ (see inset legend) and of the hard X-ray emission at $5$ and $12\un{keV}$ for models C, S and V.
The light-curves are all normalised to their peak value for easier representation (absolute photon fluxes decrease with increasing photon energy).
The emission in lower energy bands is sustained for longer times (compared to non-thermal emission), as commonly observed in flares.
Photon fluxes at lower energy bands (\emph{e.g}, $1-3\un{keV}$) systematically have a slow rise and decay, while flux at higher energy bands (\emph{e.g}, $6-12\un{keV}$) always displays a more rapid evolution \citep[as in][]{pinto_soft_2015}.
The peak photon fluxes do not occur simultaneously at all photon energy bands: the peaks in higher energy light-curves tends to precede those on lower energy light-curves.
This is particularly noticeable in models S and V.
A few post-flare oscillations are visible in all light-curves represented (hence, in an energy range between $1$ and $25\un{keV}$).
As one would expect, non-thermal emission evolves much faster than thermal X-ray emission. 
In model S, for example, the HXR flux peaks at about $t = 60\un{s}$, already after the saturation of the kink instability and the triggering of the fast reconnection, when the initial large-scale helical current sheet has already started fragmenting.
This is visible by comparison with Figure \ref{fig:blines_current}; the HXR peak occurs just in between the second and the third panels ($t = 30$ and $120 \un{s}$, respectively).
The current density distribution is indeed very much fragmented during this whole time interval, even though its large-scale spatial organisation is a remnant of the initial helical current sheet.
Both $5$ and $12\un{keV}$ bremsstrahlung emission decay thereafter and nearly disappear after $\sim100-200\un{s}$, when thermal emission is still reaching its peak values.

\section{Discussion}
\label{sec:disc}

We study here the consequences of the triggering of the kink instability in twisted coronal loops in terms of X-ray emission (thermal continuum SXR and bremsstrahlung HXR) by means of numerical MHD simulations combined with test-particle methods.
The MHD model consists of a semi-circular kink-unstable coronal loop embedded in a stratified atmosphere composed of coronal and chromospheric layers.
The resulting plasma density and temperature are used to calculate emission measures and thermal continuum emission (soft X-rays), and the electric and magnetic fields are used to compute test-particle trajectories.
Non-thermal (hard X-ray) emission is estimated based on the interaction between these test-particles and the stratified background.
Heating caused by these collisions and currents generated by the accelerated particles are not fed back into the MHD simulation.
That would require a different type of approach based on hybrid methods, which is clearly beyond the scope of this manuscript.

%
The vertical grid resolution -- $\sim 0.16\un{Mm}$ and $\sim 0.04\un{Mm}$ for the "large" and "small" models, respectively -- under-resolves the steep transition region (TR) between the chromosphere and the corona, even though they are at the forefront of current day simulations of the solar corona.
  For comparison, our vertical resolution is similar or higher than the landmark three-dimensional simulations by \citet{gudiksen_ab_2005} and \citet{bingert_intermittent_2011}, respectively $0.15\un{Mm}$ and $0.23\un{Mm}$, and approach the vertical resolution requirement at the TR of $20\un{km}$ suggested by \cite{klimchuk_solving_2006}.
  Under-resolving the TR can have an impact on the properties of the simulated corona (density, temperature and/or heat fluxes), hence possibly affecting both the synthetic SXR and HXR emissions.
  In fact, \citet{bradshaw_influence_2013} have shown that under-resolving the transition region may significantly underestimate the peak density of coronal loops undergoing impulsive heating episodes (and being loaded with plasma by means of chromospheric evaporation), but without much effect on the plasma temperatures they reach.
In spite of these limitations, our results are consistent with the main common features of observed solar flares, and furthermore allow us to clearly establish links between different types of X-ray emission.

  Our models represent confined (non-eruptive) flares in single loops.
  The cases studied here represent smaller -- and also thus more frequently-occurring --  events.
  Larger energy releases can be obtained by considering stronger magnetic fields and larger loops (Gordovskyy et al 2015 submitted). 
  However, whilst the kink instability in a loop may play a role in the largest flares (e.g. by triggering a filament eruption), such events would involve larger-scale and more complex field configurations than considered here.
  Our results provide a prediction of what is expected to be observed in kink-unstable loops, and thus new observational campaigns are required to look for such signatures. 
  Detailed comparison between simulations and observations is a task for future work. 
  However, much work already published provides evidence of events which appear to be consistent with our models. 
  The simulated loops rise somewhat during the main reconnection phase, then contract back down, so may be observed as "failed eruptions", which appear sometimes to be associated with kink instability \citep[\emph{e.g.}][]{alexander_hard_2006,song_temperature_2014}.
  Flares can  be directly associated with apparent instability of observed twisted flux ropes \citep[\emph{e.g}][]{wang_witnessing_2015}.
  \citet{liu_impulsive_2013} analyse a flare within a loop, suggesting that the x-ray emission is generated by fragmented current sheets in the coronal part of the loop. 
  Furthermore, there is evidence for an association of a period of helicity build up prior to flare onset \citep{reinard_evidence_2010}, consistent with our scenario.

Our results suggest that highly twisted coronal loops may be more abundant than what current observations seem to indicate.
In fact, individual magnetic field-lines in flaring coronal loops will only be visible as emission threads if (and only if) the enclosed plasma has already been considerably heated in respect to the background plasma during the course of the flare.
At that point, magnetic field-lines twisted well above the threshold for the kink instability will already have lost an important fraction of their initial twist.
The effect is here even stronger than in the simulations reported by \citet{pinto_soft_2015}, in that part of the emission threads nearly disappear in the line-of-sight projections.
Comparing Figure \ref{fig:thermal_los}, which shows the photon fluxes integrated along the line-of-sight for the three-dimensional emission structures shown in Fig. \ref{fig:thermal_vol}, makes this particularly evident.
The threads which cross the loop above its apex nearly disappear when integrated along a vertical line of sight, for example.
These emission patterns could even be interpreted as coming from a coronal structure with a slight large-scale torsion but with no twist at all.
This result removes one of the main obstacles faced by the kink instability scenario for the theoretical understanding of the triggering and evolution of solar flares.

The temporal ordering of the soft and hard X-ray light-curves we obtained are consistent with those typically observed.
The light-curves in Fig. \ref{fig:lightcurves} show similarities with GOES and RHESSI light-curves, such as those reported by \citet{sylwester_solar_2014} for an M1.0 class flare.
Lower energy thermal (SXR) light-curves grow and decay more slowly than higher energy thermal light-curves \citep[in a way similar to that in]{pinto_soft_2015}.
Furthermore, the hard X-ray (HXR) light-curves due to the bremsstrahlung of non-thermal electrons reach their peak value when the growth-rate of the SXR curves are maximal.
This behaviour is consistent across all the models we tested (C, S, V and Y), and approximately reproduces the so-called Neupert effect.
The most commonly accepted interpretation of this effect invokes heating of chromospheric plasma by the supra-thermal electron beams which produce HXR emission, followed by the up-flow of hot plasma into the coronal loops and subsequent SXR emission.
However, our simulations suggest a different cause for this effect, as the electron beams we consider are not allowed to heat up the background plasma by construction (the supra-thermal electrons are test-particles).
Here, the instant at which a given HXR light-curve peaks is very close to the time when the current sheet that forms around the twisted coronal loop reaches its maximum amplitude and spatial extent (\emph{i.e}, just before the main magnetic reconnection episode stars).
Particle acceleration is maximal at this instant, and so are the flux of particle reaching the dense lower layers of the atmosphere and the consequent HXR photon flux.
Then, actual plasma heating leading to SXR emission keeps occurring afterwards during the relaxation phase (even though the ohmic heating is proportional to $\eta j^2$, and hence maximal at that instant).
This result is in agreement with some of the conclusions of \citet{veronig_physics_2005}, namely that fast electron beams may not be the main source of SXR plasma supply and heating, and that other heating sources are necessary to explain Neupert effect quantitatively.

The ensemble of results presented in this manuscript support the view that plasma heating and particle acceleration processes in solar flares may occur in large volumes of the respective flaring loops rather than in a small volumes located near their apexes.
The flare scenario based on the triggering of the kink-instability of coronal loops with magnetic helicity (twist) naturally provides the physical mechanisms necessary for volume-distributed heating and acceleration.

\section{Summary}\label{summa}

The aim of this study is to determine observational manifestations of solar flares occurring in kink-unstable twisted coronal loops.
Specifically, we focus on thermal soft X-ray (SXR) continuum emission together with non-thermal hard X-ray (HXR) bremsstrahlung emission deduced from a set of MHD simulations.
The thermal SXR emission is calculated from the distributions of density and temperature of the compressible plasma, and the HXR emission is derived from the interaction between test-particles and the time-evolving magneto-hydrodynamical stratified background.

Our results show that the geometry of the emission produced during flares does not necessarily correspond to that of the underlying magnetic field.
In particular, the twist angles revealed by the emission threads (soft X-ray thermal emission; SXR) are consistently lower than the actual field-line twist present at the onset of the kink-instability.
Hard X-ray (HXR) emission due to the collisions of energetic electrons with the stratified background are concentrated at the loop foot-points (quasi-simultaneously), event though the electrons are accelerated everywhere within the coronal volume of the loop.
The maximum of HXR emission consistently precedes that of SXR emission
The HXR light-curve being approximately proportional to the temporal derivative of the SXR light-curve, hence reproducing the so-called Neupert effect.

Future work will cover other aspects of thermal emission in flares such as thermal emission in the extreme ultraviolet energy bands (continuum and line emission), and also properties of micro-wave gyro-synchrotron emission in flares triggered the kink instability.

\begin{acknowledgements}
We thank the anonymous referee, whose insightful remarks helped us improving this manuscript considerably.
R. F. P. was funded by the French National Space Centre (CNES) and by the FP7 project \#606692 (HELCATS).
M. G. and P. K. B. are funded by the Science and Technology Facilities Council (UK). 
Numerical simulations have been performed using UKMHD computational facilities (part of the STFC DiRAC HPC).
\end{acknowledgements}


\bibliographystyle{aa}
\bibliography{/data/rpinto/BIBTEX/refs,temp}

\begin{thebibliography}{44}
\expandafter\ifx\csname natexlab\endcsname\relax\def\natexlab#1{#1}\fi

\bibitem[{Alexander {et~al.}(2006)Alexander, Liu, \&
  Gilbert}]{alexander_hard_2006}
Alexander, D., Liu, R., \& Gilbert, H.~R. 2006, ApJ, 653,
  719

\bibitem[{Arber {et~al.}(2001)Arber, Longbottom, Gerrard, \&
  Milne}]{arber_staggered_2001}
Arber, T.~D., Longbottom, A.~W., Gerrard, C.~L., \& Milne, A.~M. 2001, J. Comput. Phys., 171, 151

\bibitem[{Archontis {et~al.}(2013)Archontis, Hood, \&
  Tsinganos}]{archontis_emergence_2013}
Archontis, V., Hood, A.~W., \& Tsinganos, K. 2013, ApJ,
  778, 42

\bibitem[{Aulanier {et~al.}(2005)Aulanier, D{\'e}moulin, \&
  Grappin}]{aulanier_equilibrium_2005}
Aulanier, G., D{\'e}moulin, P., \& Grappin, R. 2005, A\&A, 430, 1067

\bibitem[{Bareford {et~al.}(2013)Bareford, Hood, \&
  Browning}]{bareford_coronal_2013}
Bareford, M.~R., Hood, A.~W., \& Browning, P.~K. 2013, A\&A, 550, 40

\bibitem[{Battaglia \& Kontar(2012)}]{battaglia_rhessi_2012}
Battaglia, M. \& Kontar, E.~P. 2012, ApJ, 760, 142

\bibitem[{Benz(2008)}]{benz_flare_2008}
Benz, A.~O. 2008, Liv. Rev. Sol. Phys., 5, 1

\bibitem[{Bingert \& Peter(2011)}]{bingert_intermittent_2011}
Bingert, S. \& Peter, H. 2011, A\&A, 530, A112

\bibitem[{Botha {et~al.}(2012)Botha, Arber, \&
  Srivastava}]{botha_observational_2012}
Botha, G. J.~J., Arber, T.~D., \& Srivastava, A.~K. 2012, ApJ, 745, 53

\bibitem[{Bradshaw \& Cargill(2013)}]{bradshaw_influence_2013}
Bradshaw, S.~J. \& Cargill, P.~J. 2013, ApJ, 770, 12

\bibitem[{Brown {et~al.}(2003)Brown, Nightingale, Alexander, Schrijver,
  Metcalf, Shine, Title, \& Wolfson}]{brown_observations_2003}
Brown, D.~S., Nightingale, R.~W., Alexander, D., {et~al.} 2003, Sol. Phys.,
  216, 79

\bibitem[{Brown(1976)}]{brown_interpretation_1976}
Brown, J.~C. 1976, Royal Society of London Philosophical Transactions Series A,
  281, 473

\bibitem[{Brown {et~al.}(2009)Brown, Turkmani, Kontar, MacKinnon, \&
  Vlahos}]{brown_local_2009}
Brown, J.~C., Turkmani, R., Kontar, E.~P., MacKinnon, A.~L., \& Vlahos, L.
  2009, A\&A, 508, 993

\bibitem[{Browning {et~al.}(2008)Browning, Gerrard, Hood, Kevis, \& van~der
  Linden}]{browning_heating_2008}
Browning, P.~K., Gerrard, C., Hood, A.~W., Kevis, R., \& van~der Linden, R.
  A.~M. 2008, A\&A, 485, 837

\bibitem[{Cargill {et~al.}(2006)Cargill, Vlahos, Turkmani, Galsgaard, \&
  Isliker}]{cargill_particle_2006}
Cargill, P.~J., Vlahos, L., Turkmani, R., Galsgaard, K., \& Isliker, H. 2006,
  S. S. Rev., 124, 249

\bibitem[{Dalmasse {et~al.}(2014)Dalmasse, Pariat, D{\'e}moulin, \&
  Aulanier}]{dalmasse_photospheric_2014}
Dalmasse, K., Pariat, E., D{\'e}moulin, P., \& Aulanier, G. 2014, Sol. Phys., 289, 107

\bibitem[{Emslie(1978)}]{emslie_collisional_1978}
Emslie, A.~G. 1978, ApJ, 224, 241

\bibitem[{Fletcher {et~al.}(2011)Fletcher, Dennis, Hudson, Krucker, Phillips,
  Veronig, Battaglia, Bone, Caspi, Chen, Gallagher, Grigis, Ji, Liu, Milligan,
  \& Temmer}]{fletcher_observational_2011}
Fletcher, L., Dennis, B.~R., Hudson, H.~S., {et~al.} 2011, Space Science Reviews, 159, 19

\bibitem[{Gordovskyy {et~al.}(2013)Gordovskyy, Browning, Kontar, \&
  Bian}]{m._gordovskyy_particle_2013}
Gordovskyy, M., Browning, P., Kontar, E., \& Bian, N. 2013, A\&A

\bibitem[{Gordovskyy \& Browning(2011)}]{gordovskyy_particle_2011}
Gordovskyy, M. \& Browning, P.~K. 2011, ApJ, 729, 101

\bibitem[{Gordovskyy \& Browning(2012)}]{gordovskyy_magnetic_2012}
Gordovskyy, M. \& Browning, P.~K. 2012, Sol. Phys., 277, 299

\bibitem[{Gordovskyy {et~al.}(2012)Gordovskyy, Browning, Kontar, \&
  Bian}]{gordovskyy_effect_2012}
Gordovskyy, M., Browning, P.~K., Kontar, E.~P., \& Bian, N.~H. 2012, Sol. Phys.

\bibitem[{Gordovskyy {et~al.}(2014)Gordovskyy, Browning, Kontar, \&
  Bian}]{gordovskyy_particle_2014}
Gordovskyy, M., Browning, P.~K., Kontar, E.~P., \& Bian, N.~H. 2014, A\&A, 561, 72

\bibitem[{Gordovskyy {et~al.}(2010)Gordovskyy, Browning, \&
  Vekstein}]{gordovskyy_particle_2010-1}
Gordovskyy, M., Browning, P.~K., \& Vekstein, G.~E. 2010, A\&A, 519, 21

\bibitem[{Gudiksen \& Nordlund(2005)}]{gudiksen_ab_2005}
Gudiksen, B.~V. \& Nordlund, {\r A}. 2005, Astrophysical Journal, 618, 1031

\bibitem[{Hood {et~al.}(2009)Hood, Browning, \& Linden}]{hood_coronal_2009}
Hood, A.~W., Browning, P.~K., \& Linden, R. A. M. V.~d. 2009, A\&A, 506, 13 pages

\bibitem[{Klimchuk(2006)}]{klimchuk_solving_2006}
Klimchuk, J.~A. 2006, Sol. Phys., 234, 41

\bibitem[{Klimchuk {et~al.}(2008)Klimchuk, Patsourakos, \&
  Cargill}]{klimchuk_highly_2008}
Klimchuk, J.~A., Patsourakos, S., \& Cargill, P.~J. 2008, ApJ, 682, 1351

\bibitem[{Kontar {et~al.}(2002)Kontar, Brown, \&
  McArthur}]{kontar_nonuniform_2002}
Kontar, E.~P., Brown, J.~C., \& McArthur, G.~K. 2002, Sol. Phys., 210, 419

\bibitem[{Landau(1937)}]{landau_statistical_1937}
Landau, L.~D. 1937, Zh. Eksp. Teor. Fiz., 819

\bibitem[{Liu {et~al.}(2013)Liu, Li, \& Fletcher}]{liu_impulsive_2013}
Liu, S., Li, Y., \& Fletcher, L. 2013, ApJ, 769, 135

\bibitem[{Luoni {et~al.}(2011)Luoni, D{\'e}moulin, Mandrini, \& van
  Driel-Gesztelyi}]{luoni_twisted_2011}
Luoni, M.~L., D{\'e}moulin, P., Mandrini, C.~H., \& van Driel-Gesztelyi, L.
  2011, Sol. Phys., 270, 45

\bibitem[{Pinto \& Brun(2013)}]{pinto_flux_2013}
Pinto, R.~F. \& Brun, A.~S. 2013, ApJ, 772, 55

\bibitem[{Pinto {et~al.}(2015)Pinto, Vilmer, \& Brun}]{pinto_soft_2015}
Pinto, R.~F., Vilmer, N., \& Brun, A.~S. 2015, A\&A, 576,
  A37

\bibitem[{Reale {et~al.}(2009)Reale, Testa, Klimchuk, \&
  Parenti}]{reale_evidence_2009}
Reale, F., Testa, P., Klimchuk, J.~A., \& Parenti, S. 2009, ApJ, 698, 756

\bibitem[{Reinard {et~al.}(2010)Reinard, Henthorn, Komm, \&
  Hill}]{reinard_evidence_2010}
Reinard, A.~A., Henthorn, J., Komm, R., \& Hill, F. 2010, ApJL, 710, L121

\bibitem[{Song {et~al.}(2014)Song, Zhang, Cheng, Chen, Liu, Wang, \&
  Li}]{song_temperature_2014}
Song, H.~Q., Zhang, J., Cheng, X., {et~al.} 2014, ApJ,
  784, 48

\bibitem[{Srivastava {et~al.}(2010)Srivastava, Zaqarashvili, Kumar, \&
  Khodachenko}]{srivastava_observation_2010}
Srivastava, A.~K., Zaqarashvili, T.~V., Kumar, P., \& Khodachenko, M.~L. 2010,
  ApJ, 715, 292

\bibitem[{Sylwester {et~al.}(2014)Sylwester, Sylwester, Phillips, Kepa, \&
  Mrozek}]{sylwester_solar_2014}
Sylwester, B., Sylwester, J., Phillips, K. J.~H., Kepa, A., \& Mrozek, T. 2014,
  ApJ, 787, 122

\bibitem[{T{\"o}r{\"o}k \& Kliem(2003)}]{torok_evolution_2003}
T{\"o}r{\"o}k, T. \& Kliem, B. 2003, A\&A, 406, 1043

\bibitem[{Tucker(1975)}]{tucker_radiation_1975}
Tucker, W.~H. 1975, Radiation {Processes} in {Astrophysics} (MIT Press)

\bibitem[{Turkmani {et~al.}(2006)Turkmani, Cargill, Galsgaard, Vlahos, \&
  Isliker}]{turkmani_particle_2006}
Turkmani, R., Cargill, P.~J., Galsgaard, K., Vlahos, L., \& Isliker, H. 2006,
  A\&A, 449, 749

\bibitem[{Veronig {et~al.}(2005)Veronig, Brown, Dennis, Schwartz, Sui, \&
  Tolbert}]{veronig_physics_2005}
Veronig, A.~M., Brown, J.~C., Dennis, B.~R., {et~al.} 2005, ApJ, 621, 482

\bibitem[{Wang {et~al.}(2015)Wang, Cao, Liu, Xu, Liu, Zeng, Chae, \&
  Ji}]{wang_witnessing_2015}
Wang, H., Cao, W., Liu, C., {et~al.} 2015, Nature Comm., 6, 7008

\end{thebibliography}

\end{document}